\documentclass[aps,draft,showpacs,preprintnumbers,amsmath,amssymb,nofootinbib]{revtex4}
\usepackage[final]{graphics}
\usepackage{amsmath}
\usepackage{amsfonts,amsbsy}
\usepackage{amssymb}
\usepackage{color}
\usepackage[colorlinks=true]{hyperref}

\usepackage[utf8]{inputenc} 

\usepackage{mathtools}
\usepackage{accents}

\def\lsim{ \,\, \vcenter{\hbox{$\buildrel{\displaystyle <}\over\sim$}}
 \,\,}
\def\be{\begin{equation}}
\def\ee{\end{equation}}
\def\bea{\begin{eqnarray}}
\def\eea{\end{eqnarray}}

\def\beq{\begin{equation}}
\def\eeq{\end{equation}}

\usepackage{graphics,graphicx}
\usepackage{dcolumn}
\usepackage{bm}
\usepackage{epsfig}
\usepackage{mathbbol}
\usepackage{epstopdf}
\usepackage{simplewick}
\usepackage[utf8]{inputenc} 
\usepackage{slashed}
\usepackage{stackrel}

\def\eq#1{{Eq.~(\ref{#1})}}
\def\fig#1{{Fig.~\ref{#1}}}
\newcommand{\ben}{\begin{eqnarray*}}
\newcommand{\een}{\end{eqnarray*}}
\newcommand{\un}[1]{\underline{#1}}

\newcommand{\pd}{\partial}

\newcommand{\as}{\alpha_s}

\newcommand{\dhd}{{\textstyle d}
\lower.03ex\hbox{\kern-0.38em$^{\scriptstyle-}$}\kern-0.05em{}}
\newcommand{\dbar}{{\textstyle \delta}
\lower.03ex\hbox{\kern-0.38em$^{\scriptstyle-}$}\kern-0.05em{}}

\newcommand{\myref}{}
\newcommand{\myrefs}{}

\begin{document}

\title{ When gluons go odd: how classical gluon fields generate
  odd azimuthal harmonics for the two-gluon correlation function in
  high-energy collisions }

\author{Yuri~V.~Kovchegov}
\affiliation{Department of Physics, The Ohio State University,
  Columbus, OH 43210, USA}

\author{Vladimir~V.~Skokov}
\affiliation{RIKEN/BNL, Brookhaven National Laboratory, Upton, NY 11973}

\begin{abstract}
  We show that, in the saturation/Color Glass Condensate framework,
  odd azimuthal harmonics of the two-gluon correlation function with a
  long-range separation in rapidity are generated by the higher-order
  saturation corrections in the interactions with the projectile and
  the target. At the very least, the odd harmonics require three
  scatterings in the projectile and three scatterings in the
  target. We derive the leading-order expression for the two-gluon
  production cross section which generates odd harmonics: the
  expression includes all-order interactions with the target and three
  interactions with the projectile. We evaluate the obtained
  expression both analytically and numerically, confirming that the
  odd-harmonics contribution to the two-gluon production in the
  saturation framework is non-zero.
\end{abstract}

\date{\today} 


\maketitle


\section{Introduction}

Over the past decade long-range rapidity correlations between the
produced hadrons were observed in heavy ion (AA) and high-multiplicity
proton-nucleus (pA) and proton-proton (pp) collisions at RHIC
\cite{Adams:2005ph,Adare:2008cqb,Alver:2009id,Abelev:2009af} and LHC
\cite{Khachatryan:2010gv,CMS:2012qk,Abelev:2012aa,Chatrchyan:2013nka}. The
novel di-hadron correlations enhance the $\Delta \phi =0$ and $\Delta
\phi = \pi$ regions of the azimuthal opening angle between the hadrons
and stretch over several units in the rapidity interval $\Delta \eta$
between the hadrons. Finding an explanation of these previously
unobserved correlations is important for the understanding of the
strong interactions dynamics in high energy hadronic and nuclear
collisions.

Since the long-range rapidity correlations were first discovered in
heavy ion collisions, it is natural to ascribe their origin to the
dynamics of quark-gluon plasma (QGP) produced in such collisions. A
vigorous activity is under way to account for the long-range rapidity
di-hadron correlations within hydrodynamic models of
QGP~\cite{Bzdak:2013zma,Bozek:2011if,Shuryak:2013ke,Bozek:2013uha,Shen:2016zpp,Weller:2017tsr,Song:2017wtw,Zhao:2017rgg}. However,
these approaches suffer from one conceptual problem: long-range
rapidity correlations cannot originate at later proper times in the
collision, when plasma is produced. There exists a causality argument
\cite{Dumitru:2008wn} demonstrating that the later the proper time of
the interaction, the narrower in $\Delta \eta$ the corresponding
correlation will be. Therefore, hydrodynamics, being applicable at a
relatively late times, is not likely to generate these long-range
rapidity correlations: rather, the rapidity correlations have to be
included into the initial conditions for hydrodynamic evolution. Thus,
the question about the origin of the long-range rapidity correlations
remains, with the initial-state early-proper-time dynamics being the
most probable suspect.

Long-range rapidity correlations were first advocated in the
saturation/Color Glass Condensate (CGC) framework in
\myref\cite{Dumitru:2008wn} (see also
\myref\cite{Kovchegov:1999ep}). (See
\cite{Gribov:1984tu,Iancu:2003xm,Jalilian-Marian:2005jf,Weigert:2005us,Gelis:2010nm,Albacete:2014fwa,KovchegovLevin}
for reviews of saturation/CGC physics.) The weakly-coupled CGC
dynamics can generate long-range correlations which at large rapidity
intervals $\Delta \eta$ are independent of $\Delta \eta$, in
qualitative agreement with the experimentally observed
correlations. Efforts to obtain similar correlation function behavior
originating in a different initial-state dynamics scenario so far came
up short: at strong coupling, calculations employing the anti-de
Sitter/Conformal Field Theory (AdS/CFT) correspondence
\cite{Maldacena:1997re,Witten:1998zw} so far give a correlation
function that grows with $\Delta \eta$ \cite{Grigoryan:2010pz}, in
disagreement with the experiment.

The ridge correlation may result from the CGC dynamics giving
long-range correlations in rapidity, with the subsequent hydrodynamic
evolution generating the azimuthal collimation observed in the data
\cite{Gavin:2008ev}. This explanation of the ridge phenomenon requires
a thermal medium to be generated in high-multiplicity pp and pA
collisions. Thermalization of the medium produced in heavy ion
collisions requires multiple interactions between the quarks and
gluons: it is natural to assume that thermalization in
high-multiplicity pp and pA collisions, if it does take place, would
require similar multi-parton interactions. Perhaps a more conservative
way of generating the ridge correlation is due to the so-called
``glasma graphs" proposed in the saturation/CGC framework in
\myref\cite{Dumitru:2008wn}: they require only a double interaction in
both the projectile and the target. The ``glasma graphs" generate both
the long-range correlation in rapidity
\cite{Armesto:2006bv,Armesto:2007ia,Dumitru:2008wn,Gavin:2008ev,Gelis:2008sz,Dusling:2009ni,Dumitru:2010iy,Dumitru:2010mv,Kovner:2010xk,Kovner:2011pe,Kovner:2012jm,Dusling:2012cg,Dusling:2012wy,Dusling:2012iga},
along with the near- and away-side ridge correlations
\cite{Kovchegov:2012nd,Kovchegov:2013ewa}. Since the ``glasma graphs"
had been originally put forward in \myref\cite{Dumitru:2008wn}, the
corresponding two-gluon production cross section was improved by
including multiple interactions (saturation effects) with one of the
nuclei (the target) in \myrefs\cite{Kovner:2012jm,Kovchegov:2012nd}. As
usual in the saturation physics, the effect of extra rescatterings
appears to be mainly in screening the infrared (IR) divergences
\cite{Kovchegov:2013ewa}. In addition, one can show that multiple
rescatterings violate the $k_T$-factorization formula one could
conjecture for two-gluon production based on the ``glasma graphs''
alone \cite{Kovchegov:2013ewa}.

The approach based on the ``glasma graphs'' with the infrared
screening provided by the saturation scale $Q_s$ enjoyed successful
phenomenology
\cite{Dusling:2012cg,Dusling:2012wy,Dusling:2012iga}. However, the
two-gluon production cross section given by the ``glasma graphs''
\cite{Dumitru:2008wn} along with the saturation corrections in the
target hadron/nucleus \cite{Kovner:2012jm,Kovchegov:2012nd} turned out
to be invariant under ${\un k}_1 \leftrightarrow {\un k}_2$
interchange and, also, is separately invariant under ${\un k}_1 \to -
{\un k}_1$ or ${\un k}_2 \to - {\un k}_2$ replacements. Here ${\un
  k}_1$ and ${\un k}_2$ are the transverse momenta of the two produced
gluons. The result of this symmetry is that the corresponding
gluon-gluon correlation function only contains even azimuthal
harmonics $v_{2n}^2\{2\}$, with all the odd harmonics
$v_{2n+1}^2\{2\}$ being zero. At the same time, odd azimuthal
harmonics have been observed in the di-hadron correlators measured at
RHIC and at
LHC~\cite{Abelev:2012ola,Chatrchyan:2013nka,Aad:2014lta}. Odd
harmonics are somewhat smaller than the even ones, but are
non-zero. This experimental result presented a conundrum for the
saturation community: can the saturation dynamics account for the
observed long-range rapidity correlations with the non-zero odd
azimuthal harmonics?

To resolve this ambiguity, several observations have been put
forward. In \myref\cite{Kovner:2016jfp} the authors observed that the
symmetry of the di-gluon correlator under ${\un k}_1 \leftrightarrow
{\un k}_2$, ${\un k}_1 \to - {\un k}_1$ and ${\un k}_2 \to - {\un
  k}_2$ is ``accidental'', and is not required by the symmetries in
the problem. They have then argued that odd harmonics may arise if one
includes saturation effects in the wave function of the projectile:
thus to find odd harmonics one needs to augment the existing
calculations of the two-gluon production cross section
\cite{Kovner:2012jm,Kovchegov:2012nd}, in which the saturation effects
were only included in the interaction with the target. The idea of
saturation corrections in the projectile being responsible for the odd
harmonics was developed in \myref\cite{McLerran:2016snu}, where it was
shown that such corrections indeed have the potential to generate odd
harmonics by violating the ${\un k}_1 \to - {\un k}_1$ and ${\un k}_2
\to - {\un k}_2$ symmetries of the two-gluon correlation function. In
addition, a numerical simulation of the classical gluon fields
produced in heavy ion collisions in the McLerran--Venugopalan (MV)
model \cite{McLerran:1994vd,McLerran:1993ka,McLerran:1993ni} appears
to generate odd harmonics as well \cite{Lappi:2009xa,Schenke:2015aqa}:
since the difference between this numerical result for inclusive
two-gluon production and the expressions obtained in
\myrefs\cite{Kovner:2012jm,Kovchegov:2012nd} is due to the saturation
effects in the projectile, and since the calculation in
\myref\cite{Kovner:2012jm,Kovchegov:2012nd} only gave even harmonics,
it is natural to conclude that the odd harmonics likely originate in
the higher-order projectile interactions.

Our goal in the present paper is to construct a complete expression
for the part of the two-gluon inclusive production cross section
responsible for the odd harmonics by calculating the first saturation
correction in the interactions with the projectile. Our goal is
complicated by the fact that even for the single inclusive gluon
production cross section the first saturation correction in the
projectile has not been found yet. However, partial results exist in
\myrefs\cite{Balitsky:2004rr,Chirilli:2015tea}. Let us first consider
the single inclusive gluon production cross section. Suppressing the
transverse momentum dependence, the cross section in the classical MV
model can be written as (in the MV model power counting, using the
approach to it from \cite{Kovchegov:1996ty,Kovchegov:1997pc})
 \begin{align}
   \frac{d\sigma}{d^2 k \, d^2 b \, d^2 B} = \left\langle
     \frac{d\sigma}{d^2 k \, d^2 b \, d^2 B} (\rho_p, \rho_T)
   \right\rangle_{\rho_p, \rho_T} = \frac{1}{\as} \, f \left( \as^2 \,
     A_1^{1/3} , \as^2 \, A_2^{1/3} \right)\, ,
 \end{align}
 where $A_1$ and $A_2$ are the atomic numbers of the projectile and
 target nuclei respectively, $\as$ is the strong coupling constant,
 $B$ is the impact parameter between the nuclei, $b$ is the transverse
 position of the gluon, and the angle brackets denote averaging in the
 wave functions of the projectile and target nuclei, which is
 equivalent in the MV model to averaging over their color charge
 densities $\rho_p$ and $\rho_T$
 \cite{McLerran:1994vd,McLerran:1993ka,McLerran:1993ni,Kovchegov:1996ty,Kovchegov:1997pc}. The
 function $f$ was only studied numerically
 \cite{Krasnitz:1999wc,Krasnitz:2003jw,Lappi:2003bi,Blaizot:2010kh}. Analytically
 we only know its expansion in either one of its arguments. Assuming
 that the projectile is a dilute object with $\as^2 \, A_1^{1/3} \lsim
 1$ one can expand the cross section in this parameter
\begin{align}
  \frac{d\sigma}{d^2 k \, d^2 b \, d^2 B} = \frac{1}{\as} \, \left[
    \as^2 \, A_1^{1/3} \ f_1 \! \left( \as^2 \, A_2^{1/3} \right) +
    \left( \as^2 \, A_1^{1/3} \right)^2 \ f_2 \! \left( \as^2 \,
      A_2^{1/3} \right) + \ldots \right].
\end{align}
The function $f_1$ is known analytically from the gluon production
cross section in the proton--nucleus (pA) collisions
\cite{Kovchegov:1998bi,Dumitru:2001ux}, since it comes in with the
term involving only one power of $A_1^{1/3}$, that is, only one
nucleon in the projectile, making the projectile effectively a
``proton" in this power counting. Functions $f_2 , f_3 , \ldots$ are
not known analytically at present.

The function $f_2$ gives the first saturation correction in the
projectile, since it comes in with two powers of $A_1^{1/3}$,
corresponding to interactions with two nucleons in the projectile
nucleus. The efforts to calculate $f_2$ analytically was started in
\myref\cite{Balitsky:2004rr} and more recently revisited in
\myref\cite{Chirilli:2015tea}. The calculation of the order-$\left(
  \as^2 \, A_1^{1/3} \right)^2$ correction implies including an
order-$\as^2$ correction to the projectile interaction as compared to
the leading order-$\as^2 \, A_1^{1/3}$ term from
\myref\cite{Kovchegov:1998bi}. This order-$\as^2$ correction involves
interaction with the extra nucleon in the projectile, which brings in
an additional $A_1^{1/3}$ factor. With the help of the retarded gluon
Green function one can rearrange the diagrams such that the
order-$\as^2$ correction enters in two different ways: it may enter as an
order-$\as$ correction in the amplitude {\sl and} in the complex
conjugate amplitude, or as an order-$\as^2$ correction either in the
amplitude or in the complex conjugate amplitude. The former case was
calculated in \myrefs\cite{Balitsky:2004rr,Chirilli:2015tea}, where the
order-$\as$ correction to the leading-order (pA) gluon production
amplitude was found. No one has yet analytically calculated the
order-$\as^2$ correction to the same amplitude to complete the efforts
to determine $f_2$!

The same philosophy applies to the two-gluon production. For the
classical two-gluon production cross section one can write
\begin{align}
  \frac{d\sigma}{d^2 k_1 \, d^2 b_1 \, d^2 k_2 \, d^2 b_2 \, d^2 B} =
  \left\langle \frac{d\sigma}{d^2 k_1 \, d^2 b_1 \, d^2 B} (\rho_p,
    \rho_T) \ \frac{d\sigma}{d^2 k_2 \, d^2 b_2 \, d^2 B} (\rho_p,
    \rho_T) \right\rangle_{\rho_p, \rho_T} = \frac{1}{\as^2} \, h
  \left( \as^2 \, A_1^{1/3} , \as^2 \, A_2^{1/3} \right)
 \end{align}
 with the new unknown function $h$. Here $k_1$ and $k_2$ are the
 gluons' transverse momenta, while $b_1$ and $b_2$ are their
 transverse positions. Again, assuming a dilute projectile we expand
 in $\as^2 \, A_1^{1/3}$ getting
\begin{align}
  \frac{d\sigma}{d^2 k_1 \, d^2 b_1 \, d^2 k_2 \, d^2 b_2 \, d^2 B} =
  \frac{1}{\as^2} \, \left[ \left( \as^2 \, A_1^{1/3} \right)^2 \ h_1
    \!\left( \as^2 \, A_2^{1/3} \right) + \left( \as^2 \, A_1^{1/3}
    \right)^3 \ h_2 \! \left( \as^2 \, A_2^{1/3} \right) + \ldots
  \right] .
\end{align}
The function $h_1$ can be found from the results of
\myrefs\cite{Kovner:2012jm,Kovchegov:2012nd}. As described above, this
part of the two-gluon production cross section generates only even
harmonics. Finding the function $h_2$ requires a rather lengthy
calculation. However, as we will argue below (basing our argument on
\myref\cite{McLerran:2016snu}), the part of $h_2$ responsible for the
odd harmonics can be found using the results of
\myref\cite{Chirilli:2015tea}. The corresponding diagrams are shown
below in \fig{all_graphs}. We thus find the part of the two-gluon
production cross section responsible for odd harmonics at the order
$\left( \as^2 \, A_1^{1/3} \right)^3$: it is given by
\eq{full_expr}. This is the leading contribution to the odd harmonics
resulting from the two-gluon correlation function.

The paper is structured as follows: in Sec.~\ref{sec:appearance} we
derive the contribution \eqref{full_expr} to the two-gluon production
cross section giving the odd harmonics. We then proceed in
Sec.~\ref{Sec:Eval} by evaluating the expression \eqref{full_expr}
analytically to the lowest order in the interactions with the target
using the Golec-Biernat--Wusthoff (GBW)
\cite{GolecBiernat:1998js,GolecBiernat:1999qd} approximation to the
full MV interaction with the target. The resulting two-gluon
production cross section is given in \eq{full_result}. The most
important conclusion we draw here is that \eq{full_result} is
non-zero: hence the saturation dynamics does generate non-trivial odd
harmonics. Interestingly, a prominent contribution to the correlation
function comes from the $\delta$-functions resulting from the gluon
Hanbury-Brown--Twiss (HBT) \cite{HanburyBrown:1956pf} diagrams, of the
same general type as those advocated in
\myrefs\cite{Capella:1991mp,Kovchegov:2013ewa}. The odd-harmonics
correlation function resulting from \eq{full_expr} with all-orders
interactions with the target is evaluated in Sec.~\ref{sec:num}
numerically, with the resulting odd harmonic coefficients plotted in
\fig{vodd} and the odd part of the two-gluon correlation function
shown in \fig{Codd}. Keeping the interaction with the target to the
lowest non-trivial order in the numerical simulations we observe
qualitative and even quantitative agreement with \eq{full_result}. We
conclude in Sec.~\ref{sec:conc} by observing that saturation/CGC
dynamics does lead to the odd harmonics in di-gluon correlation
functions, which may allow for further successes of the saturation
approach to the correlation function phenomenology.


\section{Appearance of odd harmonics in gluon production diagrams}
\label{sec:appearance}


\subsection{General discussion}
\label{sec:disc}

Let us begin by analyzing how the two-gluon production cross section
involving two classical gluon fields can in principle generate the odd
azimuthal harmonics. As discussed in the Introduction, we need to
violate the ${\un k}_1 \leftrightarrow {\un k}_2$, ${\un k}_1 \to -
{\un k}_1$ and ${\un k}_2 \to - {\un k}_2$ symmetry of the the
two-gluon production cross section. For two gluons originating from
two identical classical fields the ${\un k}_1 \leftrightarrow {\un
  k}_2$ symmetry appears impossible to break: we thus conclude that
the classical two-gluon production cross section should always be
${\un k}_1 \leftrightarrow {\un k}_2$ symmetric. This implies that the
dependence on the azimuthal angle $\Delta \phi$ between the two
produced gluon enters the cross section via terms proportional to
$\cos ( n \Delta \phi)$ with the integer values of $n$, while terms
proportional to $\sin ( n \Delta \phi)$ do not enter the cross
section.

We thus see that the only way to generate odd harmonics is to violate
the ${\un k}_1 \to - {\un k}_1$ and/or ${\un k}_2 \to - {\un k}_2$
symmetries. Let us describe how this violation may happen in
general. Imagine a production cross section for a particle with the
transverse momentum $\un k = (k^1, k^2)$, possibly alongside with a
number of other particles whose momenta we do not explicitly display
below for brevity. The production cross section is proportional to
\begin{align}\label{cross_sect1}
  \frac{d \sigma}{d^2 k} \sim |M ({\un k})|^2 = \int d^2 x \, d^2 y \,
  e^{- i {\un k} \cdot ({\un x} - {\un y})} \, M ({\un x}) \, M^*
  ({\un y}).
\end{align} 
Here $M ({\un x})$ is the Fourier transform of the scattering
amplitude $M ({\un k})$ into transverse coordinate space and the
asterisk denotes complex conjugation. We want to find a condition
under which this cross section has a contribution that changes sign
under ${\un k} \to - {\un k}$. To do this, imagine that the scattering
amplitude in the transverse coordinate space can be written as
\begin{align}\label{Mdecomp}
M ({\un x}) = M_1 ({\un x}) + M_3 ({\un x}) + \ldots,
\end{align}
for instance due to an expansion in the coupling constant (that is,
$M_1$ is the leading contribution, and $M_3$ is one of the
higher-order corrections with the ellipsis denoting other higher-order
corrections). Using \eq{Mdecomp} in \eq{cross_sect1} (while keeping
only $M_1$ and $M_3$) one can easily see that only the interference
terms between $M_1$ and $M_3$ may lead to a contribution odd under
${\un k} \to - {\un k}$. Keeping only the interference terms we write
their contribution to \eq{cross_sect1} as
\begin{align}\label{interf1}
  \int d^2 x \, d^2 y \, e^{- i {\un k} \cdot ({\un x} - {\un y})} \,
  \left[ M_1 ({\un x}) \, M_3^* ({\un y}) + M_3 ({\un x}) \, M_1^*
    ({\un y}) \right].
\end{align}
Requiring that the expression \eqref{interf1} changes sign under ${\un
  k} \to - {\un k}$ results in the following condition on $M_1$ and
$M_3$:
\begin{align}\label{interf_cond1}
  M_1 ({\un x}) \, M_3^* ({\un y}) + M_3 ({\un x}) \, M_1^* ({\un y})
  = - M_1 ({\un y}) \, M_3^* ({\un x}) - M_3 ({\un y}) \, M_1^* ({\un
    x}).
\end{align}
Since $M_1$ and $M_3$, in general, are very different functions of
their arguments, \eq{interf_cond1} is most easily satisfied by
requiring that
\begin{align}\label{interf_cond2}
  M_1 ({\un x}) \, M_3^* ({\un y}) = - M_3 ({\un y}) \, M_1^* ({\un
    x}),
\end{align}
which, in turn, means that $M_1 ({\un x}) \, M_3^* ({\un y})$ is
imaginary. Therefore, we conclude that the phases of $M_1$ and $M_3$
(in transverse coordinate space) should be off by $\pi$. More
generally, for the higher-order corrections to the leading-order
amplitude $M_1$ to generate ${\un k} \to - {\un k}$ odd contribution
one needs a phase difference between $M_1$ and the full higher-order
amplitude $M_3 + \ldots$, where the ellipsis denote other higher-order
corrections, some of them possibly of the same order as $M_3$. Hence,
to find the odd harmonics we need to find a higher-order correction to
the results of \myref\cite{Kovner:2012jm,Kovchegov:2012nd} that
generates a phase difference between the higher- and leading-order
amplitudes (see also Appendix~\ref{Sec:CFL}).

The situation is not dissimilar to the single transverse spin
asymmetry (STSA) which is observed in the scattering of the
transversely polarized protons on the unpolarized ones and in
semi-inclusive deep inelastic scattering (SIDIS) on a transversely
polarized proton
\cite{Airapetian:2004tw,Bradamante:2011xu,Alekseev:2010rw,Bradamante:2009zz,Adolph:2012sp,Avakian:2010ae,Gao:2010av}. There,
generating the asymmetry requires the amplitude dependence on the
transverse proton polarization and a phase difference between the
leading-order amplitude and the higher-order correction to it: the
asymmetry is then given by the interference between the two amplitude
contributions
\cite{Qiu:1991pp,Qiu:1998ia,Brodsky:2002cx,Brodsky:2002rv,Brodsky:2013oya}. The
difference between the STSA case and the odd harmonics problem at hand
is that the phase difference in the former case is between the
momentum-space amplitudes, while for the latter, \eq{interf_cond2}
implies a phase difference between the Fourier transforms of the
amplitudes into the transverse coordinate space.


\subsection{Order $g^3$ amplitudes}

Let us now apply the insight we obtained in the previous section to
the calculation of gluon production in the saturation framework. As
discussed in the Introduction, we will assume that our calculation
resums all orders in the interaction with the target via the Wilson
lines for quarks and gluons traversing the target. We will perform
expansion of the gluon production amplitude to the first two
non-trivial orders in the interaction with the projectile.

\begin{figure}[ht]
\begin{center}
\includegraphics[width= 0.7 \textwidth]{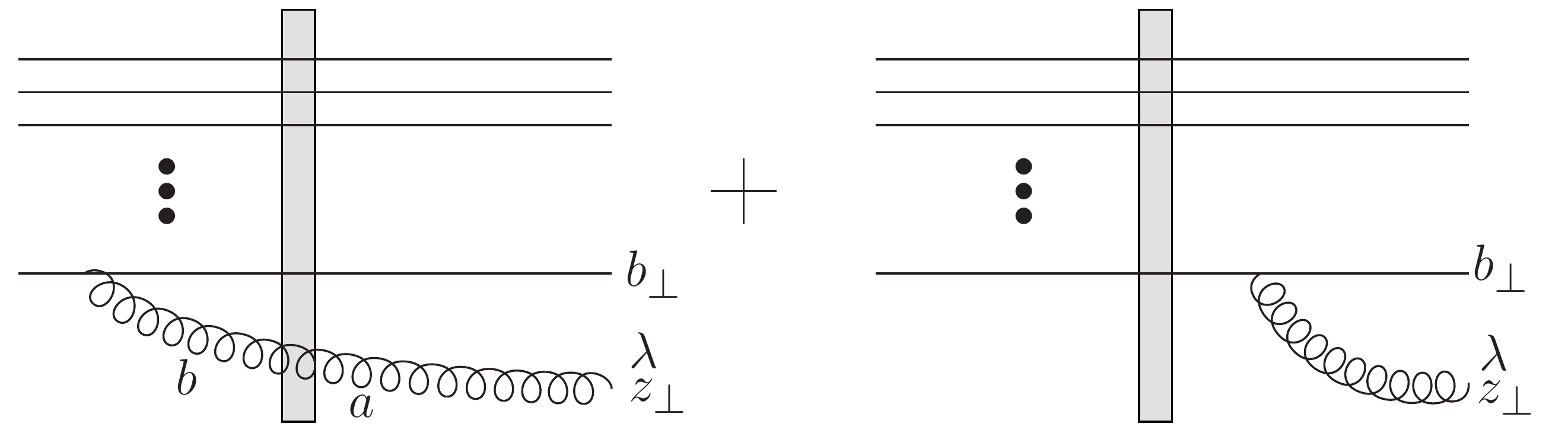} 
\caption{The diagrams contributing to the amplitude $M_1$ given by \eq{M1}.}
\label{fig:M1}
\end{center}
\end{figure}

The leading-order gluon production is described by the amplitude ${\un
  \epsilon}^*_\lambda \cdot {\un M}_1$ which is given by the diagrams
shown in \fig{fig:M1}. They yield~\cite{Kovchegov:1998bi}
\begin{align}\label{M1}
  {\un \epsilon}^*_\lambda \cdot {\un M}_1 ({\un z}, {\un b}) =
  \frac{i \, g}{\pi} \, \frac{{\un \epsilon}^*_\lambda \cdot ({\un z}
    - {\un b})}{|{\un z} - {\un b}|^2} \, \left[ U^{ab}_{\un z} -
    U^{ab}_{\un b} \right] \, (V_{\un b} \, t^b)
\end{align}
with the adjoint
\begin{align}
  \label{eq:Wadjoint}
  U^{ab}_{{\un z}} = \left( \mbox{P} \exp \left\{ i \, g \,
      \int\limits_{-\infty}^\infty d x^+ T^c \, A^{c \, -} (x^+,
      z^-=0, {\un z}) \right\} \right)^{ab}
\end{align}
and fundamental 
\begin{align}
  \label{eq:Wfund}
  V_{{\un b}} = \mbox{P} \exp \left\{ i \, g \,
    \int\limits_{-\infty}^\infty d x^+ t^a \, A^{a \, -} (x^+, b^-=0,
    {\un b}) \right\}
\end{align}
Wilson lines directed along the ``$+$" light cone, defined by the
direction of the projectile motion. All the notation is explained in
\fig{fig:M1}: the outgoing gluon has polarization $\lambda$ with ${\un
  \epsilon}_\lambda = - (\lambda, i)/\sqrt{2} $ and color $a$. The
horizontal straight lines in \fig{fig:M1} represent (valence) quarks
in the nucleons of the projectile nucleus. The vertical bar denotes
the target, which is Lorentz-contracted to a shock wave, whose
interaction with the projectile quark and emitted gluon is very fast,
practically instantaneous on the time scales of the projectile wave
functions: hence the gluon produced in \fig{fig:M1} can be emitted
either before or after the interaction with the target, but not during
that interaction \cite{Kovchegov:1998bi}. The gluon can be emitted by
either one of the many projectile quarks: we show the emission by only
one of the quarks, with the sum over other emissions implied
implicitly.

Following the power counting from
\myrefs\cite{Chirilli:2015tea,Kovchegov:1997pc} we assume that the
interaction with the target is strong, $\as^2 \, A_2^{1/3} = {\cal O}
(1)$, and, therefore, all the Wilson lines are also of order one, $U
\sim V = {\cal O} (1)$. Hence we see that $M_1 = {\cal O} (g)$.

\begin{figure}[ht]
\begin{center}
\includegraphics[width= 0.8 \textwidth]{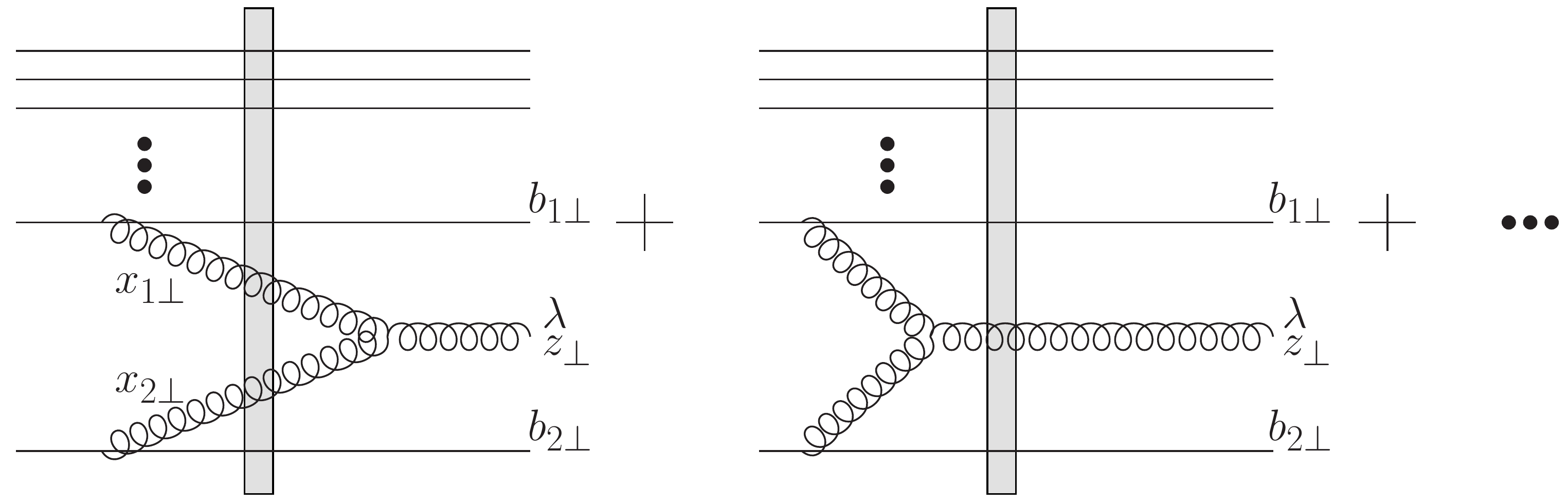} 
\caption{A sample of diagrams contributing to the amplitude $M_3$
  involving interactions with two nucleons in the projectile, given by
  \eq{eq:ABCsum_coord}.}
\label{fig:M3}
\end{center}
\end{figure}

The next order correction to $M_1$ could be an order-$g^2$ amplitude
with two gluons emitted by the quarks (with only one of these gluons
being tagged on if one wants to calculate the single inclusive gluon
production cross section). However, such contribution in the counting
of powers of $\as^2 \, A_1^{1/3}$ can be rearranged and absorbed into
the complex conjugate amplitude where one uses retarded Green
functions instead of Feynman propagators for the gluons (see
\myrefs\cite{Balitsky:2004rr,Chirilli:2015tea}). This way, the first
correction to $M_1$ is ${\cal O} (g^3)$. Some of (the large number of)
the relevant diagrams are shown in \fig{fig:M3}: in the notation of
the previous sub-section, these are the diagrams contributing to
$M_3$.

\begin{figure}[ht]
\begin{center}
\includegraphics[width= 0.8 \textwidth]{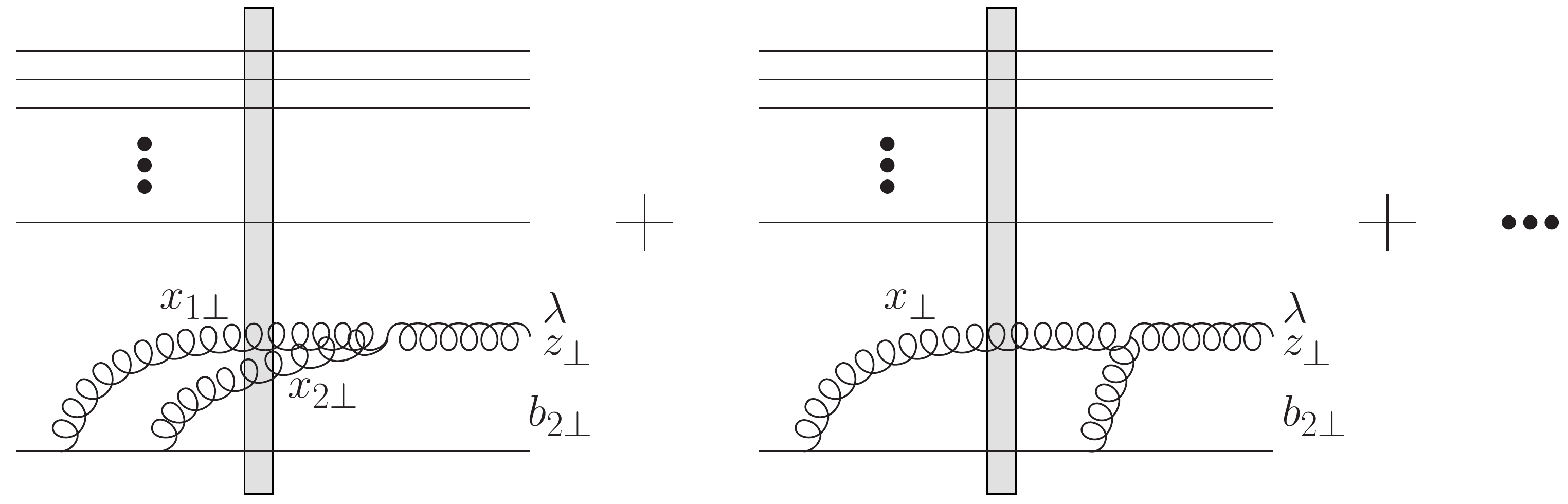} 
\caption{A sample of diagrams contributing to the amplitude $M_3$
  involving interactions with one nucleon in the projectile, given by
  \eq{eq:Dall_coord}.}
\label{fig:D3}
\end{center}
\end{figure}

The diagrams contributing to $M_3$ and involving two nucleons from the
projectile, as shown in \fig{fig:M3}, were calculated in
\myref\cite{Chirilli:2015tea}, with the result
\begin{align}
  \label{eq:ABCsum_coord}
  &
  	{\un \epsilon}^*_\lambda \cdot 
	{\un M}_3^{ABC} = - \frac{g^3}{4 \, \pi^4} \int d^2 x_1 \, d^2 x_2
  \, \delta [({\un z} - {\un x}_{1}) \times ({\un z} - {\un x}_{2})] \left[ \frac{{\un \epsilon}_\lambda{*} \cdot ({\un x}_{2} -
      {\un x}_{1}) }{|{\un x}_{2} - {\un x}_{1}|^2}
    \, \frac{{\un x}_{1} - {\un b}_{1}}{|{\un x}_{1}
      - {\un b}_{1}|^2} \cdot \frac{{\un x}_{2} - {\un b}_{2}}{|{\un x}_{2} - {\un b}_{2}|^2}
  \right. \notag \\ & \left. - \frac{{\un \epsilon}_\lambda^{*} \cdot ({\un x}_{1} - {\un b}_{1}) }{|{\un x}_{1} - {\un b}_{1}|^2} \, \frac{{\un z}
      - {\un x}_{1}}{|{\un z} - {\un x}_{1}|^2}
    \cdot \frac{{\un x}_{2} - {\un b}_{2}}{|{\un x}_{2} - {\un b}_{2}|^2} + \frac{{\un \epsilon}_\lambda^{*} \cdot ({\un x}_{2} -
      {\un b}_{2}) }{|{\un x}_{2} - {\un b}_{2}|^2}
    \, \frac{{\un x}_{1} - {\un b}_{1}}{|{\un x}_{1}
      - {\un b}_{1}|^2} \cdot \frac{{\un z} - {\un x}_{2}}{|{\un z} - {\un x}_{2}|^2} \right]
  \notag \\ & \times \, f^{abc} \, \left[ U^{bd}_{{\un x}_{1}} -
    U^{bd}_{{\un b}_{1}} \right] \, \left[ U^{ce}_{{\un x}_{2}} - U^{ce}_{{\un b}_{2}} \right] \, \left(
    V_{{\un b}_{1}} t^d \right)_1 \, \left( V_{{\un b}_{2}} t^e \right)_2 \notag \\ & + \frac{i \, g^3}{4 \,
    \pi^3} \, f^{abc} \, \left( V_{{\un b}_{1}} t^d \right)_1
  \, \left( V_{{\un b}_{2}} t^e \right)_2 \int d^2 x \, \left[
    U^{bd}_{{\un b}_{1}} \, \left( U^{ce}_{{\un x}} -
      U^{ce}_{{\un b}_{2}} \right) \, \left( \frac{{\un \epsilon}_\lambda^{*} \cdot ({\un z} -
        {\un x}) }{|{\un z} - {\un x}|^2}
      \, \frac{{\un x} - {\un b}_{1}}{|{\un x}
        - {\un b}_{1}|^2} \cdot \frac{{\un x} - {\un b}_{2}}{|{\un x} - {\un b}_{2}|^2}
    \right. \right. \notag \\ & \left. - \frac{{\un \epsilon}_\lambda^* \cdot ({\un z} - {\un b}_{1}) }{|{\un z} - {\un b}_{1}|^2} \,
    \frac{{\un z} - {\un x}}{|{\un z} -
      {\un x}|^2} \cdot \frac{{\un x} - {\un b}_{2}}{|{\un x} - {\un b}_{2}|^2} -
    \frac{{\un \epsilon}_\lambda^* \cdot ({\un z}
      - {\un b}_{1}) }{|{\un z} - {\un b}_{1}|^2}
    \, \frac{{\un x} - {\un b}_{1}}{|{\un x} -
      {\un b}_{1}|^2} \cdot \frac{{\un x} - {\un b}_{2}}{|{\un x} - {\un b}_{2}|^2} \right)
  \notag \\ & - \left( U^{bd}_{{\un x}} - U^{bd}_{{\un b}_{1}} \right) \, U^{ce}_{{\un b}_{2}} \, \left(
    \frac{{\un \epsilon}_\lambda^* \cdot ({\un z}
      - {\un x}) }{|{\un z} - {\un x}|^2}
    \, \frac{{\un x} - {\un b}_{1}}{|{\un x} -
      {\un b}_{1}|^2} \cdot \frac{{\un x} - {\un b}_{2}}{|{\un x} - {\un b}_{2}|^2} -
    \frac{{\un \epsilon}_\lambda^* \cdot ({\un z}
      - {\un b}_{2}) }{|{\un z} - {\un b}_{2}|^2}
    \, \frac{{\un z} - {\un x}}{|{\un z} -
      {\un x}|^2} \cdot \frac{{\un x} - {\un b}_{1}}{|{\un x} - {\un b}_{1}|^2}
  \right. \notag \\ & \left. \left. - \frac{{\un \epsilon}^{\,
          \lambda *} \cdot ({\un z} - {\un b}_{2})
      }{|{\un z} - {\un b}_{2}|^2} \, \frac{{\un x} - {\un b}_{1}}{|{\un x} - {\un b}_{1}|^2} \cdot \frac{{\un x} - {\un b}_{2}}{|{\un x} - {\un b}_{2}|^2}
    \right) \right] \notag \\
  & - \frac{i \, g^3}{4 \, \pi^2} \, f^{abc} \, \left( V_{{\un b}_{1}} t^d \right)_1 \, \left( V_{{\un b}_{2}} t^e
  \right)_2 \notag \\ & \times \, \left[ (U^{bd}_{{\un z}} -
    U^{bd}_{{\un b}_{1}}) \, U^{ce}_{{\un b}_{2}} \,
    \frac{{\un \epsilon}_\lambda^* \cdot ({\un z}
      - {\un b}_{1})}{|{\un z} - {\un b}_{1}|^2}
    \, \ln \frac{1}{|{\un z} - {\un b}_{2}| \, \Lambda}
    - U^{bd}_{{\un b}_{1}} \, (U^{ce}_{{\un z}} -
    U^{ce}_{{\un b}_{2}}) \, \frac{{\un \epsilon}_\lambda^* \cdot ({\un z} - {\un b}_{2})}{|{\un z} - {\un b}_{2}|^2} \, \ln
    \frac{1}{|{\un z} - {\un b}_{1}| \, \Lambda} \right]  \notag \\
  & - \frac{i \, g^3}{4 \, \pi^3} \, \int d^2 x \, \left[
    U_{\un x}^{ab} - U_{\un z}^{ab} \right] \, f^{bde} \, \left(
    V_{{\un b}_{1}} t^d \right)_1 \, \left( V_{{\un b}_{2}} t^e \right)_2 \notag \\ & \times \, \frac{{\un \epsilon}_\lambda^* \cdot ({\un z} - {\un x})}{|{\un z} - {\un x}|^2} \
  \frac{{\un x} - {\un b}_{1}}{|{\un x} -
    {\un b}_{1}|^2} \cdot \frac{{\un x} - {\un b}_{2}}{|{\un x} - {\un b}_{2}|^2} \,
  \mbox{Sign} (b_2^- - b_1^-)
  .
\end{align}
The label $M_3^{ABC}$ reflects the fact that \eq{eq:ABCsum_coord}
contains contributions of the diagrams A, B and C in the notation of
\myref\cite{Chirilli:2015tea}. Here $\Lambda$ is the IR cutoff.

There are also order-$g^3$ diagrams involving interaction with only
one nucleon in the projectile, labeled D and E in
\myref\cite{Chirilli:2015tea}: those are shown in \fig{fig:D3} and
their sum is given by
\begin{align}
  \label{eq:Dall_coord}
 	{\un \epsilon}^*_\lambda \cdot 
	{\un M}_3^{DE} & =  - \frac{g^3}{8 \, \pi^4} \, \int
  d^2 x_1 \, d^2 x_2 \, \delta [({\un z} - {\un x}_{1})
  \times ({\un z} - {\un x}_{2})] \left[ \frac{{\un \epsilon}_\lambda^* \cdot ({\un x}_{2} -
      {\un x}_{1}) }{|{\un x}_{2} - {\un x}_{1}|^2}
    \, \frac{{\un x}_{1} - {\un b}_{2}}{|{\un x}_{1}
      - {\un b}_{2}|^2} \cdot \frac{{\un x}_{2} - {\un b}_{2}}{|{\un x}_{2} - {\un b}_{2}|^2}
  \right. \notag \\ & \left. - \frac{{\un \epsilon}^{\, \lambda
        *} \cdot ({\un x}_{1} - {\un b}_{2}) }{|{\un x}_{1} - {\un b}_{2}|^2} \, \frac{{\un z}
      - {\un x}_{1}}{|{\un z} - {\un x}_{1}|^2}
    \cdot \frac{{\un x}_{2} - {\un b}_{2}}{|{\un x}_{2} - {\un b}_{2}|^2} + \frac{{\un \epsilon}_\lambda^* \cdot ({\un x}_{2} -
      {\un b}_{2}) }{|{\un x}_{2} - {\un b}_{2}|^2}
    \, \frac{{\un x}_{1} - {\un b}_{2}}{|{\un x}_{1}
      - {\un b}_{2}|^2} \cdot \frac{{\un z} - {\un x}_{2}}{|{\un z} - {\un x}_{2}|^2} \right]
  \notag \\ & \times \, f^{abc} \, \left[ U^{bd}_{{\un x}_{1}} -
    U^{bd}_{{\un b}_{2}} \right] \, \left[ U^{ce}_{{\un x}_{2}} - U^{ce}_{{\un b}_{2}} \right]\, \left(
    V_{{\un b}_{1}} \right)_1 \, \left( V_{{\un b}_{2}}
    t^e \, t^d \right)_2 \notag \\ & + \frac{i \, g^3}{4 \, \pi^3} \,
  \int d^2 x \, f^{abc} \, U^{bd}_{{\un b}_{2}} \, \left[
    U^{ce}_{{\un x}} - U^{ce}_{{\un b}_{2}} \right] \,
  \left( V_{{\un b}_{1}} \right)_1 \, \left( V_{{\un b}_{2}} t^e \, t^d \right)_2 \, \left( \frac{{\un \epsilon}_\lambda^* \cdot ({\un z} - {\un x}) }{|{\un z} - {\un x}|^2} \,
    \frac{1}{|{\un x} - {\un b}_{2}|^2} \right. \notag
  \\ & \left. - \frac{{\un \epsilon}_\lambda^* \cdot
      ({\un z} - {\un b}_{2}) }{|{\un z} -
      {\un b}_{2}|^2} \, \frac{{\un z} - {\un x}}{|{\un z} - {\un x}|^2} \cdot
    \frac{{\un x} - {\un b}_{2}}{|{\un x} -
      {\un b}_{2}|^2} - \frac{{\un \epsilon}_\lambda^* \cdot ({\un z} - {\un b}_{2}) }{|{\un z} - {\un b}_{2}|^2} \, \frac{1}{|{\un x} - {\un b}_{2}|^2} \right) \notag \\ & +
  \frac{i \, g^3}{4 \, \pi^2} \, f^{abc} \, U^{bd}_{{\un b}_{2}}
  \, \left[ U^{ce}_{{\un z}} - U^{ce}_{{\un b}_{2}}
  \right] \, \left( V_{{\un b}_{1}} \right)_1 \, \left(
    V_{{\un b}_{2}} t^e \, t^d \right)_2 \, \frac{{\un \epsilon}_\lambda^* \cdot ({\un z} - {\un b}_{2}) }{|{\un z} - {\un b}_{2}|^2} \, \ln
  \frac{1}{|{\un z} - {\un b}_{2}| \, \Lambda}.
\end{align}

Both equations \eqref{eq:ABCsum_coord} and \eqref{eq:Dall_coord} are
written in transverse coordinate space. Therefore, the formalism of
Sec.~\ref{sec:disc} applies. Since ${\un M}_1$ in \eq{M1} contains a
factor of $i$, and we require a phase shift of $\pi$ between ${\un
  M}_1$ and ${\un M}_3$, we need the ``real'' part of ${\un M}_3^{ABC}
+ {\un M}_3^{DE}$ to obtain the odd harmonics via the formula
\eq{interf1}. We conclude that the odd harmonics should be given by
the interference between ${\un M}_1$ from \eq{M1} with the ``real''
part of ${\un M}_3^{ABC} + {\un M}_3^{DE}$ from
Eqs.~\eqref{eq:ABCsum_coord} and \eqref{eq:Dall_coord}. Note that
correlators of Wilson lines in the MV model are all real, if one keeps
the leading $C$-even parts: hence by the real part of ${\un M}_3^{ABC}
+ {\un M}_3^{DE}$ we mean only the real part of the associated
light-cone wave functions, such that the Re sign does not apply to the
Wilson lines. With this caveat, the real parts are
\begin{align}\label{M3}
	{\un \epsilon}^*_\lambda \cdot 
	{\un M}_3 ({\un z} , {\un b}_1 , {\un b}_2) \equiv 
		{\un \epsilon}^*_\lambda \cdot 
	\mbox{Re} \, \left[ 
	{\un M}_3^{ABC} \right] = & - \frac{g^3}{4 \pi^4} \, \int d^2 x_1 \, d^2 x_2 \ \delta\left[ ({\un z} - {\un x}_1) \times ({\un z} - {\un x}_2) \right] \times \notag \\ & \left[ \frac{{\un \epsilon}^*_\lambda \cdot ({\un x}_2 - {\un x}_1)}{|{\un x}_2 - {\un x}_1|^2} \, \frac{{\un x}_1 - {\un b}_1}{|{\un x}_1 - {\un b}_1|^2} \cdot \frac{{\un x}_2 - {\un b}_2}{|{\un x}_2 - {\un b}_2|^2}  \right. \notag \\ & \left. - \frac{{\un \epsilon}^*_\lambda \cdot ({\un x}_1 - {\un b}_1)}{|{\un x}_1 - {\un b}_1|^2} \, \frac{{\un z} - {\un x}_1}{|{\un z} - {\un x}_1|^2} \cdot \frac{{\un x}_2 - {\un b}_2}{|{\un x}_2 - {\un b}_2|^2} + \frac{{\un \epsilon}^*_\lambda \cdot ({\un x}_2 - {\un b}_2)}{|{\un x}_2 - {\un b}_2|^2} \, \frac{{\un x}_1 - {\un b}_1}{|{\un x}_1 - {\un b}_1|^2} \cdot \frac{{\un z} - {\un x}_2}{|{\un z} - {\un x}_2|^2}\right] \notag \\ & \times f^{abc} \, \left[ U^{bd}_{{\un x}_1} - U^{bd}_{{\un b}_1} \right] \, \left[ U^{ce}_{{\un x}_2} - U^{ce}_{{\un b}_2} \right] \, (V_{{\un b}_1} \, t^d)_1 \, (V_{{\un b}_2} \, t^e)_2 ,
\end{align}
and
\begin{align}\label{D3}
	{\un \epsilon}^*_\lambda \cdot 
	{\un D}_3 ({\un z} , {\un b}_2) \equiv
		{\un \epsilon}^*_\lambda \cdot 
	\mbox{Re} \, \left[ 
	{\un M}_3^{DE} \right] = & - \frac{g^3}{8 \pi^4} \, \int d^2 x_1 \, d^2 x_2 \ \delta\left[ ({\un z} - {\un x}_1) \times ({\un z} - {\un x}_2) \right] \left[ \frac{{\un \epsilon}^*_\lambda \cdot ({\un x}_2 - {\un x}_1)}{|{\un x}_2 - {\un x}_1|^2} \, \frac{{\un x}_1 - {\un b}_2}{|{\un x}_1 - {\un b}_2|^2} \cdot \frac{{\un x}_2 - {\un b}_2}{|{\un x}_2 - {\un b}_2|^2}  \right. \notag \\ & \left. - \frac{{\un \epsilon}^*_\lambda \cdot ({\un x}_1 - {\un b}_2)}{|{\un x}_1 - {\un b}_2|^2} \, \frac{{\un z} - {\un x}_1}{|{\un z} - {\un x}_1|^2} \cdot \frac{{\un x}_2 - {\un b}_2}{|{\un x}_2 - {\un b}_2|^2} + \frac{{\un \epsilon}^*_\lambda \cdot ({\un x}_2 - {\un b}_2)}{|{\un x}_2 - {\un b}_2|^2} \, \frac{{\un x}_1 - {\un b}_2}{|{\un x}_1 - {\un b}_2|^2} \cdot \frac{{\un z} - {\un x}_2}{|{\un z} - {\un x}_2|^2}\right] \notag \\ & \times f^{abc} \, \left[ U^{bd}_{{\un x}_1} - U^{bd}_{{\un b}_2} \right] \, \left[ U^{ce}_{{\un x}_2} - U^{ce}_{{\un b}_2} \right] \, (V_{{\un b}_1}) \, (V_{{\un b}_2} \, t^e \, t^d) .
\end{align}
The factor of $(V_{{\un b}_1})$ in \eq{D3} denotes the Wilson line of
the spectator quark line which simply cancels the same (but conjugate)
Wilson line in the diagrams of \fig{fig:M1} contributing to ${\un
  M}_1$ (but not shown explicitly in the expression for ${\un M}_1$)
when ${\un D}_3$ is used in place of ${\un M}_3$ in the interference
formula \eqref{interf1}: due to such a trivial role, the dependence on
${\un b}_1$ is not explicitly included in the arguments of ${\un
  D}_3$.

According to \eq{interf1}, the odd azimuthal harmonics can be given by
the interference between ${\un M}_1$ from \eq{M1} with ${\un M}_3$ and
${\un D}_3$ from Eqs.~\eqref{M3} and \eqref{D3} respectively.


\subsection{Odd-harmonic part of the two-gluon production cross
  section at the order $\as^4$}

Equation \eqref{interf1} is written down for the single-gluon
production, where it is impossible to have a cross section
contribution which is odd under ${\un k} \to - {\un k}$ in the case of
unpolarized initial and final states: indeed, if we take the
interference of a single ${\un M}_1$ with a single ${\un M}_3$
contribution from above, as shown in \fig{all_graphs_num} below, the
color averaging would give zero. 

\begin{figure}[ht]
\begin{center}
\includegraphics[width= 0.95 \textwidth]{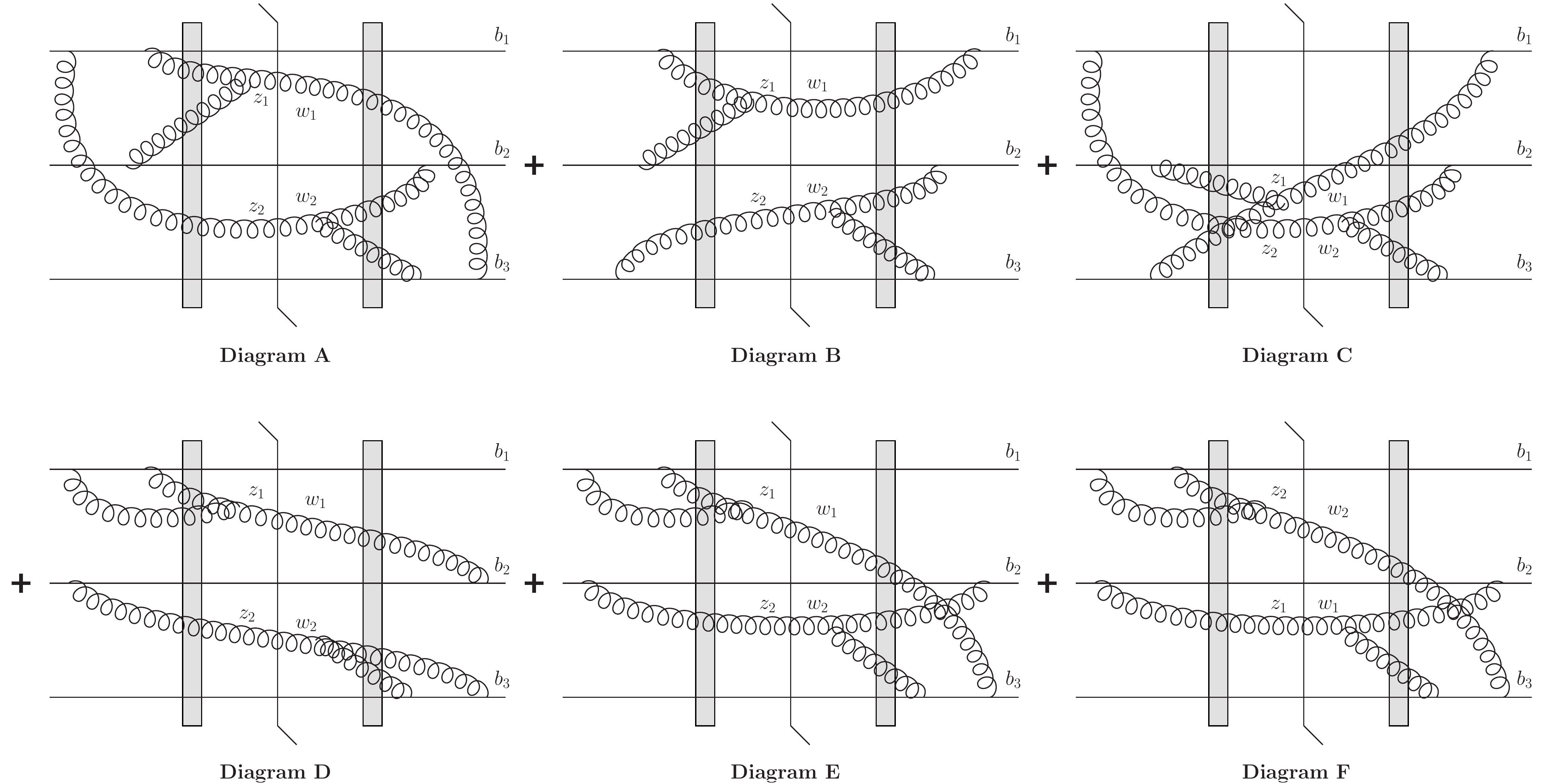} 
\caption{The contributing diagrams for the odd-harmonic part of the
  two-gluon production cross section. Diagram F is different from diagram E by ${\un z}_1, {\un w}_1 \leftrightarrow {\un z}_2, {\un w}_2$.}
\label{all_graphs}
\end{center}
\end{figure}

However, we are interested in odd harmonics in two-gluon
production. Hence we need to iterate \eq{interf1} twice, once for one
produced gluon, one for another. The corresponding diagrams we need to
calculate are given in \fig{all_graphs}: they involve quarks from
three nucleons in the projectile. Each diagram in \fig{all_graphs}
denotes a class of diagrams including all the diagrams obtained from
it by reflecting either one (or both) of the connected gluon
interactions with respect to the final-state cut (the vertical solid
line in \fig{all_graphs}). Given the above ingredients, the diagram
calculation is straightforward. The part of their contribution to the
two-gluon production cross section that gives odd harmonics is
\begin{align}\label{full_expr}
  \frac{d \sigma_{odd}}{d^2 k_1 \, d y_1 \, d^2 k_2 \, d y_2} =
  \frac{1}{[2 (2 \pi)^3]^2} \, \int d^2 B \, d^2 b_1 \, d^2 b_2 \, d^2
  b_3 \, T_1 ({\un B} - {\un b}_1) \, T_1 ({\un B} - {\un b}_2) \, T_1
  ({\un B} - {\un b}_3) \\ \times \, d^2 z_1 \, d^2 w_1 \, d^2 z_2 \,
  d^2 w_2 \, e^{- i {\un k}_1 \cdot ({\un z}_1 - {\un w}_1) - i {\un
      k}_2 \cdot ({\un z}_2 - {\un w}_2) } \, \langle {\hat A}
  \rangle_{\rho_T} \notag
\end{align}
with
\begin{align}\label{Ahat}
{\hat A} & = {\un M}_3 ({\un z}_1 , {\un b}_1 , {\un b}_2) \cdot {\un M}_1^* ({\un w}_1, {\un b}_3) \, {\un M}_1 ({\un z}_2 , {\un b}_1) \cdot {\un M}_3^* ({\un w}_2 , {\un b}_2 , {\un b}_3) +  {\un M}_1 ({\un z}_1, {\un b}_3) \cdot {\un M}_3^* ({\un w}_1 , {\un b}_1 , {\un b}_2) \, {\un M}_1 ({\un z}_2 , {\un b}_1) \cdot {\un M}_3^* ({\un w}_2 , {\un b}_2 , {\un b}_3) \notag \\ 
& + {\un M}_3 ({\un z}_1 , {\un b}_1 , {\un b}_2) \cdot {\un M}_1^* ({\un w}_1, {\un b}_3) \, {\un M}_3 ({\un z}_2 , {\un b}_2 , {\un b}_3) \cdot {\un M}_1^* ({\un w}_2 , {\un b}_1)  + {\un M}_1 ({\un z}_1, {\un b}_3) \cdot {\un M}_3^* ({\un w}_1 , {\un b}_1 , {\un b}_2) \, {\un M}_3 ({\un z}_2 , {\un b}_2 , {\un b}_3) \cdot {\un M}_1^* ({\un w}_2 , {\un b}_1) \notag \\ 
& + {\un M}_3 ({\un z}_1 , {\un b}_1 , {\un b}_2) \cdot {\un M}_1^* ({\un w}_1, {\un b}_1) \, {\un M}_1 ({\un z}_2 , {\un b}_3) \cdot {\un M}_3^* ({\un w}_2 , {\un b}_2 , {\un b}_3) +  {\un M}_1 ({\un z}_1, {\un b}_1) \cdot {\un M}_3^* ({\un w}_1 , {\un b}_1 , {\un b}_2) \, {\un M}_1 ({\un z}_2 , {\un b}_3) \cdot {\un M}_3^* ({\un w}_2 , {\un b}_2 , {\un b}_3) \notag \\ 
& + {\un M}_3 ({\un z}_1 , {\un b}_1 , {\un b}_2) \cdot {\un M}_1^* ({\un w}_1, {\un b}_1) \, {\un M}_3 ({\un z}_2 , {\un b}_2 , {\un b}_3) \cdot {\un M}_1^* ({\un w}_2 , {\un b}_3)  + {\un M}_1 ({\un z}_1, {\un b}_1) \cdot {\un M}_3^* ({\un w}_1 , {\un b}_1 , {\un b}_2) \, {\un M}_3 ({\un z}_2 , {\un b}_2 , {\un b}_3) \cdot {\un M}_1^* ({\un w}_2 , {\un b}_3) \notag \\ 
& + {\un M}_3 ({\un z}_1 , {\un b}_2 , {\un b}_3) \cdot {\un M}_1^* ({\un w}_1, {\un b}_1) \, {\un M}_1 ({\un z}_2 , {\un b}_1) \cdot {\un M}_3^* ({\un w}_2 , {\un b}_2 , {\un b}_3) +  {\un M}_1 ({\un z}_1, {\un b}_1) \cdot {\un M}_3^* ({\un w}_1 , {\un b}_2 , {\un b}_3) \, {\un M}_1 ({\un z}_2 , {\un b}_1) \cdot {\un M}_3^* ({\un w}_2 , {\un b}_2 , {\un b}_3) \notag \\ 
& + {\un M}_3 ({\un z}_1 , {\un b}_2 , {\un b}_3) \cdot {\un M}_1^* ({\un w}_1, {\un b}_1) \, {\un M}_3 ({\un z}_2 , {\un b}_2 , {\un b}_3) \cdot {\un M}_1^* ({\un w}_2 , {\un b}_1)  + {\un M}_1 ({\un z}_1, {\un b}_1) \cdot {\un M}_3^* ({\un w}_1 , {\un b}_2 , {\un b}_3) \, {\un M}_3 ({\un z}_2 , {\un b}_2 , {\un b}_3) \cdot {\un M}_1^* ({\un w}_2 , {\un b}_1) \notag \\ 
& + {\un D}_3 ({\un z}_1 , {\un b}_1) \cdot {\un M}_1^* ({\un w}_1, {\un b}_2) \, {\un M}_1 ({\un z}_2 , {\un b}_2) \cdot {\un D}_3^* ({\un w}_2 , {\un b}_3) +  {\un M}_1 ({\un z}_1, {\un b}_2) \cdot {\un D}_3^* ({\un w}_1 , {\un b}_1) \, {\un M}_1 ({\un z}_2 , {\un b}_2) \cdot {\un D}_3^* ({\un w}_2 , {\un b}_3) \notag \\ 
& + {\un D}_3 ({\un z}_1 , {\un b}_1) \cdot {\un M}_1^* ({\un w}_1, {\un b}_2) \, {\un D}_3 ({\un z}_2 , {\un b}_3) \cdot {\un M}_1^* ({\un w}_2 , {\un b}_2)  + {\un M}_1 ({\un z}_1, {\un b}_2) \cdot {\un D}_3^* ({\un w}_1 , {\un b}_1) \, {\un D}_3 ({\un z}_2 , {\un b}_3) \cdot {\un M}_1^* ({\un w}_2 , {\un b}_2)  \\ 
& + {\un D}_3 ({\un z}_1 , {\un b}_1 ) \cdot {\un M}_1^* ({\un w}_1, {\un b}_2) \, {\un M}_3 ({\un z}_2 , {\un b}_2 , {\un b}_3) \cdot {\un M}_1^* ({\un w}_2 , {\un b}_3)  + {\un M}_1 ({\un z}_1, {\un b}_2) \cdot {\un D}_3^* ({\un w}_1 , {\un b}_1) \, {\un M}_3 ({\un z}_2 , {\un b}_2 , {\un b}_3) \cdot {\un M}_1^* ({\un w}_2 , {\un b}_3) \notag \\ 
& + {\un D}_3 ({\un z}_1 , {\un b}_1) \cdot {\un M}_1^* ({\un w}_1, {\un b}_2) \, {\un M}_1 ({\un z}_2 , {\un b}_3) \cdot {\un M}_3^* ({\un w}_2 , {\un b}_2 , {\un b}_3) +  {\un M}_1 ({\un z}_1, {\un b}_2) \cdot {\un D}_3^* ({\un w}_1 , {\un b}_1) \, {\un M}_1 ({\un z}_2 , {\un b}_3) \cdot {\un M}_3^* ({\un w}_2 , {\un b}_2 , {\un b}_3) \notag \\ 
& + {\un D}_3 ({\un z}_2 , {\un b}_1 ) \cdot {\un M}_1^* ({\un w}_2, {\un b}_2) \, {\un M}_3 ({\un z}_1 , {\un b}_2 , {\un b}_3) \cdot {\un M}_1^* ({\un w}_1 , {\un b}_3)  + {\un M}_1 ({\un z}_2, {\un b}_2) \cdot {\un D}_3^* ({\un w}_2 , {\un b}_1) \, {\un M}_3 ({\un z}_1 , {\un b}_2 , {\un b}_3) \cdot {\un M}_1^* ({\un w}_1 , {\un b}_3) \notag \\ 
& + {\un D}_3 ({\un z}_2 , {\un b}_1) \cdot {\un M}_1^* ({\un w}_2, {\un b}_2) \, {\un M}_1 ({\un z}_1 , {\un b}_3) \cdot {\un M}_3^* ({\un w}_1 , {\un b}_2 , {\un b}_3) +  {\un M}_1 ({\un z}_2, {\un b}_2) \cdot {\un D}_3^* ({\un w}_2 , {\un b}_1) \, {\un M}_1 ({\un z}_1 , {\un b}_3) \cdot {\un M}_3^* ({\un w}_1 , {\un b}_2 , {\un b}_3) . \notag 
\end{align}
The dot-product in \eq{Ahat} arises after a sum over the final-state
gluon polarizations $\lambda$. Each diagram of A-F in \fig{all_graphs}
contributes four terms to \eq{Ahat}, with the terms in the latter
ordered in the same way as the diagrams A-F: the first four terms come
from the diagram A, the next four terms are from B, etc.

Equation \eqref{full_expr} is the main analytic result of this work:
it gives the leading contribution to the inclusive two-gluon
production cross section in the saturation framework which generates
odd azimuthal harmonics in the correlation function.  It involves
interactions with three nucleons in the projectile nucleus and
all-order interactions with the nucleons in the target nucleus. The
Wilson line correlators also include the linear and non-linear
small-$x$ evolution between the target nucleus and the produced gluon
\cite{Kuraev:1977fs,Balitsky:1978ic,Balitsky:1995ub,Balitsky:1998ya,Kovchegov:1999yj,Kovchegov:1999ua,Jalilian-Marian:1997dw,Jalilian-Marian:1997gr,Iancu:2001ad,Iancu:2000hn}.

However, we are not done yet: to convincingly demonstrate that
\eq{full_expr} does generate odd harmonics in the correlation
function, one needs to show that it is not zero. While there appear to
be no symmetries requiring \eq{full_expr} to be zero, an actual
evaluation of this expression is required to establish this with full
certainty.  Below we will evaluate \eq{full_expr} using a
quasi-classical MV-based approximation for the correlators of Wilson
lines, expanding those correlators to the lowest non-trivial order,
which involves interaction with three nucleons in the target nucleus.

We will use the following Fourier representations of the amplitude
contributions to evaluate \eq{full_expr}:
\begin{align}\label{M1mom}
{\un \epsilon}^*_\lambda \cdot 
{\un M}_1({\un z}, {\un b}) = 2 g \int \frac{d^2 k} {(2\pi)^2} \, e^{i
  {\un k} \cdot ({\un z} - {\un b})} \, \frac{{\un \epsilon}^*_\lambda
  \cdot {\un k} }{{ \un k}^2} \, \left[ U^{ab}_{\un z} - U^{ab}_{\un
    b} \right] \, (V_{{\un b}} \, t^b)
\end{align}
and 
\begin{align}\label{M3mom}
  {\un \epsilon}^*_\lambda \cdot {\un M}_3 & ({\un z}, {\un b}_1, {\un
    b}_2) = - 2 i g^3 \! \int d^2 x_1 d^2 x_2 \! \int \frac{d^2 k}{(2
    \pi)^2} \, \frac{d^2 l}{(2 \pi)^2} \, \frac{d^2 q_1}{(2 \pi)^2} \,
  \frac{d^2 q_2}{(2 \pi)^2} \, e^{i {\un q}_{1} \cdot ({\un x}_{1} -
    {\un b}_{1}) + i {\un q}_{2} \cdot ({\un x}_{2} - {\un b}_{2}) + i
    {\un l} \cdot ({\un x}_{2} - {\un x}_{1}) + i {\un k} \cdot ({\un
      z} - {\un x}_{2})} \notag \\ & \times \frac{1}{{\un q}_{1}^2 \,
    {\un q}_{2}^2} \, \left( - {\un q}_{1} \cdot {\un q}_{2} \,
    \frac{{\un \epsilon}^{\, *}_\lambda \times {\un k}}{{\un k}^2} +
    {\un \epsilon}^{\, *}_\lambda \cdot {\un q}_{1} \, \frac{{\un
        q}_{2} \times ({\un k} - {\un l})}{({\un k} - {\un l})^2} +
    {\un \epsilon}^{\, *}_\lambda \cdot {\un q}_{2} \, \frac{{\un
        q}_{1} \times {\un l}}{{\un l}^2} \right) \, \mbox{Sign} ({\un
    k} \times {\un l}) 
  \notag \\ & \times \, f^{abc} \, \left[ U^{bd}_{{\un x}_{1}} -
    U^{bd}_{{\un b}_{1}} \right] \, \left[ U^{ce}_{{\un x}_{2}} -
    U^{ce}_{{\un b}_{2}} \right] \, \left( V_{{\un b}_{1}} \, t^d
  \right)_1 \, \left( V_{{\un b}_{2}} \, t^e \right)_2 .
  \end{align}


  \section{Evaluating the odd-harmonic part of the two-gluon
    production cross section: Analytic approach}
\label{Sec:Eval}


\subsection{Diagram A}

Our goal here is to evaluate the expression \eqref{full_expr}. We
begin with the first four terms in \eq{Ahat}, whose contribution to
\eq{full_expr} is proportional to
\begin{align}\label{4M}
  & \int d^2 z_1 \, d^2 w_1 \, e^{- i {\un k}_1 \cdot ({\un z}_1 - {\un w}_1) }  \, \left [ {\un M}_3 ({\un z}_1 , {\un b}_1 , {\un b}_2) \cdot {\un M}_1^* ({\un w}_1, {\un b}_3) + {\un M}_1 ({\un z}_1, {\un b}_3) \cdot {\un M}_3^* ({\un w}_1 , {\un b}_1 , {\un b}_2) \right] \notag \\
  & \times \int d^2 z_2 \, d^2 w_2 \, e^{- i {\un k}_2 \cdot ({\un
      z}_2 - {\un w}_2) } \, \left[ {\un M}_1 ({\un z}_2 , {\un b}_1)
    \cdot {\un M}_3^* ({\un w}_2 , {\un b}_2 , {\un b}_3) + {\un M}_3
    ({\un z}_2 , {\un b}_2 , {\un b}_3) \cdot {\un M}_1^* ({\un w}_2 ,
    {\un b}_1) \right]
\end{align}
and corresponds to the diagram A in \fig{all_graphs} along with all
the gluon reflections with respect to the final-state cut. Let us
evaluate the first line of \eq{4M} first. Disregarding the conjugation
of the $V's$, since they cancel in the net expression \eq{4M} anyway,
we write
\begin{align}
  & \int d^2 z_1 \, d^2 w_1 \, e^{- i {\un k}_1 \cdot ({\un z}_1 -
    {\un w}_1) } \, \left [ {\un M}_3 ({\un z}_1 , {\un b}_1 , {\un
      b}_2) \cdot {\un M}_1^* ({\un w}_1, {\un b}_3) + {\un M}_1 ({\un
      z}_1, {\un b}_3) \cdot {\un M}_3^* ({\un w}_1 , {\un b}_1 , {\un
      b}_2) \right] \notag \\ & = \int d^2 z_1 \, d^2 w_1 \, e^{- i
    {\un k}_1 \cdot ({\un z}_1 - {\un w}_1) } \, \left[ {\un M}_3
    ({\un z}_1 , {\un b}_1 , {\un b}_2) \cdot {\un M}_1^* ({\un w}_1,
    {\un b}_3) - (z_1 \leftrightarrow w_1) \right] \notag \\ & = \int
  d^2 z_1 \, d^2 w_1 \, e^{- i {\un k}_1 \cdot ({\un z}_1 - {\un w}_1)
  } \, {\un M}_3 ({\un z}_1 , {\un b}_1 , {\un b}_2) \cdot {\un M}_1^*
  ({\un w}_1, {\un b}_3) - ({\un k}_1 \leftrightarrow - {\un k}_1).
\end{align} 
This way \eq{4M} can be written as
\begin{align}\label{4M1}
  & \int d^2 z_1 \, d^2 w_1 \, e^{- i {\un k}_1 \cdot ({\un z}_1 -
    {\un w}_1) } \, {\un M}_3 ({\un z}_1 , {\un b}_1 , {\un b}_2)
  \cdot {\un M}_1^* ({\un w}_1, {\un b}_3) \int d^2 z_2 \, d^2 w_2 \,
  e^{- i {\un k}_2 \cdot ({\un z}_2 - {\un w}_2) } \, {\un M}_1 ({\un
    z}_2 , {\un b}_1) \cdot {\un M}_3^* ({\un w}_2 , {\un b}_2 , {\un
    b}_3) \notag \\ & - ({\un k}_1 \to - {\un k}_1) - ({\un k}_2 \to -
  {\un k}_2) + ({\un k}_1 \to - {\un k}_1, \, {\un k}_2 \to - {\un
    k}_2) .
\end{align}

Plugging in Eqs.~\eqref{M1mom} and \eqref{M3mom} into the first line
of \eq{4M1} and taking all the fundamental color traces while
averaging over colors of all three quarks in the projectile yields
\begin{align}\label{4M2}
& \frac{(4 \, g^4)^2}{(2 N_c)^3} \! \int d^2 z_1 \, d^2 w_1 \, d^2 z_2 \, d^2 w_2 \, e^{- i {\un k}_1 \cdot ({\un z}_1 - {\un w}_1) - i {\un k}_2 \cdot ({\un z}_2 - {\un w}_2)} \sum_{\lambda , \lambda'} \notag \\ &   \int d^2 x_1 d^2 x_2 \! \int
  \frac{d^2 k}{(2 \pi)^2} \, \frac{d^2 l}{(2 \pi)^2} \, \frac{d^2
    q_1}{(2 \pi)^2} \, \frac{d^2 q_2}{(2 \pi)^2} \, e^{i {\un q}_{1} \cdot ({\un x}_{1} - {\un b}_{1}) + i
    {\un q}_{2} \cdot ({\un x}_{2} - {\un b}_{2}) +
    i {\un l} \cdot ({\un x}_{2} - {\un x}_{1}) +
    i {\un k} \cdot ({\un z}_{1} - {\un x}_{2})}
	\notag \\ & \times \frac{1}{{\un q}_{1}^2 \, {\un q}_{2}^2} \, \left(
    - {\un q}_{1} \cdot {\un q}_{2} \, \frac{{\un \epsilon}_\lambda^* \times {\un k}}{{\un k}^2} + {\un \epsilon}_\lambda^*
    \cdot {\un q}_{1} \, \frac{{\un q}_{2} \times ({\un k} - {\un l})}{({\un k} - {\un l})^2} + {\un \epsilon}_\lambda^* \cdot
    {\un q}_{2} \, \frac{{\un q}_{1} \times {\un l}}{{\un l}^2} \right) \, \mbox{Sign} ({\un k} \times {\un l})   \notag \\ & 
  \int \frac{d^2 q_3} {(2\pi)^2} \, e^{- i {\un q}_3 \cdot ({\un w}_1 - {\un b}_3)} 
	\frac{{\un \epsilon}_\lambda \cdot {\un q}_3 }{{ \un q}_3^{2}}  \int \frac{d^2 q'_3} {(2\pi)^2} \, e^{i {\un q}'_3 \cdot ({\un z}_2 - {\un b}_1)} 
	\frac{{\un \epsilon}^*_{\lambda'} \cdot {\un q}'_3 }{{ \un q}_3^{\prime \, 2}} 
 \notag \\ & \int d^2 y_1 d^2 y_2 \! \int
  \frac{d^2 k'}{(2 \pi)^2} \, \frac{d^2 l'}{(2 \pi)^2} \, \frac{d^2
    q'_1}{(2 \pi)^2} \, \frac{d^2 q'_2}{(2 \pi)^2} \, e^{- i {\un q}'_{1} \cdot ({\un y}_{1} - {\un b}_{2}) - i
    {\un q}'_{2} \cdot ({\un y}_{2} - {\un b}_{3}) - i {\un l}^{\, \prime}_{} \cdot ({\un y}_{2} - {\un y}_{1}) -
    i {\un k}^{\, \prime}_{} \cdot ({\un w}_{2} - {\un y}_{2})}
  \notag \\ & \times \frac{1}{{\un q}_{1}^{\prime \, 2} \, {\un q}_{2}^{\prime \, 2}} \, \left(
  - {\un q}^{\, \prime}_{1} \cdot {\un q}^{\, \prime}_{2} \, \frac{{\un \epsilon}^{\, \lambda'} \times {\un k}^{\, \prime}}{{\un k}^{\prime \, 2}} + {\un \epsilon}^{\, \lambda'}
    \cdot {\un q}^{\, \prime}_{1} \, \frac{{\un q}^{\, \prime}_{2} \times ({\un k}^{\, \prime}_{} - {\un l}^{\, \prime})}{({\un k}^{\, \prime} - {\un l}^{\, \prime})^2} + {\un \epsilon}^{\, \lambda'} \cdot
    {\un q}^{\, \prime}_{2} \, \frac{{\un q}^{\, \prime}_{1} \times {\un l}^{\, \prime}_{}}{{\un l}^{\prime \, 2}} \right) \, \mbox{Sign} ({\un k}^{\, \prime} \times {\un l}^{\, \prime})   \notag \\ & \times \left\langle f^{abc} \, \left[ U^{bd}_{{\un x}_{1}} - U^{bd}_{{\un b}_{1}} \right] \, \left[ U^{ce}_{{\un x}_{2}} - U^{ce}_{{\un b}_{2}} \right] \, \left[ 
	U^{a'd}_{{\un z}_2}  -  U^{a'd}_{{\un b}_1} 
\right] \, f^{a'b'c'} \, \left[ U^{b'e}_{{\un y}_{1}} - U^{b'e}_{{\un b}_{2}}
  \right] \, \left[ U^{c'd'}_{{\un y}_{2}} - U^{c'd'}_{{\un b}_{3}} \right] \, \left[ 
	U^{ad'}_{{\un w}_1}  -  U^{ad'}_{{\un b}_3} 
\right] \right\rangle.
\end{align}
Here we took a step back and reintroduced the polarization vectors (e.g. we replaced ${\un M}_1 \cdot {\un M}_3^* \to\sum_\lambda {\un \epsilon}^*_\lambda \cdot  {\un M}_1  \ {\un \epsilon}_\lambda \cdot  {\un M}_3^* $) to simplify the derivation.  

\begin{figure}[ht]
\begin{center}
\includegraphics[width= 0.35 \textwidth]{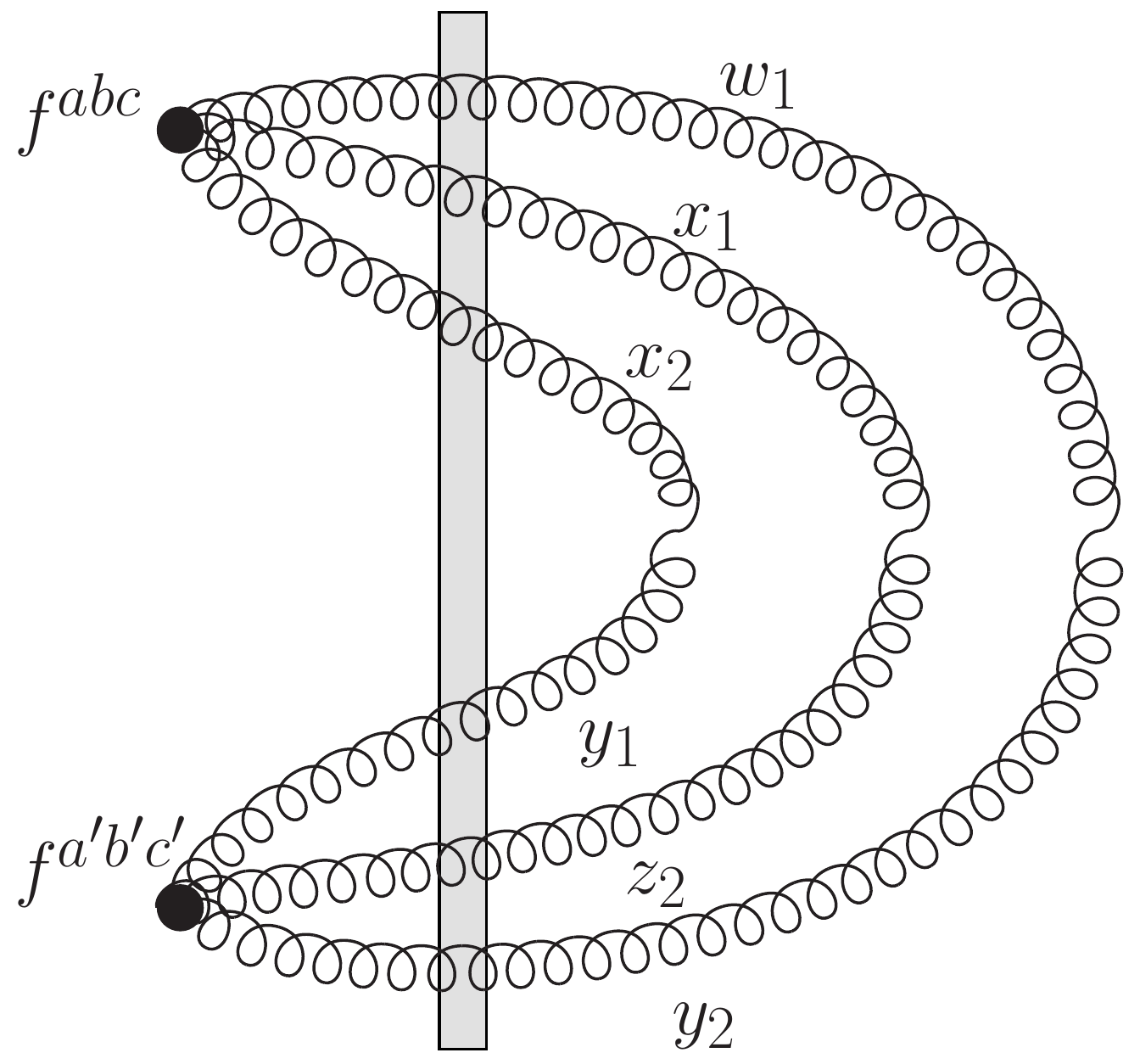} 
\caption{The interaction with the target in \eq{4M2}.}
\label{quadrupoles}
\end{center}
\end{figure}

The interaction with the target from \eq{4M2} (the expression in angle
brackets) is depicted in \fig{quadrupoles} diagrammatically: there we
have used the crossing symmetry \cite{Kovchegov:2001sc,Mueller:2012bn}
to reflect all the adjoint Wilson lines, represented by gluons, from
the complex conjugate amplitude into the actual amplitude. We evaluate
this interaction with the target in the large-$N_c$ limit, in which it
reduces to a sum of products of three fundamental quadrupoles
\cite{JalilianMarian:2004da,Dominguez:2011gc},
\begin{align}\label{int1}
  & \left\langle f^{abc} \, \left[ U^{bd}_{{\un x}_{1}} - U^{bd}_{{\un
          b}_{1}} \right] \, \left[ U^{ce}_{{\un x}_{2}} -
      U^{ce}_{{\un b}_{2}} \right] \, \left[ U^{a'd}_{{\un z}_2} -
      U^{a'd}_{{\un b}_1} \right] \, f^{a'b'c'} \, \left[
      U^{b'e}_{{\un y}_{1}} - U^{b'e}_{{\un b}_{2}} \right] \, \left[
      U^{c'd'}_{{\un y}_{2}} - U^{c'd'}_{{\un b}_{3}} \right] \,
    \left[ U^{ad'}_{{\un w}_1} - U^{ad'}_{{\un b}_3} \right]
  \right\rangle \notag \\ & = N_c^3 \, Q({\un x}_{1}, {\un x}_{2} ;
  {\un y}_{1}, {\un z}_{2}) \, Q({\un w}_{1}, {\un x}_{1} ; {\un
    z}_{2}, {\un y}_{2}) \, Q({\un w}_{1}, {\un x}_{2} ; {\un y}_{1},
  {\un y}_{2}) \pm \mbox{permutations} + {\cal O} \left(
    \frac{1}{N_c^2} \right).
\end{align}
Here ``permutations" denote 63 other 3-quadrupole products. Hence the
exact analytic evaluation of \eq{4M2} appears to be prohibitively
complicated. Instead we will expand the interaction with the target
\eqref{int1} to the lowest non-trivial order in gluon exchanges, or,
equivalently, in the saturation scale. To do this, we employ the
large-$N_c$ fundamental quadrupole amplitude evaluated in the
quasi-classical MV approximation in \myref\cite{JalilianMarian:2004da}
(see Eq. (14) there)
\begin{align}
  \label{eq:quad_quark}
  Q ({\un x}_1, {\un x}_2; {\un x}_3, {\un x}_4) = e^{D_1/2} +
  \frac{D_3 - D_2}{D_1 - D_3} \, \left[ e^{D_1/2} - e^{D_3/2} \right]
\end{align}
with
\begin{subequations}\label{Ds}
\begin{align}
  D_1 & = - \frac{Q_{s0}^2}{4} \left[ | {\un x}_1- {\un x}_2|^2 \, \ln \left( \frac{1}{| {\un x}_1- {\un x}_2| \Lambda } \right) + | {\un x}_3- {\un x}_4|^2 \, \ln \left( \frac{1}{| {\un x}_3- {\un x}_4| \Lambda } \right) \right], \\
  D_2 & = - \frac{Q_{s0}^2}{4} \left[ | {\un x}_1- {\un x}_3|^2 \, \ln \left( \frac{1}{| {\un x}_1- {\un x}_3| \Lambda } \right) + | {\un x}_2- {\un x}_4|^2 \, \ln \left( \frac{1}{| {\un x}_2- {\un x}_4| \Lambda } \right) \right], \\
  D_3 & = - \frac{Q_{s0}^2}{4} \left[ | {\un x}_1- {\un x}_4|^2 \, \ln
    \left( \frac{1}{| {\un x}_1- {\un x}_4| \Lambda } \right) + | {\un
      x}_2- {\un x}_3|^2 \, \ln \left( \frac{1}{| {\un x}_2- {\un
          x}_3| \Lambda } \right) \right].
	\end{align}
\end{subequations}
Here $Q_{s0}$ is the gluon saturation scale of the target taken in the
quasi-classical limit \cite{Mueller:1989st}.

Inserting \eq{eq:quad_quark} into \eq{int1} with all the permutations
included, and expanding in $Q_{s0}$ to the lowest non-trivial order,
which is order-$Q_{s0}^6$ corresponding to the 6-gluon exchange,
yields
\begin{align}\label{int2}
& \left\langle f^{abc} \,
  \left[ U^{bd}_{{\un x}_{1}} - U^{bd}_{{\un b}_{1}}
  \right] \, \left[ U^{ce}_{{\un x}_{2}} - U^{ce}_{{\un b}_{2}} \right] \, \left[ 
	U^{a'd}_{{\un z}_2}  -  U^{a'd}_{{\un b}_1} 
\right] \, f^{a'b'c'} \, \left[ U^{b'e}_{{\un y}_{1}} - U^{b'e}_{{\un b}_{2}}
  \right] \, \left[ U^{c'd'}_{{\un y}_{2}} - U^{c'd'}_{{\un b}_{3}} \right] \, \left[ 
	U^{ad'}_{{\un w}_1}  -  U^{ad'}_{{\un b}_3} 
\right] \right\rangle \notag \\ & \approx \frac{N_c^3}{64} \, Q_{s0}^6 \ \mbox{Int}_A 
\end{align}
with 
\begin{align}\label{intA}
\mbox{Int}_A  = & - ({\un b}_1 - {\un x}_1) \cdot ({\un b}_3 - {\un y}_2) \ \ ({\un b}_2 - {\un x}_2) \cdot ({\un b}_1 - {\un z}_2) \ \ ({\un b}_2 - {\un y}_1) \cdot ({\un b}_3 - {\un w}_1) \notag \\ & + 2 \, ({\un b}_1 - {\un x}_1) \cdot ({\un b}_1 - {\un z}_2) \ \ ({\un b}_2 - {\un x}_2) \cdot ({\un b}_3 - {\un w}_1) \ \ ({\un b}_2 - {\un y}_1) \cdot ({\un b}_3 - {\un y}_2) \notag \\ & - ({\un b}_1 - {\un x}_1) \cdot ({\un b}_3 - {\un w}_1) \ \ ({\un b}_2 - {\un x}_2) \cdot ({\un b}_1 - {\un z}_2) \ \ ({\un b}_2 - {\un y}_1) \cdot ({\un b}_3 - {\un y}_2) \notag \\ &  + 8 \, 
	({\un b}_1 - {\un z}_2)
	\cdot 
	({\un b}_1 - {\un x}_1)
	\ \ 
	({\un b}_3 - {\un w}_1)
	\cdot 
	({\un b}_3 - {\un y}_2)
	\ \ 
	({\un b}_2 - {\un x}_2)
	\cdot 
	({\un b}_2 - {\un y}_1) \notag \\ & - ({\un b}_1 - {\un x}_1) \cdot ({\un b}_2 - {\un y}_1) \ \ ({\un b}_2 - {\un x}_2) \cdot ({\un b}_3 - {\un y}_2) \ \ ({\un b}_1 - {\un z}_2)  \cdot ({\un b}_3 - {\un w}_1) \notag \\ & - ({\un b}_1 - {\un x}_1) \cdot ({\un b}_2 - {\un x}_2) \ \ ({\un b}_2 - {\un y}_1) \cdot ({\un b}_3 - {\un y}_2) \ \  ({\un b}_1 - {\un z}_2)  \cdot ({\un b}_3 - {\un w}_1) \notag \\ & - ({\un b}_1 - {\un x}_1) \cdot ({\un b}_3 - {\un y}_2) \ \ ({\un b}_2 - {\un x}_2) \cdot ({\un b}_3 - {\un w}_1) \ \  ({\un b}_1 - {\un z}_2)  \cdot ({\un b}_2 - {\un y}_1) \notag \\ & - ({\un b}_1 - {\un x}_1) \cdot ({\un b}_3 - {\un w}_1) \ \ ({\un b}_2 - {\un x}_2) \cdot ({\un b}_3 - {\un y}_2) \ \ ({\un b}_1 - {\un z}_2)  \cdot ({\un b}_2 - {\un y}_1) \notag \\ & + 2 \, ({\un b}_1 - {\un x}_1) \cdot ({\un b}_2 - {\un x}_2)  \ \ ({\un b}_3 - {\un y}_2) \cdot ({\un b}_3 - {\un w}_1) \ \ ({\un b}_1 - {\un z}_2)  \cdot ({\un b}_2 - {\un y}_1) \notag \\ & - ({\un b}_1 - {\un x}_1) \cdot ({\un b}_2 - {\un y}_1) \ \ ({\un b}_2 - {\un x}_2) \cdot ({\un b}_3 - {\un w}_1)  \ \ ({\un b}_1 - {\un z}_2)  \cdot ({\un b}_3 - {\un y}_2) \notag \\ & + 2 \, ({\un b}_1 - {\un x}_1) \cdot ({\un b}_3 - {\un w}_1) \ \ ({\un b}_2 - {\un x}_2) \cdot ({\un b}_2 - {\un y}_1)  \ \ ({\un b}_1 - {\un z}_2)  \cdot ({\un b}_3 - {\un y}_2) \notag \\ & - ({\un b}_1 - {\un x}_1) \cdot ({\un b}_2 - {\un x}_2)  \ \ ({\un b}_2 - {\un y}_1) \cdot ({\un b}_3 - {\un w}_1)  \ \ ({\un b}_1 - {\un z}_2)  \cdot ({\un b}_3 - {\un y}_2).
\end{align}
Alternatively, these results can be obtained by expanding the Wilson
lines in \eq{int2} to the first nontrivial order in the target field
and averaging with respect to the target ensemble; this leaves rather
complicated color sums involving eight anti-symmetric structure
constants, see Appendix~\ref{Ap:Color}.  This method is used to
compute the diagrams B and C below.

In arriving at \eq{intA} we have neglected the logarithms of \eq{Ds}
by putting them equal to 1. Henceforth we will refer to this
approximation as the Golec-Biernat--Wusthoff (GBW) approximation,
since it was used in
\myrefs\cite{GolecBiernat:1998js,GolecBiernat:1999qd} for the
Glauber--Mueller dipole amplitude \cite{Mueller:1989st} to
successfully describe the small-$x$ HERA data on structure
functions. Despite this phenomenological success, this is admittedly a
dangerous approximation for observables depending on transverse
momenta: the expansion to the lowest order of gluon exchanges must be
valid at high transverse momenta, where logarithms from \eq{Ds} are
most important. For a number of transverse-momentum dependent gluon
distributions (gluon TMDs) at small $x$, evaluated in the
quasi-classical approximation, neglecting such logarithms either leads
to a zero result (e.g. for $h^{(1)}_g$, as discussed in
\myref\cite{Dumitru:2016jku}) or to an incorrect high-$k_T$
asymptotics, which becomes Gaussian in $k_T$ instead of giving the
correct $\sim 1/k_T^2$ power law. We are encouraged, however, by the
fact that such problem does not arise for the single inclusive gluon
production cross section
\cite{Kovchegov:1998bi,Jalilian-Marian:2005jf}, where neglecting
transverse coordinate logarithms still leads to the correct power-law
high-$k_T$ asymptotics (but does not capture the factor of $\ln
(k_T/\Lambda)$).

Substituting \eq{int2} into \eq{4M2} we obtain
\begin{align}\label{4M3}
& \frac{g^8}{32} \, Q_{s0}^6 \, \int d^2 z_1 \, d^2 w_1 \, d^2 z_2 \, d^2 w_2 \, e^{- i {\un k}_1 \cdot ({\un z}_1 - {\un w}_1) - i {\un k}_2 \cdot ({\un z}_2 - {\un w}_2)} \sum_{\lambda , \lambda'} \notag \\ &   \int d^2 x_1 d^2 x_2 \! \int
  \frac{d^2 k}{(2 \pi)^2} \, \frac{d^2 l}{(2 \pi)^2} \, \frac{d^2
    q_1}{(2 \pi)^2} \, \frac{d^2 q_2}{(2 \pi)^2} \, e^{i {\un q}_{1} \cdot ({\un x}_{1} - {\un b}_{1}) + i
    {\un q}_{2} \cdot ({\un x}_{2} - {\un b}_{2}) +
    i {\un l} \cdot ({\un x}_{2} - {\un x}_{1}) +
    i {\un k} \cdot ({\un z}_{1} - {\un x}_{2})}
  \notag \\ & \times \frac{1}{{\un q}_{1}^2 \, {\un q}_{2}^2} \, \left(
    - {\un q}_{1} \cdot {\un q}_{2} \, \frac{{\un \epsilon}_\lambda^* \times {\un k}}{{\un k}^2} + {\un \epsilon}_\lambda^*
    \cdot {\un q}_{1} \, \frac{{\un q}_{2} \times ({\un k} - {\un l})}{({\un k} - {\un l})^2} + {\un \epsilon}_\lambda^* \cdot
    {\un q}_{2} \, \frac{{\un q}_{1} \times {\un l}}{{\un l}^2} \right) \, \mbox{Sign} ({\un k} \times {\un l})   \notag \\ & 
  \int \frac{d^2 q_3} {(2\pi)^2} \, e^{- i {\un q}_3 \cdot ({\un w}_1 - {\un b}_3)} 
	\frac{{\un \epsilon}_\lambda \cdot {\un q}_3 }{{ \un q}_3^{2}}  \int \frac{d^2 q'_3} {(2\pi)^2} \, e^{i {\un q}'_3 \cdot ({\un z}_2 - {\un b}_1)} 
	\frac{{\un \epsilon}^*_{\lambda'} \cdot {\un q}'_3 }{{ \un q}_3^{\prime \, 2}} 
 \notag \\ & \int d^2 y_1 d^2 y_2 \! \int
  \frac{d^2 k'}{(2 \pi)^2} \, \frac{d^2 l'}{(2 \pi)^2} \, \frac{d^2
    q'_1}{(2 \pi)^2} \, \frac{d^2 q'_2}{(2 \pi)^2} \, e^{-i {\un q}'_{1} \cdot ({\un y}_{1} - {\un b}_{2}) - i
    {\un q}'_{2} \cdot ({\un y}_{2} - {\un b}_{3}) -
    i {\un l}^{\, \prime}_{} \cdot ({\un y}_{2} - {\un y}_{1}) -
    i {\un k}^{\, \prime}_{} \cdot ({\un w}_{2} - {\un y}_{2})}
  \notag \\ & \times \frac{1}{{\un q}_{1}^{\prime \, 2} \, {\un q}_{2}^{\prime \, 2}} \, \left(
  - {\un q}^{\, \prime}_{1} \cdot {\un q}^{\, \prime}_{2} \, \frac{{\un \epsilon}^{\, \lambda'} \times {\un k}^{\, \prime}}{{\un k}^{\prime \, 2}} + {\un \epsilon}^{\, \lambda'}
    \cdot {\un q}^{\, \prime}_{1} \, \frac{{\un q}^{\, \prime}_{2} \times ({\un k}^{\, \prime}_{} - {\un l}^{\, \prime})}{({\un k}^{\, \prime} - {\un l}^{\, \prime})^2} + {\un \epsilon}^{\, \lambda'} \cdot
    {\un q}^{\, \prime}_{2} \, \frac{{\un q}^{\, \prime}_{1} \times {\un l}^{\, \prime}_{}}{{\un l}^{\prime \, 2}} \right) \, \mbox{Sign} ({\un k}^{\, \prime} \times {\un l}^{\, \prime})   \notag \\ & \times \mbox{Int}_A  .
\end{align}

It is more convenient to evaluate the expression \eqref{4M3} by
pieces, which is allowed by the factorized form of each term in
\eq{intA}. Using
\begin{equation}
	 x_i \int  \frac{d^2 q} {(2\pi)^2} e^{i {\un x} \cdot {\un q }}  f({\un q}) = 
	i \int  \frac{d^2 q} {(2\pi)^2} e^{i {\un x} \cdot {\un q }}  \frac{\partial f({\un q})}{\partial q_i}  
\end{equation}
for any well-behaved function $f({\un q})$ we can evaluate
\begin{align}\label{M3mod}
& \int d^2 z_1 \, e^{- i {\un k}_1 \cdot {\un z}_1} \, \int d^2 x_1 d^2 x_2 \! \int
  \frac{d^2 k}{(2 \pi)^2} \, \frac{d^2 l}{(2 \pi)^2} \, \frac{d^2
    q_1}{(2 \pi)^2} \, \frac{d^2 q_2}{(2 \pi)^2} \, ({\un b}_1 - {\un x}_1)^i \, ({\un b}_2 - {\un x}_2)^j \notag \\ & \times \, e^{i {\un q}_{1} \cdot ({\un x}_{1} - {\un b}_{1}) + i
    {\un q}_{2} \cdot ({\un x}_{2} - {\un b}_{2}) +
    i {\un l} \cdot ({\un x}_{2} - {\un x}_{1}) +
    i {\un k} \cdot ({\un z}_{1} - {\un x}_{2})}
  \notag \\ & \times \frac{1}{{\un q}_{1}^2 \, {\un q}_{2}^2} \, \left(
    - {\un q}_{1} \cdot {\un q}_{2} \, \frac{{\un \epsilon}_\lambda^* \times {\un k}}{{\un k}^2} + {\un \epsilon}_\lambda^*
    \cdot {\un q}_{1} \, \frac{{\un q}_{2} \times ({\un k} - {\un l})}{({\un k} - {\un l})^2} + {\un \epsilon}_\lambda^* \cdot
    {\un q}_{2} \, \frac{{\un q}_{1} \times {\un l}}{{\un l}^2} \right) \, \mbox{Sign} ({\un k} \times {\un l}) \notag \\ & = - \int \frac{d^2 l}{(2 \pi)^2} \, e^{i {\un l} \cdot {\un b}_{21} - i {\un k}_1 \cdot {\un b}_2} \bigg[ - \frac{{\un \epsilon}_\lambda^* \times {\un k}_1}{{\un k}_1^2} \, \left( \frac{\delta^{ij}}{{\un l}^2 \, ({\un k}_1 - {\un l})^2} - \frac{2 \, l^i \, l^j}{{\un l}^4 \, ({\un k}_1 - {\un l})^2} - \frac{2 \, (k_1-l)^i \, (k_1-l)^j}{{\un l}^2 \, ({\un k}_1 - {\un l})^4} + \frac{4 \, l^i \, (k_1-l)^j \, l \cdot (k_1-l)}{{\un l}^4 \, ({\un k}_1 - {\un l})^4} \right) \notag \\ & + \frac{\left[ {\un l}^2 \epsilon^{* \, i}_\lambda - 2 {\un \epsilon}_\lambda^* \cdot {\un l} \, l^i \right] \, \epsilon^{jm} \, (k_1-l)^m}{{\un l}^4 \, ({\un k}_1 - {\un l})^4} + \frac{\left[ ({\un k}_1 - {\un l})^2 \epsilon^{* \, j}_\lambda - 2 {\un \epsilon}_\lambda^* \cdot ({\un k}_1 - {\un l}) \, (k_1 -l)^j \right] \, \epsilon^{im} \, l^m}{{\un l}^4 \, ({\un k}_1 - {\un l})^4} \bigg]  \, \mbox{Sign} ({\un k}_{1} \times {\un l}).
\end{align}
The other term depending on the same gluon polarization $\lambda$ is
\begin{align}\label{M1mod}
\int d^2 w_1 \, e^{i {\un k}_1 \cdot {\un w}_1} \int \frac{d^2 q_3} {(2\pi)^2} \, e^{- i {\un q}_3 \cdot ({\un w}_1 - {\un b}_3)} 
	\frac{{\un \epsilon}_\lambda \cdot {\un q}_3 }{{ \un q}_3^{2}} \, ({\un b}_3 - {\un w}_1)^n = i \, e^{i {\un k}_1 \cdot {\un b}_3} \ \frac{{\un k}_1^2 \, \epsilon^{n}_\lambda - 2 {\un \epsilon}_\lambda \cdot {\un k}_1 \, k_1^n }{{\un k}_1^4}. 
\end{align}
Using 
\begin{align}
\sum_\lambda \epsilon^{* \, i}_\lambda \, \epsilon^j_\lambda = \delta^{ij}
\end{align}
we can multiply \eq{M3mod} by \eq{M1mod} and sum over polarizations obtaining
\begin{align}\label{M3M1mod}
& \int d^2 z_1 \, d^2 w_1  \, e^{- i {\un k}_1 \cdot ({\un z}_1 - {\un w}_1)} \, \int d^2 x_1 d^2 x_2 \! \int
  \frac{d^2 k}{(2 \pi)^2} \, \frac{d^2 l}{(2 \pi)^2} \, \frac{d^2
    q_1}{(2 \pi)^2} \, \frac{d^2 q_2}{(2 \pi)^2} \, \frac{d^2 q_3} {(2\pi)^2} \, e^{- i {\un q}_3 \cdot ({\un w}_1 - {\un b}_3)} 
	\frac{{\un \epsilon}_\lambda \cdot {\un q}_3 }{{ \un q}_3^{2}}  \notag \\ & \times \, e^{i {\un q}_{1} \cdot ({\un x}_{1} - {\un b}_{1}) + i
    {\un q}_{2} \cdot ({\un x}_{2} - {\un b}_{2}) +
    i {\un l} \cdot ({\un x}_{2} - {\un x}_{1}) +
    i {\un k} \cdot ({\un z}_{1} - {\un x}_{2})} \, ({\un b}_1 - {\un x}_1)^i \, ({\un b}_2 - {\un x}_2)^j \, ({\un b}_3 - {\un w}_1)^n
  \notag \\ & \times \frac{1}{{\un q}_{1}^2 \, {\un q}_{2}^2} \, \left(
    - {\un q}_{1} \cdot {\un q}_{2} \, \frac{{\un \epsilon}_\lambda^* \times {\un k}}{{\un k}^2} + {\un \epsilon}_\lambda^*
    \cdot {\un q}_{1} \, \frac{{\un q}_{2} \times ({\un k} - {\un l})}{({\un k} - {\un l})^2} + {\un \epsilon}_\lambda^* \cdot
    {\un q}_{2} \, \frac{{\un q}_{1} \times {\un l}}{{\un l}^2} \right) \, \mbox{Sign} ({\un k} \times {\un l}) \notag \\ & = - i \int \frac{d^2 l}{(2 \pi)^2} \, e^{i {\un l} \cdot {\un b}_{21} - i {\un k}_1 \cdot {\un b}_{23}} \, H^{ijn} ({\un k}_1, {\un l}),
\end{align}
where we have defined
\begin{align}\label{Hdef}
& H^{ijn} ({\un k}_1, {\un l}) \equiv  \bigg[  - \frac{\epsilon^{nm} k^m_1}{{\un k}_1^4} \, \left( \frac{\delta^{ij}}{{\un l}^2 \, ({\un k}_1 - {\un l})^2} - \frac{2 \, l^i \, l^j}{{\un l}^4 \, ({\un k}_1 - {\un l})^2} - \frac{2 \, (k_1-l)^i \, (k_1-l)^j}{{\un l}^2 \, ({\un k}_1 - {\un l})^4} + \frac{4 \, l^i \, (k_1-l)^j \, l \cdot (k_1-l)}{{\un l}^4 \, ({\un k}_1 - {\un l})^4} \right) \notag \\ & + \frac{\left[ {\un l}^2 \delta^{ni} - 2 \, l^n \, l^i \right] \, \epsilon^{jm} \, (k_1-l)^m}{{\un k}_1^2 \, {\un l}^4 \, ({\un k}_1 - {\un l})^4} + \frac{\left[ ({\un k}_1 - {\un l})^2 \delta^{nj} - 2 \, ({\un k}_1 - {\un l})^n \, (k_1 -l)^j \right] \, \epsilon^{im} \, l^m}{{\un k}_1^2 \,  {\un l}^4 \, ({\un k}_1 - {\un l})^4} - 2 \, k_1^n \frac{\left[ {\un l}^2 k_1^{i} - 2 \, {\un k}_1 \cdot {\un l} \, l^i \right] \, \epsilon^{jm} \, (k_1-l)^m}{{\un k}_1^4 \, {\un l}^4 \, ({\un k}_1 - {\un l})^4} \notag \\ & - 2 \, k_1^n \frac{\left[ ({\un k}_1 - {\un l})^2 k_1^{j} - 2 \, {\un k}_1 \cdot ({\un k}_1 - {\un l}) \, (k_1 -l)^j \right] \, \epsilon^{im} \, l^m}{{\un k}_1^4 \,  {\un l}^4 \, ({\un k}_1 - {\un l})^4} \bigg] \, \mbox{Sign} ({\un k}_{1} \times {\un l}) .
\end{align}
Here $\epsilon^{ij}$ is the Levi-Civita symbol  in two dimensions.

Using this result in \eq{4M3} yields
\begin{align}\label{4M4}
& \frac{g^8}{32} \, Q_{s0}^6 \, \int \frac{d^2 l}{(2 \pi)^2} \, e^{i {\un l} \cdot {\un b}_{21} - i {\un k}_1 \cdot {\un b}_{23}}  \, \int \frac{d^2 l'}{(2 \pi)^2} \, e^{-i {\un l}' \cdot {\un b}_{32} + i {\un k}_2 \cdot {\un b}_{31}} \, H^{ijn} ({\un k}_1, {\un l}) \, H^{* \, i'j'n'} ({\un k}_2, {\un l}') \, \big[ - \delta^{ij'} \,  \delta^{jn'} \,  \delta^{ni'} + 2 \,  \delta^{in'} \,  \delta^{jn} \,  \delta^{i'j'} \notag \\ & - \delta^{in} \,  \delta^{jn'} \,  \delta^{i'j'} + 8 \, \delta^{in'} \,  \delta^{nj'} \,  \delta^{ji'} - \delta^{ii'} \,  \delta^{jj'} \,  \delta^{nn'} - \delta^{ij} \,  \delta^{i'j'} \,  \delta^{nn'} - \delta^{ij'} \,  \delta^{jn} \,  \delta^{n'i'} - \delta^{in} \,  \delta^{jj'} \,  \delta^{i'n'} + 2 \, \delta^{ij} \,  \delta^{nj'} \,  \delta^{i'n'} \notag \\ &  - \delta^{ii'} \,  \delta^{jn} \,  \delta^{j'n'} + 2 \, \delta^{in} \,  \delta^{ji'} \,  \delta^{n'j'} -  \delta^{ij} \,  \delta^{i'n} \,  \delta^{n'j'} \big]  .
\end{align}

Next, in \eq{full_expr} we approximate (cf. \myref\cite{Kovchegov:2012nd})
\begin{align}\label{appr}
\int d^2 b_1 \, d^2 b_2 \, d^2 b_3 \, T_1 ({\un B} - {\un b}_1) \, T_1 ({\un B} - {\un b}_2) \, T_1 ({\un B} - {\un b}_3) \approx \int d^2 b \, [T_1 ({\un B} - {\un b})]^3 \, d^2 b_{12} \, d^2 b_{23}
\end{align}
where ${\un b}_{ij} = {\un b}_i - {\un b}_j$. The rationale here is
that the nuclear profile function $T_1 ({\un b})$ does not vary much
across the size of a single nucleon. At the same time, from the
standpoint of our perturbative calculation, integrating over the
impact parameter over the distances comparable to the nucleon radius
is approximately equivalent to integrating to infinity.

Integrating \eq{4M4} over $b_{12}$ and $b_{23}$, and over $l$ with
$l'$, we arrive at
\begin{align}\label{HH}
& \frac{g^8}{32} \, Q_{s0}^6 \, H^{ijn} ({\un k}_1, - {\un k}_2) \, H^{* \, i'j'n'} ({\un k}_2, {\un k}_1 + {\un k}_2 ) \, \big[ - \delta^{ij'} \,  \delta^{jn'} \,  \delta^{ni'} + 2 \,  \delta^{in'} \,  \delta^{jn} \,  \delta^{i'j'} \notag \\ & - \delta^{in} \,  \delta^{jn'} \,  \delta^{i'j'} + 8 \, \delta^{in'} \,  \delta^{nj'} \,  \delta^{ji'} - \delta^{ii'} \,  \delta^{jj'} \,  \delta^{nn'} - \delta^{ij} \,  \delta^{i'j'} \,  \delta^{nn'} - \delta^{ij'} \,  \delta^{jn} \,  \delta^{n'i'} - \delta^{in} \,  \delta^{jj'} \,  \delta^{i'n'} + 2 \, \delta^{ij} \,  \delta^{nj'} \,  \delta^{i'n'} \notag \\ &  - \delta^{ii'} \,  \delta^{jn} \,  \delta^{j'n'} + 2 \, \delta^{in} \,  \delta^{ji'} \,  \delta^{n'j'} -  \delta^{ij} \,  \delta^{i'n} \,  \delta^{n'j'} \big]  .
\end{align}

Summing over all the indices in \eq{HH} yields
\begin{align}\label{HH1}
& \frac{g^8}{2} \, Q_{s0}^6 \, \frac{({\un k}_1^2 + {\un k}_2^2 + {\un k}_1 \cdot {\un k}_2)^2}{{\un k}_1^6 \, {\un k}_2^6 \, ({\un k}_1 + {\un k}_2)^6} .
\end{align}

Finally, anti-symmetrizing \eq{HH1} under ${\un k}_1 \to -{\un k}_1$
and ${\un k}_2 \to -{\un k}_2$ according to the prescription
\eqref{4M1} we obtain the expression for the contribution to the
two-gluon production cross section \eqref{full_expr} of the first four
M-terms corresponding to the diagram A in \fig{all_graphs},
\begin{align}\label{Adiag}
\int d^2 b_{12} \, d^2 b_{23} \, A =  g^8 \, \frac{Q_{s0}^6}{{\un k}_1^6 \, {\un k}_2^6} \, \left[ \frac{({\un k}_1^2 + {\un k}_2^2 + {\un k}_1 \cdot {\un k}_2)^2}{({\un k}_1 + {\un k}_2)^6} - \frac{({\un k}_1^2 + {\un k}_2^2 - {\un k}_1 \cdot {\un k}_2)^2}{({\un k}_1 - {\un k}_2)^6} \right].
\end{align}
For brevity we are omitting the factor of $\int d^2 B \, d^2 b \,
\left[ T_1 ({\un B} - {\un b}) \right]^3 /[2 (2 \pi)^3]^2$ which will
be reinstated later.


\subsection{Diagram B}

Now let us evaluate the diagram B from \fig{all_graphs}, along with
all of its reflections with respect to the final-state cut, in the
same GBW approximation. The first of the next four M-terms from
\eq{Ahat} gives
\begin{align}\label{second4M}
& \frac{(4 \, g^4)^2}{(2 N_c)^3} \! \int d^2 z_1 \, d^2 w_1 \, d^2 z_2 \, d^2 w_2 \, e^{- i {\un k}_1 \cdot ({\un z}_1 - {\un w}_1) - i {\un k}_2 \cdot ({\un z}_2 - {\un w}_2)} \sum_{\lambda , \lambda'} \notag \\ &   \int d^2 x_1 d^2 x_2 \! \int
  \frac{d^2 k}{(2 \pi)^2} \, \frac{d^2 l}{(2 \pi)^2} \, \frac{d^2
    q_1}{(2 \pi)^2} \, \frac{d^2 q_2}{(2 \pi)^2} \, e^{i {\un q}_{1} \cdot ({\un x}_{1} - {\un b}_{1}) + i
    {\un q}_{2} \cdot ({\un x}_{2} - {\un b}_{2}) +
    i {\un l} \cdot ({\un x}_{2} - {\un x}_{1}) +
    i {\un k} \cdot ({\un z}_{1} - {\un x}_{2})}
  \notag \\ & \times \frac{1}{{\un q}_{1}^2 \, {\un q}_{2}^2} \, \left(
    - {\un q}_{1} \cdot {\un q}_{2} \, \frac{{\un \epsilon}_\lambda^* \times {\un k}}{{\un k}^2} + {\un \epsilon}_\lambda^*
    \cdot {\un q}_{1} \, \frac{{\un q}_{2} \times ({\un k} - {\un l})}{({\un k} - {\un l})^2} + {\un \epsilon}_\lambda^* \cdot
    {\un q}_{2} \, \frac{{\un q}_{1} \times {\un l}}{{\un l}^2} \right) \, \mbox{Sign} ({\un k} \times {\un l})   \notag \\ & 
  \int \frac{d^2 q_3} {(2\pi)^2} \, e^{- i {\un q}_3 \cdot ({\un w}_1 - {\un b}_1)} 
	\frac{{\un \epsilon}_\lambda \cdot {\un q}_3 }{{ \un q}_3^{2}}  \int \frac{d^2 q'_3} {(2\pi)^2} \, e^{i {\un q}'_3 \cdot ({\un z}_2 - {\un b}_3)} 
	\frac{{\un \epsilon}^*_{\lambda'} \cdot {\un q}'_3 }{{ \un q}_3^{\prime \, 2}} 
 \notag \\ & \int d^2 y_1 d^2 y_2 \! \int
  \frac{d^2 k'}{(2 \pi)^2} \, \frac{d^2 l'}{(2 \pi)^2} \, \frac{d^2
    q'_1}{(2 \pi)^2} \, \frac{d^2 q'_2}{(2 \pi)^2} \, e^{- i {\un q}'_{1} \cdot ({\un y}_{1} - {\un b}_{2}) - i
    {\un q}'_{2} \cdot ({\un y}_{2} - {\un b}_{3}) -
    i {\un l}^{\, \prime}_{} \cdot ({\un y}_{2} - {\un y}_{1}) -
    i {\un k}^{\, \prime}_{} \cdot ({\un w}_{2} - {\un y}_{2})}
  \notag \\ & \times \frac{1}{{\un q}_{1}^{\prime \, 2} \, {\un q}_{2}^{\prime \, 2}} \, \left(
  - {\un q}^{\, \prime}_{1} \cdot {\un q}^{\, \prime}_{2} \, \frac{{\un \epsilon}^{\, \lambda'} \times {\un k}^{\, \prime}}{ {\un k}^{\prime \, 2}} + {\un \epsilon}^{\, \lambda'}
    \cdot {\un q}^{\, \prime}_{1} \, \frac{{\un q}^{\, \prime}_{2} \times ({\un k}^{\, \prime}_{} - {\un l}^{\, \prime})}{({\un k}^{\, \prime} - {\un l}^{\, \prime})^2} + {\un \epsilon}^{\, \lambda'} \cdot
    {\un q}^{\, \prime}_{2} \, \frac{{\un q}^{\, \prime}_{1} \times {\un l}^{\, \prime}_{}}{{\un l}^{\prime \, 2}} \right) \, \mbox{Sign} ({\un k}^{\, \prime} \times {\un l}^{\, \prime})   \notag \\ & \times \left\langle f^{abc} \,
  \left[ U^{bd}_{{\un x}_{1}} - U^{bd}_{{\un b}_{1}}
  \right] \, \left[ U^{ce}_{{\un x}_{2}} - U^{ce}_{{\un b}_{2}} \right] \, \left[ 
	U^{a'd'}_{{\un z}_2}  -  U^{a'd'}_{{\un b}_3} 
\right] \, f^{a'b'c'} \, \left[ U^{b'e}_{{\un y}_{1}} - U^{b'e}_{{\un b}_{2}}
  \right] \, \left[ U^{c'd'}_{{\un y}_{2}} - U^{c'd'}_{{\un b}_{3}} \right] \, \left[ 
	U^{ad}_{{\un w}_1}  -  U^{ad}_{{\un b}_1} 
\right] \right\rangle.
\end{align}
Color structure of the interaction with the target is different from
the case of diagram A,
\begin{align}
& \left\langle f^{abc} \,
  \left[ U^{bd}_{{\un x}_{1}} - U^{bd}_{{\un b}_{1}}
  \right] \, \left[ U^{ce}_{{\un x}_{2}} - U^{ce}_{{\un b}_{2}} \right] \, \left[ 
	U^{a'd'}_{{\un z}_2}  -  U^{a'd'}_{{\un b}_3} 
\right] \, f^{a'b'c'} \, \left[ U^{b'e}_{{\un y}_{1}} - U^{b'e}_{{\un b}_{2}}
  \right] \, \left[ U^{c'd'}_{{\un y}_{2}} - U^{c'd'}_{{\un b}_{3}} \right] \, \left[ 
	U^{ad}_{{\un w}_1}  -  U^{ad}_{{\un b}_1} \right] \right\rangle \notag \\ & = \frac{N_c^3}{64} \, Q_{s0}^6 \ \mbox{Int}_B
\end{align}
with 
\begin{align}
\mbox{Int}_B  = & \ ({\un b}_1 - {\un x}_1) \cdot ({\un b}_3 - {\un z}_2) \ \ ({\un b}_2 - {\un x}_2) \cdot ({\un b}_3 - {\un y}_2) \ \ ({\un b}_2 - {\un y}_1) \cdot ({\un b}_1 - {\un w}_1) \notag \\ & - ({\un b}_1 - {\un x}_1) \cdot ({\un b}_3 - {\un y}_2) \ \ ({\un b}_2 - {\un x}_2) \cdot ({\un b}_3 - {\un z}_2) \ \ ({\un b}_2 - {\un y}_1) \cdot ({\un b}_1 - {\un w}_1) \notag \\ & + 2 ({\un b}_1 - {\un x}_1) \cdot ({\un b}_3 - {\un z}_2) \ \ ({\un b}_2 - {\un x}_2) \cdot ({\un b}_1 - {\un w}_1) \ \ ({\un b}_2 - {\un y}_1) \cdot ({\un b}_3 - {\un y}_2) \notag \\ & + 2 ({\un b}_1 - {\un x}_1) \cdot ({\un b}_3 - {\un z}_2) \ \ ({\un b}_2 - {\un x}_2) \cdot ({\un b}_2 - {\un y}_1) \ \ ({\un b}_3 - {\un y}_2) \cdot ({\un b}_1 - {\un w}_1) \notag \\ & + ({\un b}_1 - {\un x}_1) \cdot ({\un b}_2 - {\un y}_1) \ \ ({\un b}_2 - {\un x}_2) \cdot ({\un b}_3 - {\un z}_2) \ \ ({\un b}_3 - {\un y}_2) \cdot ({\un b}_1 - {\un w}_1) \notag \\ & - 2 ({\un b}_1 - {\un x}_1) \cdot ({\un b}_3 - {\un y}_2) \ \ ({\un b}_2 - {\un x}_2) \cdot ({\un b}_2 - {\un y}_1) \ \ ({\un b}_3 - {\un z}_2) \cdot ({\un b}_1 - {\un w}_1) \notag \\ & - ({\un b}_1 - {\un x}_1) \cdot ({\un b}_2 - {\un y}_1) \ \ ({\un b}_2 - {\un x}_2) \cdot ({\un b}_3 - {\un y}_2) \ \ ({\un b}_3 - {\un z}_2) \cdot ({\un b}_1 - {\un w}_1) \notag \\ & - 2 ({\un b}_1 - {\un x}_1) \cdot ({\un b}_2 - {\un x}_2) \ \ ({\un b}_2 - {\un y}_1) \cdot ({\un b}_3 - {\un y}_2) \ \ ({\un b}_3 - {\un z}_2) \cdot ({\un b}_1 - {\un w}_1) \notag \\ & - 2 ({\un b}_1 - {\un x}_1) \cdot ({\un b}_3 - {\un y}_2) \ \ ({\un b}_2 - {\un x}_2) \cdot ({\un b}_1 - {\un w}_1) \ \ ({\un b}_3 - {\un z}_2) \cdot ({\un b}_2 - {\un y}_1) \notag \\ & + 2 ({\un b}_1 - {\un x}_1) \cdot ({\un b}_2 - {\un x}_2) \ \ ({\un b}_3 - {\un y}_2) \cdot ({\un b}_1 - {\un w}_1) \ \ ({\un b}_3 - {\un z}_2) \cdot ({\un b}_2 - {\un y}_1) 
\end{align}
again in the GBW approximation.

Employing the technique from the previous Subsection to evaluate
equation \eqref{second4M} yields
\begin{align}\label{BB1}
& \frac{g^8 \, Q_{s0}^6}{32} \int \frac{d^2 l}{(2 \pi)^2} \, e^{i ({\un l} -{\un k}_1) \cdot {\un b}_{21}}  \, \int \frac{d^2 l'}{(2 \pi)^2} \, e^{-i {\un l}' \cdot {\un b}_{32} } \, H^{ijn} ({\un k}_1, {\un l}) \, H^{* \, i'j'n'} ({\un k}_2, {\un l}') \ [ \delta^{in'} \, \delta^{jj'} \, \delta^{ni'} -  \delta^{ij'} \, \delta^{jn'} \, \delta^{ni'} \notag \\ &  + 2 \, \delta^{in'} \, \delta^{jn} \, \delta^{i'j'} + 2 \, \delta^{in'} \, \delta^{ji'} \, \delta^{j'n} + \delta^{ii'} \, \delta^{jn'} \, \delta^{nj'} - 2 \, \delta^{ij'} \, \delta^{ji'} \, \delta^{nn'} - \delta^{ii'} \, \delta^{jj'} \, \delta^{nn'} - 2 \, \delta^{ij} \, \delta^{i'j'} \, \delta^{nn'} - 2 \, \delta^{ij'} \, \delta^{jn} \, \delta^{i'n'} \notag \\ &  + 2 \, \delta^{ij} \, \delta^{nj'} \, \delta^{i'n'}],
\end{align}
where $H^{ijn}$ and $H^{* \, i'j'n'}$ are defined by
\eq{Hdef}. Integrating \eq{BB1} over $b_{12}$ and $b_{23}$ has to be
done with care: naive integration over these variables up to infinity
puts ${\un l} = {\un k}_1$ and ${\un l}' =0$, which are ill-defined
limits in $H^{ijn}$ and $H^{* \, i'j'n'}$. To deal with this issue we
integrate over $b_{12}$ and $b_{23}$ in the range $0 < b_{ij} < R$
with $R$ the projectile radius obtaining~\footnote{In order to
  simplify the notation, we define the magnitude of a two-dimensional
  vector ${\un x}$ as $x = |{\un x}|$.}
\begin{align}\label{BB11}
& \frac{g^8 \, Q_{s0}^6}{32} \int \frac{d^2 l}{2 \pi} \, \frac{R}{|{\un l} -{\un k}_1|} \, J_1 (|{\un l} -{\un k}_1| \, R) \, \int \frac{d^2 l'}{2 \pi} \, \frac{R}{l'} \, J_1 (l' \, R) \, H^{ijn} ({\un k}_1, {\un l}) \, H^{* \, i'j'n'} ({\un k}_2, {\un l}') \ [ \delta^{in'} \, \delta^{jj'} \, \delta^{ni'} -  \delta^{ij'} \, \delta^{jn'} \, \delta^{ni'} \notag \\ &  + 2 \, \delta^{in'} \, \delta^{jn} \, \delta^{i'j'} + 2 \, \delta^{in'} \, \delta^{ji'} \, \delta^{j'n} + \delta^{ii'} \, \delta^{jn'} \, \delta^{nj'} - 2 \, \delta^{ij'} \, \delta^{ji'} \, \delta^{nn'} - \delta^{ii'} \, \delta^{jj'} \, \delta^{nn'} - 2 \, \delta^{ij} \, \delta^{i'j'} \, \delta^{nn'} - 2 \, \delta^{ij'} \, \delta^{jn} \, \delta^{i'n'} \notag \\ &  + 2 \, \delta^{ij} \, \delta^{nj'} \, \delta^{i'n'}].
\end{align}

Note that the rest of the four terms in \eq{Ahat} that correspond to
diagram B in \fig{all_graphs} anti-symmetrize \eq{BB11} under ${\un
  k}_1 \to -{\un k}_1$ and ${\un k}_2 \to -{\un k}_2$, giving
\begin{align}\label{BB12}
& \frac{g^8 \, Q_{s0}^6}{32} \int \frac{d^2 l}{2 \pi} \, \frac{R}{|{\un l} -{\un k}_1|} \, J_1 (|{\un l} -{\un k}_1| \, R) \, \int \frac{d^2 l'}{2 \pi} \, \frac{R}{l'} \, J_1 (l' \, R) \, \left[ H^{ijn} ({\un k}_1, {\un l}) - H^{ijn} (-{\un k}_1, {\un l})  \right] \notag \\ & \times \, \left[ H^{* \, i'j'n'} ({\un k}_2, {\un l}') - H^{* \, i'j'n'} (-{\un k}_2, {\un l}') \right] \ [ \delta^{in'} \, \delta^{jj'} \, \delta^{ni'} -  \delta^{ij'} \, \delta^{jn'} \, \delta^{ni'}   + 2 \, \delta^{in'} \, \delta^{jn} \, \delta^{i'j'} + 2 \, \delta^{in'} \, \delta^{ji'} \, \delta^{j'n} \notag \\ & + \delta^{ii'} \, \delta^{jn'} \, \delta^{nj'} - 2 \, \delta^{ij'} \, \delta^{ji'} \, \delta^{nn'} - \delta^{ii'} \, \delta^{jj'} \, \delta^{nn'} - 2 \, \delta^{ij} \, \delta^{i'j'} \, \delta^{nn'} - 2 \, \delta^{ij'} \, \delta^{jn} \, \delta^{i'n'} + 2 \, \delta^{ij} \, \delta^{nj'} \, \delta^{i'n'}].
\end{align}

Employing
\begin{align}
H^{ijn} (-{\un k}, -{\un l}) = - H^{ijn} ({\un k}, {\un l}) .
\end{align}
we can write an essential part of \eq{BB12} as
\begin{align}\label{HHangles}
	\int \frac{d^2 l'}{2 \pi} \, \frac{R}{l'} \, J_1 (l' \, R) \,  \left[ H^{* \, i'j'n'} ({\un k}_2, {\un l}') - H^{* \, i'j'n'} (-{\un k}_2, {\un l}') \right] = \int \frac{d^2 l'}{2 \pi} \, \frac{R}{l'} \, J_1 ( l'  R) \, \left[ H^{* \, i'j'n'} ({\un k}_2, {\un l}') + H^{* \, i'j'n'} ({\un k}_2, -{\un l}') \right] .
\end{align}
We will first integrate over the angles of ${\un l}'$ and then take
the $R\to\infty$ limit. To integrate \eq{HHangles} over the angles we
note that
\begin{align}\label{phiH} 
\int\limits_0^{2 \pi} d \phi_l \, H^{ijn} ({\un k}, {\un l}) =  \frac{k^i \, \delta^{jn}}{k^4} \, \frac{4}{l^4} \, \mbox{arctanh} \left( \frac{q_<}{q_>} \right) - \frac{k^j \, \delta^{in}}{k^3}  \, \frac{4}{l} \, \frac{1}{(k^2 - l^2)^2} ,
\end{align}
where $\phi_l$ is the azimuthal angle of the transverse momentum
vector ${\un l}$, while $q_> = \max \{ k, l \}$ and $q_< = \min \{ k,
l \}$. In arriving at \eq{phiH} we have used the integrals listed in
Appendix~\ref{Ap:Int} and employed the identity
\begin{align}
	\epsilon^{jp} \epsilon^{nm} = 
	\delta^{pm} \delta^{jn} - 
	\delta^{jm} \delta^{pn} .
\end{align}

Write \eq{phiH} as
\begin{align}\label{phiHA} 
\int\limits_0^{2 \pi} d \phi_l \, H^{ijn} ({\un k}, {\un l}) =  k^i \, \delta^{jn} \, {\cal A} (k,l) + k^j \, \delta^{in} \, {\cal B} (k,l).
\end{align}
Using this along with
\begin{align}
H^{ijn} ({\un k}, {\un k}-{\un l}) = - H^{jin} ({\un k}, {\un l})
\end{align}
in \eq{BB12} yields
\begin{align}\label{BB13}
& - \frac{g^8 \, Q_{s0}^6}{8} \int\limits_{\Lambda}^{\infty} \frac{d |{\un l} -{\un k}_1| \, R}{2 \pi} \, J_1 (|{\un l} -{\un k}_1| \, R) \, \int\limits_{\Lambda}^{\infty} \frac{d l' \, R}{2 \pi}  \, J_1 (l'R) \, \left[ k_1^{j} \, \delta^{in} \, {\cal A}  (k_1, |{\un l} -{\un k}_1|) + k_1^{i} \, \delta^{jn} \, {\cal B} (k_1,|{\un l} -{\un k}_1|) \right] \notag \\  & \times \, \left[ k_2^{i'} \, \delta^{j'n'} \, {\cal A}  (k_2,l') + k_2^{j'} \, \delta^{i'n'} \, {\cal B} (k_2,l') \right]  \ [ \delta^{in'} \, \delta^{jj'} \, \delta^{ni'} -  \delta^{ij'} \, \delta^{jn'} \, \delta^{ni'}   + 2 \, \delta^{in'} \, \delta^{jn} \, \delta^{i'j'} + 2 \, \delta^{in'} \, \delta^{ji'} \, \delta^{j'n} + \delta^{ii'} \, \delta^{jn'} \, \delta^{nj'} \notag \\ & - 2 \, \delta^{ij'} \, \delta^{ji'} \, \delta^{nn'} - \delta^{ii'} \, \delta^{jj'} \, \delta^{nn'} - 2 \, \delta^{ij} \, \delta^{i'j'} \, \delta^{nn'} - 2 \, \delta^{ij'} \, \delta^{jn} \, \delta^{i'n'}  + 2 \, \delta^{ij} \, \delta^{nj'} \, \delta^{i'n'}].
\end{align}
Performing the summation over all repeated indices we arrive at the
contribution of the diagram B
\begin{align}\label{BB14}
  \frac{5 \, g^8 \, Q_{s0}^6}{8} \int\limits_{\Lambda}^{\infty}
  \frac{d |{\un l} -{\un k}_1| \, R}{2 \pi} \, J_1 (|{\un l} -{\un
    k}_1| \, R) \, \int\limits_{\Lambda}^{\infty} \frac{d l' \, R}{2
    \pi} \, J_1 (l'R) \, {\cal B} (k_1,|{\un l} -{\un k}_1|) \, {\cal
    B} (k_2,l') \, {\un k}_1 \cdot {\un k}_2
\end{align}
with
\begin{align}
{\cal B} (k,l) = - \frac{4}{l \, k^3} \, \frac{1}{(k^2 - l^2)^2} ,
\end{align}
as follows from comparing Eqs.~\eqref{phiH} and
\eqref{phiHA}. Equation \eqref{BB14} contributes only to the first
azimuthal harmonic coefficient in the two-gluon correlation function.

The leading IR divergent part of the $|{\un l} -{\un k}_1|$ and $l'$
integrals can be readily extracted from the above results, yielding
\begin{align}\label{BB15}
  \int d^2 b_{12} \, d^2 b_{23} \, B = \frac{5 \, g^8 \, Q_{s0}^6}{2
    \pi^2} \, R^2 \, c^2 \, \frac{{\un k}_1 \cdot {\un k}_2}{k_1^7 \,
    k_2^7} = \frac{5 \, g^8 \, Q_{s0}^6}{2 \pi^2} \,
  \frac{1}{\Lambda^2} \, c^2 \, \frac{{\un k}_1 \cdot {\un k}_2}{k_1^7
    \, k_2^7}
\end{align}
with the constant
\begin{align}
 c = \int\limits_1^\infty dv \, \frac{J_1 (v)}{v} \approx 0.52\, ,
\end{align}
where $v = l' R$ or $v =|{\un l} -{\un k}_1| R$ depending on the
integral and $R = 1/\Lambda$.

In arriving at \eq{BB15} we have assumed that the IR divergence in the
$|{\un l} -{\un k}_1|$ and $l'$ integrals is regulated by $\Lambda =
1/R$. Similar power-law divergences arise in the leading-order
calculations of two-gluon production in the saturation framework
\cite{Kovchegov:2012nd,Kovchegov:2013ewa}. In
\myref\cite{Kovchegov:2013ewa} it was argued that they are regulated by
the saturation effects in the projectile. If this is the case in our
calculation as well, then we should go back to \eq{BB1} and subtract
from it the ${\un k}_1 \to -{\un k}_1$ and ${\un k}_2 \to -{\un k}_2$
terms from it while adding the ${\un k}_1 \to -{\un k}_1, \, {\un k}_2
\to -{\un k}_2$ term to fully account for the diagram B and its mirror
reflections, obtaining
\begin{align}\label{BB16}
& \frac{g^8 \, Q_{s0}^6}{32} \int \frac{d^2 l}{(2 \pi)^2} \, e^{i ({\un l} -{\un k}_1) \cdot {\un b}_{21}}  \, \int \frac{d^2 l'}{(2 \pi)^2} \, e^{-i {\un l}' \cdot {\un b}_{32} } \, \left[ H^{ijn} ({\un k}_1, {\un l}) +  H^{ijn} ({\un k}_1, - {\un l}) \right] \, \left[ H^{* \, i'j'n'} ({\un k}_2, {\un l}') + H^{* \, i'j'n'} ({\un k}_2, -{\un l}') \right] \notag \\ & \times \, [ \delta^{in'} \, \delta^{jj'} \, \delta^{ni'} -  \delta^{ij'} \, \delta^{jn'} \, \delta^{ni'}  + 2 \, \delta^{in'} \, \delta^{jn} \, \delta^{i'j'} + 2 \, \delta^{in'} \, \delta^{ji'} \, \delta^{j'n} + \delta^{ii'} \, \delta^{jn'} \, \delta^{nj'} - 2 \, \delta^{ij'} \, \delta^{ji'} \, \delta^{nn'} - \delta^{ii'} \, \delta^{jj'} \, \delta^{nn'} \notag \\ &   - 2 \, \delta^{ij} \, \delta^{i'j'} \, \delta^{nn'} - 2 \, \delta^{ij'} \, \delta^{jn} \, \delta^{i'n'} + 2 \, \delta^{ij} \, \delta^{nj'} \, \delta^{i'n'}].
\end{align}
Next we regulate all the IR momentum singularities in $H$-factors of \eq{BB16} by
\begin{align}
\frac{1}{{\un l}^2} \to \frac{1}{{\un l}^2 + Q_{s}^2}.
\end{align}
This is indeed not an exact way to account for the saturation effects
in the projectile and should be understood in a qualitative way. After
such regularization, integrating \eq{BB16} over $b_{21}$ and $b_{23}$
to infinity and one gets zero,
\begin{align}\label{BB17}
\int d^2 b_{12} \, d^2 b_{23} \, B = 0.
\end{align}

Hence the contribution of diagram B is either given by \eq{BB15} or is
zero depending on whether the power-law IR divergences are regulated
by the IR cutoff $\Lambda$ or by the saturation scale $Q_s$ due to
higher-order interactions with the projectile. A more careful analysis
of our main result \eqref{full_expr} along with the inclusion of even
higher-order corrections in the interaction with the projectile are
needed to resolve this ambiguity. While the former can be accomplished
with sufficient amount of hard work applied to \eq{full_expr}, the
latter would require diagram calculations beyond those done in
\myref\cite{Chirilli:2015tea}, which is a significantly larger
effort. We leave this investigation for further work, noting here that
the diagram B, even if it is not zero and is given by \eq{BB15}, would
only contribute to the first azimuthal harmonic, and is not going to
cancel the contribution \eqref{Adiag} of the diagram A, and, as we
will shortly see, the contributions of other diagrams in
\fig{all_graphs}.


\subsection{Diagram C}

Moving on to the diagram C along with the mirror reflections of its
gluon lines with respect to the final-state cut we see that the first
of the next four M-terms from \eq{Ahat} gives
\begin{align}\label{third4M}
& \frac{(4 \, g^4)^2}{(2 N_c)^3} \! \int d^2 z_1 \, d^2 w_1 \, d^2 z_2 \, d^2 w_2 \, e^{- i {\un k}_1 \cdot ({\un z}_1 - {\un w}_1) - i {\un k}_2 \cdot ({\un z}_2 - {\un w}_2)} \sum_{\lambda , \lambda'} \notag \\ &   \int d^2 x_1 d^2 x_2 \! \int
  \frac{d^2 k}{(2 \pi)^2} \, \frac{d^2 l}{(2 \pi)^2} \, \frac{d^2
    q_1}{(2 \pi)^2} \, \frac{d^2 q_2}{(2 \pi)^2} \, e^{i {\un q}_{1} \cdot ({\un x}_{1} - {\un b}_{2}) + i
    {\un q}_{2} \cdot ({\un x}_{2} - {\un b}_{3}) +
    i {\un l} \cdot ({\un x}_{2} - {\un x}_{1}) +
    i {\un k} \cdot ({\un z}_{1} - {\un x}_{2})}
  \notag \\ & \times \frac{1}{{\un q}_{1}^2 \, {\un q}_{2}^2} \, \left(
    - {\un q}_{1} \cdot {\un q}_{2} \, \frac{{\un \epsilon}_\lambda^* \times {\un k}}{{\un k}^2} + {\un \epsilon}_\lambda^*
    \cdot {\un q}_{1} \, \frac{{\un q}_{2} \times ({\un k} - {\un l})}{({\un k} - {\un l})^2} + {\un \epsilon}_\lambda^* \cdot
    {\un q}_{2} \, \frac{{\un q}_{1} \times {\un l}}{{\un l}^2} \right) \, \mbox{Sign} ({\un k} \times {\un l})   \notag \\ & 
  \int \frac{d^2 q_3} {(2\pi)^2} \, e^{- i {\un q}_3 \cdot ({\un w}_1 - {\un b}_1)} 
	\frac{{\un \epsilon}_\lambda \cdot {\un q}_3 }{{ \un q}_3^{2}}  \int \frac{d^2 q'_3} {(2\pi)^2} \, e^{i {\un q}'_3 \cdot ({\un z}_2 - {\un b}_1)} 
	\frac{{\un \epsilon}^*_{\lambda'} \cdot {\un q}'_3 }{{ \un q}_3^{\prime \, 2}} 
 \notag \\ & \int d^2 y_1 d^2 y_2 \! \int
  \frac{d^2 k'}{(2 \pi)^2} \, \frac{d^2 l'}{(2 \pi)^2} \, \frac{d^2
    q'_1}{(2 \pi)^2} \, \frac{d^2 q'_2}{(2 \pi)^2} \, e^{- i {\un q}'_{1} \cdot ({\un y}_{1} - {\un b}_{2}) - i
    {\un q}'_{2} \cdot ({\un y}_{2} - {\un b}_{3}) -
    i {\un l}^{\, \prime}_{} \cdot ({\un y}_{2} - {\un y}_{1}) -
    i {\un k}^{\, \prime}_{} \cdot ({\un w}_{2} - {\un y}_{2})}
  \notag \\ & \times \frac{1}{{\un q}_{1}^{\prime \, 2} \, {\un q}_{2}^{\prime \, 2}} \, \left(
  - {\un q}^{\, \prime}_{1} \cdot {\un q}^{\, \prime}_{2} \, \frac{{\un \epsilon}^{\, \lambda'} \times {\un k}^{\, \prime}}{ {\un k}^{\prime \, 2}} + {\un \epsilon}^{\, \lambda'}
    \cdot {\un q}^{\, \prime}_{1} \, \frac{{\un q}^{\, \prime}_{2} \times ({\un k}^{\, \prime}_{} - {\un l}^{\, \prime})}{({\un k}^{\, \prime} - {\un l}^{\, \prime})^2} + {\un \epsilon}^{\, \lambda'} \cdot
    {\un q}^{\, \prime}_{2} \, \frac{{\un q}^{\, \prime}_{1} \times {\un l}^{\, \prime}_{}}{{\un l}^{\prime \, 2}} \right) \, \mbox{Sign} ({\un k}^{\, \prime} \times {\un l}^{\, \prime})   \notag \\ & \times \left\langle f^{abc} \,
  \left[ U^{bd}_{{\un x}_{1}} - U^{bd}_{{\un b}_{2}}
  \right] \, \left[ U^{ce}_{{\un x}_{2}} - U^{ce}_{{\un b}_{3}} \right] \, \left[ 
	U^{a'd'}_{{\un z}_2}  -  U^{a'd'}_{{\un b}_1} 
\right] \, f^{a'b'c'} \, \left[ U^{b'd}_{{\un y}_{1}} - U^{b'd}_{{\un b}_{2}}
  \right] \, \left[ U^{c'e}_{{\un y}_{2}} - U^{c'e}_{{\un b}_{3}} \right] \, \left[ 
	U^{ad'}_{{\un w}_1}  -  U^{ad'}_{{\un b}_1} 
\right] \right\rangle.
\end{align}

Evaluating the interaction with the target in the GBW approximation
and performing a number of integrals along the lines used for other
diagram above, we rewrite this as
\begin{align}
& \frac{g^8 \, Q_{s0}^6}{32} \, \int \frac{d^2 l}{(2 \pi)^2} \, e^{i {\un l} \cdot {\un b}_{32} + i {\un k}_1 \cdot {\un b}_{13}}  \, \int \frac{d^2 l'}{(2 \pi)^2} \, e^{-i {\un l}' \cdot {\un b}_{32} - i {\un k}_2 \cdot {\un b}_{13} } \, H^{ijn} ({\un k}_1, {\un l}) \, H^{* \, i'j'n'} ({\un k}_2, {\un l}') \ [ 2 \, \delta _{i,n} \delta _{i^\prime,n^\prime} \delta _{j,j^\prime} \notag \\ & +8 \, \delta _{i,i^\prime} \delta _{j,j^\prime} \delta _{n,n^\prime}-\delta
   _{i,n^\prime} \delta _{i^\prime,j^\prime} \delta _{j,n}-\delta _{i,j^\prime} \delta _{i^\prime,n^\prime} \delta _{j,n}-\delta _{i,n}
   \delta _{i^\prime,j^\prime} \delta _{j,n^\prime}-\delta _{i,j^\prime} \delta _{i^\prime,n} \delta _{j,n^\prime}-\delta _{i,n^\prime}
   \delta _{i^\prime,j} \delta _{j^\prime,n}-\delta _{i,j} \delta _{i^\prime,n^\prime} \delta _{j^\prime,n} \notag \\ &  -\delta _{i,n} \delta
   _{i^\prime,j} \delta _{j^\prime,n^\prime}-\delta _{i,j} \delta _{i^\prime,n} \delta _{j^\prime,n^\prime}+2 \delta _{i,i^\prime} \delta
   _{j,n} \delta _{j^\prime,n^\prime}+2 \delta _{i,j} \delta _{i^\prime,j^\prime} \delta _{n,n^\prime} ].
\end{align}
Integrating over ${\un b}_{32}$ and ${\un b}_{13}$ we get
\begin{align}\label{CC3}
& \frac{g^8 \, Q_{s0}^6}{32}  \, \delta^2 ({\un k}_1 - {\un k}_2) \int \frac{d^2 l}{(2 \pi)^2} \, H^{ijn} ({\un k}_1, {\un l}) \, H^{* \, i'j'n'} ({\un k}_2, {\un l}) \ [ 2 \, \delta _{i,n} \delta _{i^\prime,n^\prime} \delta _{j,j^\prime} +8 \, \delta _{i,i^\prime} \delta _{j,j^\prime} \delta _{n,n^\prime}-\delta_{i,n^\prime} \delta _{i^\prime,j^\prime} \delta _{j,n}  \notag \\ & -\delta _{i,j^\prime} \delta _{i^\prime,n^\prime} \delta _{j,n}-\delta _{i,n}
   \delta _{i^\prime,j^\prime} \delta _{j,n^\prime}-\delta _{i,j^\prime} \delta _{i^\prime,n} \delta _{j,n^\prime}-\delta _{i,n^\prime}
   \delta _{i^\prime,j} \delta _{j^\prime,n}-\delta _{i,j} \delta _{i^\prime,n^\prime} \delta _{j^\prime,n}  -\delta _{i,n} \delta
   _{i^\prime,j} \delta _{j^\prime,n^\prime}-\delta _{i,j} \delta _{i^\prime,n} \delta _{j^\prime,n^\prime} \notag \\ & +2 \delta _{i,i^\prime} \delta
   _{j,n} \delta _{j^\prime,n^\prime}+2 \delta _{i,j} \delta _{i^\prime,j^\prime} \delta _{n,n^\prime} ].
\end{align}
Subtracting the ${\un k}_1 \to -{\un k}_1$ and ${\un k}_2 \to -{\un
  k}_2$ contributions from \eq{CC3} and adding the ${\un k}_1 \to
-{\un k}_1, \, {\un k}_2 \to -{\un k}_2$ contribution as well yields
\begin{align}\label{CC4}
& \frac{g^8 \, Q_{s0}^6}{16} \int \frac{d^2 l}{(2 \pi)^2} \ \left[ \delta^2 ({\un k}_1 - {\un k}_2)  \, H^{ijn} ({\un k}_1, {\un l}) \, H^{* \, i'j'n'} ({\un k}_2, {\un l}) - \delta^2 ({\un k}_1 + {\un k}_2) \, H^{ijn} ({\un k}_1, {\un l}) \, H^{* \, i'j'n'} ({\un k}_1, {\un l}) \right] \notag \\ & \times \, [ 2 \, \delta _{i,n} \delta _{i^\prime,n^\prime} \delta _{j,j^\prime} +8 \, \delta _{i,i^\prime} \delta _{j,j^\prime} \delta _{n,n^\prime}-\delta_{i,n^\prime} \delta _{i^\prime,j^\prime} \delta _{j,n}  -\delta _{i,j^\prime} \delta _{i^\prime,n^\prime} \delta _{j,n}-\delta _{i,n}
   \delta _{i^\prime,j^\prime} \delta _{j,n^\prime}-\delta _{i,j^\prime} \delta _{i^\prime,n} \delta _{j,n^\prime}-\delta _{i,n^\prime}
   \delta _{i^\prime,j} \delta _{j^\prime,n} \notag \\ & -\delta _{i,j} \delta _{i^\prime,n^\prime} \delta _{j^\prime,n}  -\delta _{i,n} \delta
   _{i^\prime,j} \delta _{j^\prime,n^\prime}-\delta _{i,j} \delta _{i^\prime,n} \delta _{j^\prime,n^\prime} +2 \delta _{i,i^\prime} \delta
   _{j,n} \delta _{j^\prime,n^\prime}+2 \delta _{i,j} \delta _{i^\prime,j^\prime} \delta _{n,n^\prime} ].
\end{align}
The delta-functions indicate that these are indeed gluon HBT and
anti-HBT diagrams in the terminology of \myref\cite{Kovchegov:2012nd}:
such contributions were previously observed in
\myrefs\cite{Dumitru:2008wn,Kovchegov:2012nd} in the leading-order
(even-harmonics) two-gluon production calculations. Summing over all
the indices we arrive at
\begin{align}\label{CC5}
g^8 \, Q_{s0}^6 \ \left[ \delta^2 ({\un k}_1 - {\un k}_2)   - \delta^2 ({\un k}_1 + {\un k}_2) \right] \int \frac{d^2 l}{(2 \pi)^2}  \frac{(k_1^2 + l^2 - {\un k}_1 \cdot {\un l})^2}{k_1^6 \, l^6 \, ({\un k}_1 - {\un l})^6}  \approx \frac{g^8}{4 \pi } \, \frac{Q_{s0}^6}{k_1^8 \, \Lambda^4} \,\left[ \delta^2 ({\un k}_1 - {\un k}_2)   - \delta^2 ({\un k}_1 + {\un k}_2) \right]\, ,  
\end{align}
where the integral is dominated by the IR divergences, which we
regulated by the IR cutoff $\Lambda$. (Once again, it is likely that
$\Lambda$ should be replaced by the saturation scale of the projectile
\cite{Kovchegov:2013ewa}.) We thus have for the diagram C,
\begin{align}\label{CC6}
\int d^2 b_{12} \, d^2 b_{23} \, C = \frac{g^8}{4 \pi } \, \frac{Q_{s0}^6}{k_1^8 \, \Lambda^4} \,\left[ \delta^2 ({\un k}_1 - {\un k}_2)   - \delta^2 ({\un k}_1 + {\un k}_2) \right].
\end{align}
In the GBW approximation the diagram gives only HBT (and
anti-HBT)-type correlations. It appears likely that in the full MV
model, without the GBW simplification, diagram C would also lead to
non-HBT types of terms (that is, the contribution of the full diagram
C is probably not limited to the delta-functions of \eq{CC6}).


\subsection{Diagram D}

Let us evaluate diagram D next. The first of the next four M-terms
from \eq{Ahat} gives
\begin{align}\label{third4M}
& D_1 =  \frac{4 \, g^8}{(2 N_c)^3} \! \int d^2 z_1 \, d^2 w_1 \, d^2 z_2 \, d^2 w_2 \, e^{- i {\un k}_1 \cdot ({\un z}_1 - {\un w}_1) - i {\un k}_2 \cdot ({\un z}_2 - {\un w}_2)} \sum_{\lambda , \lambda'} \notag \\ &   \int d^2 x_1 d^2 x_2 \! \int
  \frac{d^2 k}{(2 \pi)^2} \, \frac{d^2 l}{(2 \pi)^2} \, \frac{d^2
    q_1}{(2 \pi)^2} \, \frac{d^2 q_2}{(2 \pi)^2} \, e^{i {\un q}_{1} \cdot ({\un x}_{1} - {\un b}_{1}) + i
    {\un q}_{2} \cdot ({\un x}_{2} - {\un b}_{1}) +
    i {\un l} \cdot ({\un x}_{2} - {\un x}_{1}) +
    i {\un k} \cdot ({\un z}_{1} - {\un x}_{2})}
  \notag \\ & \times \frac{1}{{\un q}_{1}^2 \, {\un q}_{2}^2} \, \left(
    - {\un q}_{1} \cdot {\un q}_{2} \, \frac{{\un \epsilon}_\lambda^* \times {\un k}}{{\un k}^2} + {\un \epsilon}_\lambda^*
    \cdot {\un q}_{1} \, \frac{{\un q}_{2} \times ({\un k} - {\un l})}{({\un k} - {\un l})^2} + {\un \epsilon}_\lambda^* \cdot
    {\un q}_{2} \, \frac{{\un q}_{1} \times {\un l}}{{\un l}^2} \right) \, \mbox{Sign} ({\un k} \times {\un l})   \notag \\ & 
  \int \frac{d^2 q_3} {(2\pi)^2} \, e^{- i {\un q}_3 \cdot ({\un w}_1 - {\un b}_2)} 
	\frac{{\un \epsilon}_\lambda \cdot {\un q}_3 }{{ \un q}_3^{2}}  \int \frac{d^2 q'_3} {(2\pi)^2} \, e^{i {\un q}'_3 \cdot ({\un z}_2 - {\un b}_2)} 
	\frac{{\un \epsilon}^*_{\lambda'} \cdot {\un q}'_3 }{{ \un q}_3^{\prime \, 2}} 
 \notag \\ & \int d^2 y_1 d^2 y_2 \! \int
  \frac{d^2 k'}{(2 \pi)^2} \, \frac{d^2 l'}{(2 \pi)^2} \, \frac{d^2
    q'_1}{(2 \pi)^2} \, \frac{d^2 q'_2}{(2 \pi)^2} \, e^{- i {\un q}'_{1} \cdot ({\un y}_{1} - {\un b}_{3}) - i
    {\un q}'_{2} \cdot ({\un y}_{2} - {\un b}_{3}) -
    i {\un l}^{\, \prime}_{} \cdot ({\un y}_{2} - {\un y}_{1}) -
    i {\un k}^{\, \prime}_{} \cdot ({\un w}_{2} - {\un y}_{2})}
  \notag \\ & \times \frac{1}{{\un q}_{1}^{\prime \, 2} \, {\un q}_{2}^{\prime \, 2}} \, \left(
  - {\un q}^{\, \prime}_{1} \cdot {\un q}^{\, \prime}_{2} \, \frac{{\un \epsilon}^{\, \lambda'} \times {\un k}^{\, \prime}}{ {\un k}^{\prime \, 2}} + {\un \epsilon}^{\, \lambda'}
    \cdot {\un q}^{\, \prime}_{1} \, \frac{{\un q}^{\, \prime}_{2} \times ({\un k}^{\, \prime}_{} - {\un l}^{\, \prime})}{({\un k}^{\, \prime} - {\un l}^{\, \prime})^2} + {\un \epsilon}^{\, \lambda'} \cdot
    {\un q}^{\, \prime}_{2} \, \frac{{\un q}^{\, \prime}_{1} \times {\un l}^{\, \prime}_{}}{{\un l}^{\prime \, 2}} \right) \, \mbox{Sign} ({\un k}^{\, \prime} \times {\un l}^{\, \prime})   \notag \\ & \times \left\langle f^{abc} \,
  \left[ U^{bd}_{{\un x}_{1}} - U^{bd}_{{\un b}_{1}}
  \right] \, \left[ U^{cd}_{{\un x}_{2}} - U^{cd}_{{\un b}_{1}} \right] \, \left[ 
	U^{a'd'}_{{\un z}_2}  -  U^{a'd'}_{{\un b}_2} 
\right] \, f^{a'b'c'} \, \left[ U^{b'e}_{{\un y}_{1}} - U^{b'e}_{{\un b}_{3}}
  \right] \, \left[ U^{c'e}_{{\un y}_{2}} - U^{c'e}_{{\un b}_{3}} \right] \, \left[ 
	U^{ad'}_{{\un w}_1}  -  U^{ad'}_{{\un b}_2} 
\right] \right\rangle.
\end{align}
Evaluating the interaction with the target and performing a number of
integrals we rewrite this as
\begin{align}
D_1 = \frac{g^8 \, Q_{s0}^6}{128} \, e^{i {\un k}_1 \cdot {\un b}_{21} + i {\un k}_2 \cdot {\un b}_{32}}  \, \int \frac{d^2 l}{(2 \pi)^2} \, \frac{d^2 l'}{(2 \pi)^2} \, H^{ijn} ({\un k}_1, {\un l}) \, H^{* \, i'j'n'} ({\un k}_2, {\un l}') \,
[ \ldots \mbox{Kronecker} \ \mbox{deltas} \ldots  ],
\end{align}
with the exact structure of the Kronecker delta functions not being
important for what follows. Therefore, we do not show it
explicitly. Integrating over the impact parameters $b_{21}$ and
$b_{32}$ we arrive at
\begin{align}\label{DD4}
\int d^2 b_{21} \, d^2 b_{32} \, D_1 = \frac{g^8 \, Q_{s0}^6}{128} \, \delta^2 ({\un k}_1) \, \delta^2 ({\un k}_2)   \, \int d^2 l \, d^2 l' \, H^{ijn} ({\un k}_1, {\un l}) \, H^{* \, i'j'n'} ({\un k}_2, {\un l}') \,
[ \ldots \mbox{Kronecker} \ \mbox{deltas} \ldots  ].
\end{align}
Subtracting from \eq{DD4} the ${\un k}_1 \to -{\un k}_1$ and ${\un
  k}_2 \to -{\un k}_2$ terms and adding the ${\un k}_1 \to -{\un k}_1,
\, {\un k}_2 \to -{\un k}_2$ term gives zero:
\begin{align}\label{DD5}
  \int d^2 b_{21} \, d^2 b_{32} \, D = \int d^2 b_{21} \, d^2 b_{32}
  \, \left[ D_1 + D_2 + D_3 + D_4 \right] = 0.
\end{align}
(Here $D_1, \ldots , D_4$ denote the four contributions one could get
from the diagram D in \fig{all_graphs} by reflecting the gluons with
respect to the final-state cut.)


\subsection{Diagram E and F}

We conclude by evaluating diagrams E and F in \fig{all_graphs}. The
next four M-terms in \eq{Ahat} give
\begin{align}\label{third4M}
& E_1 =  \frac{8 \, g^8}{(2 N_c)^3} \! \int d^2 z_1 \, d^2 w_1 \, d^2 z_2 \, d^2 w_2 \, e^{- i {\un k}_1 \cdot ({\un z}_1 - {\un w}_1) - i {\un k}_2 \cdot ({\un z}_2 - {\un w}_2)} \sum_{\lambda , \lambda'} \notag \\ &   \int d^2 x_1 d^2 x_2 \! \int
  \frac{d^2 k}{(2 \pi)^2} \, \frac{d^2 l}{(2 \pi)^2} \, \frac{d^2
    q_1}{(2 \pi)^2} \, \frac{d^2 q_2}{(2 \pi)^2} \, e^{i {\un q}_{1} \cdot ({\un x}_{1} - {\un b}_{1}) + i
    {\un q}_{2} \cdot ({\un x}_{2} - {\un b}_{1}) +
    i {\un l} \cdot ({\un x}_{2} - {\un x}_{1}) +
    i {\un k} \cdot ({\un z}_{1} - {\un x}_{2})}
  \notag \\ & \times \frac{1}{{\un q}_{1}^2 \, {\un q}_{2}^2} \, \left(
    - {\un q}_{1} \cdot {\un q}_{2} \, \frac{{\un \epsilon}_\lambda^* \times {\un k}}{{\un k}^2} + {\un \epsilon}_\lambda^*
    \cdot {\un q}_{1} \, \frac{{\un q}_{2} \times ({\un k} - {\un l})}{({\un k} - {\un l})^2} + {\un \epsilon}_\lambda^* \cdot
    {\un q}_{2} \, \frac{{\un q}_{1} \times {\un l}}{{\un l}^2} \right) \, \mbox{Sign} ({\un k} \times {\un l})   \notag \\ & 
  \int \frac{d^2 q_3} {(2\pi)^2} \, e^{- i {\un q}_3 \cdot ({\un w}_1 - {\un b}_3)} 
	\frac{{\un \epsilon}_\lambda \cdot {\un q}_3 }{{ \un q}_3^{2}}  \int \frac{d^2 q'_3} {(2\pi)^2} \, e^{i {\un q}'_3 \cdot ({\un z}_2 - {\un b}_2)} 
	\frac{{\un \epsilon}^*_{\lambda'} \cdot {\un q}'_3 }{{ \un q}_3^{\prime \, 2}} 
 \notag \\ & \int d^2 y_1 d^2 y_2 \! \int
  \frac{d^2 k'}{(2 \pi)^2} \, \frac{d^2 l'}{(2 \pi)^2} \, \frac{d^2
    q'_1}{(2 \pi)^2} \, \frac{d^2 q'_2}{(2 \pi)^2} \, e^{- i {\un q}'_{1} \cdot ({\un y}_{1} - {\un b}_{2}) - i
    {\un q}'_{2} \cdot ({\un y}_{2} - {\un b}_{3}) -
    i {\un l}^{\, \prime}_{} \cdot ({\un y}_{2} - {\un y}_{1}) -
    i {\un k}^{\, \prime}_{} \cdot ({\un w}_{2} - {\un y}_{2})}
  \notag \\ & \times \frac{1}{{\un q}_{1}^{\prime \, 2} \, {\un q}_{2}^{\prime \, 2}} \, \left(
  - {\un q}^{\, \prime}_{1} \cdot {\un q}^{\, \prime}_{2} \, \frac{{\un \epsilon}^{\, \lambda'} \times {\un k}^{\, \prime}}{ {\un k}^{\prime \, 2}} + {\un \epsilon}^{\, \lambda'} \cdot {\un q}^{\, \prime}_{1} \, \frac{{\un q}^{\, \prime}_{2} \times ({\un k}^{\, \prime}_{} - {\un l}^{\, \prime})}{({\un k}^{\, \prime} - {\un l}^{\, \prime})^2} + {\un \epsilon}^{\, \lambda'} \cdot {\un q}^{\, \prime}_{2} \, \frac{{\un q}^{\, \prime}_{1} \times {\un l}^{\, \prime}_{}}{{\un l}^{\prime \, 2}} \right) \, \mbox{Sign} ({\un k}^{\, \prime} \times {\un l}^{\, \prime})   \notag \\ & \times \left\langle f^{abc} \,
  \left[ U^{bd}_{{\un x}_{1}} - U^{bd}_{{\un b}_{1}}
  \right] \, \left[ U^{cd}_{{\un x}_{2}} - U^{cd}_{{\un b}_{1}} \right] \, \left[ 
	U^{a'd'}_{{\un z}_2}  -  U^{a'd'}_{{\un b}_2} 
\right] \, f^{a'b'c'} \, \left[ U^{b'd'}_{{\un y}_{1}} - U^{b'd'}_{{\un b}_{2}}
  \right] \, \left[ U^{c'e}_{{\un y}_{2}} - U^{c'e}_{{\un b}_{3}} \right] \, \left[ 
	U^{ae}_{{\un w}_1}  -  U^{ae}_{{\un b}_3} 
\right] \right\rangle.
\end{align}
Evaluating the interaction with the target and performing a number of
integrals we rewrite this as
\begin{align}
E_1 = \frac{g^8 \, Q_{s0}^6}{64} \, e^{i {\un k}_1 \cdot {\un b}_{31}}  \, \int \frac{d^2 l}{(2 \pi)^2} \, \frac{d^2 l'}{(2 \pi)^2} \,  e^{i ({\un k}_2 - {\un l}') \cdot {\un b}_{32}}  \, H^{ijn} ({\un k}_1, {\un l}) \, H^{* \, i'j'n'} ({\un k}_2, {\un l}') \,
[ \ldots \mbox{Kronecker} \ \mbox{deltas} \ldots  ].
\end{align}
Integrating over the impact parameters $b_{31}$ and $b_{32}$ we arrive at
\begin{align}\label{EE3}
\int d^2 b_{31} \, d^2 b_{32} \, E_1 = \frac{g^8 \, Q_{s0}^6}{64} \, \delta^2 ( {\un k}_1)  \, \int d^2 l \, H^{ijn} ({\un k}_1, {\un l}) \, H^{* \, i'j'n'} ({\un k}_2, {\un k}_2) \,
[ \ldots \mbox{Kronecker} \ \mbox{deltas} \ldots  ].
\end{align}
While the $H^{* \, i'j'n'} ({\un k}_2, {\un k}_2)$ in \eq{EE3}
requires proper regularization, like it was done above in evaluating
the diagram B, it is clear that anti-symmetrization of \eqref{EE3}
under ${\un k}_1 \to -{\un k}_1$ gives zero,
\begin{align}\label{EE4}
\int d^2 b_{31} \, d^2 b_{32} \, \left[ E_1 + E_2 + E_3 + E_4 \right] = 0 . 
\end{align}

Diagram F is obtained from the diagram E by interchanging ${\un k}_1
\leftrightarrow {\un k}_2$. Therefore, the sum of the four F-graphs is
also zero,
\begin{align}\label{FF1}
\int d^2 b_{31} \, d^2 b_{32} \, \left[ F_1 + F_2 + F_3 + F_4 \right] = 0 . 
\end{align}


\subsection{Sum of all diagrams}

Adding all the above results for diagrams A, B, $\ldots$ , F together we get
\begin{align}\label{full_result}
& \frac{d \sigma_{odd}}{d^2 k_1 \, d y_1 \, d^2 k_2 \, d y_2} = \frac{1}{[2 (2 \pi)^3]^2} \, \int d^2 B \, d^2 b \left[ T_1 ({\un B} - {\un b}) \right]^3  \, g^8 \, Q_{s0}^6 (b) \, \frac{1}{{\un k}_1^6 \, {\un k}_2^6} \\ & \times \, \left\{ \left[ \frac{({\un k}_1^2 + {\un k}_2^2 + {\un k}_1 \cdot {\un k}_2)^2}{({\un k}_1 + {\un k}_2)^6} - \frac{({\un k}_1^2 + {\un k}_2^2 - {\un k}_1 \cdot {\un k}_2)^2}{({\un k}_1 - {\un k}_2)^6} \right] +  \frac{10 \, c^2}{(2 \pi)^2} \, \frac{1}{\Lambda^2} \, \frac{{\un k}_1 \cdot {\un k}_2}{k_1 \, k_2} +  \frac{1}{4 \pi } \, \frac{k_1^4}{\Lambda^4} \,\left[ \delta^2 ({\un k}_1 - {\un k}_2)   - \delta^2 ({\un k}_1 + {\un k}_2) \right]  \right\}. \notag
\end{align}
The most important conclusion of our approximate analytical
calculation that led to \eq{full_result} is that we get a non-zero
contribution, which would yield odd azimuthal harmonics in the
two-gluon (and, hence, di-hadron) correlation function. Hence, our
main exact result \eqref{full_expr} is not zero either. We conclude
that we have identified a source of odd harmonics in the two-gluon
correlation function in the saturation framework.

Interestingly, the HBT term in \eq{full_result} has the largest IR
divergence and, therefore, dominates \eq{full_result} and the
corresponding correlation function.  As was already suggested in
\myref\cite{Dumitru:2008wn}, fragmentation may ``wash out" the
delta-functions in the HBT term to some degree, such that the hadronic
correlation function would not contain the literal delta-function
correlations from \eq{full_result}. While the delta-function shape may
not survive fragmentation, the HBT-type gluon correlations should
still manifest itself in the hadronic correlation function, and may
also dominate in it just like the HBT correlations dominate the
two-gluon correlator.

Comparing \eq{full_result} to the leading-order two-gluon production
cross section in the classical formalism, e.g., to Eqs.~(49), (58) and
(59) in \cite{Kovchegov:2012nd}, we observe the following:
\eq{full_result} has an extra power of $Q_{s0}^2$ as compared to the
leading-order cross section, corresponding to an extra rescattering in
the target. Similarly, \eq{full_result} includes an extra factor of
$\as^2$ as compared to the leading-order cross section: this factor
arises through the extra interaction with the projectile. In addition,
the IR divergence in \eq{full_result} is $\sim 1/\Lambda^4$, which is
a higher degree of divergence than $\sim 1/\Lambda^2$ observed in the
leading-order result \cite{Kovchegov:2013ewa}. This is an indication
that the higher-order rescatterings in the target and in the
projectile may screen the IR divergence in the leading-order
expression for two-gluon production cross section, effectively
replacing $1/\Lambda^2$ by $1/Q_s^2$. Further work is needed to firmly
establish this conclusion.


\section{Evaluating the odd-harmonic part of the two-gluon production
  cross section: Numerical approach}
\label{sec:num}

In order to model the distribution of the color charges in the
projectile numerically, instead of the point-charge approach used in
Sec.~\ref{Sec:Eval} it is more convenient to use an alternative
one 
based on the introduction of a continuous (light-cone) color density
$\rho_p(\un{x})$.  To compute an observable, one has to averaged the
corresponding operator with the weight functional $W[\rho_p]$, similar
to the way we account for the target ensemble.  Nevertheless, the
treatment of the projectile here is still different from the treatment
of the target; the projectile charge density is considered to be
dilute facilitating the expansion of the projectile Wilson lines.  In
\cite{Kovchegov:1997pc}, it was demonstrated that, in the classical
approximation, the approach based on the continuous color density is
completely equivalent to the one used in preceding sections.

Here we introduce the density-dependent operator describing production
of a gluon with momentum $\un{k}_1$, $E_1 \frac{dN}{d^2 k_1}
\big[\rho_p,\rho_T\big]$, see Appendix~\ref{Sec:Comparing} for
details.  The single and double inclusive gluon multiplicities are
then given by
\begin{align}
	 E_1 \frac{dN}{d^3 k_1}  &= 
	 \Big\langle  E_1 \frac{dN}{d^3 k_1} \big[\rho_p,\rho_T\big]  \Big\rangle_{\rho_p,\rho_T} , \\ 
	E_1 E_2 \frac{d^2N}{d^3 k_1 d^3 k_2} &=
	\Big\langle  E_1 \frac{dN}{d^3 k_1} \big[\rho_p,\rho_T\big]\, \,  E_2 \frac{dN}{d^3 k_2} \big[\rho_p,\rho_T\big] \Big\rangle_{\rho_p,\rho_T}. 
\end{align}
Thus, in the classical approximation, the double inclusive production
factorizes on the configuration-by-configuration basis.  This leads to
the following interesting fact: the two-particle cumulants of
azimuthal anisotropy are always positive if the magnitudes of the
momenta of the produced gluons coincide, i.e.,
\begin{equation}
  \left. \int\limits_0^{2 \pi} d\phi_1 d\phi_2 \, e^{i n (\phi_1-\phi_2 )} \, \frac{d^2N}{d^3 k_1 d^3 k_2} \right|_{ |\un{k}_1|=|\un{k}_2|=k} =
  \Big\langle 
  V_n (k) 
  V_n^* (k)
  \Big\rangle_{\rho_p,\rho_T}  \ge 0\,,   
\end{equation}
where 
\begin{equation}
  V_n (k) =   \int\limits_0^{2 \pi} d\phi \, e^{i n \phi} \,  \frac{dN}{d^3 k} \big[\rho_p,\rho_T\big]\,. 
\end{equation}

Our goal is to compute the odd two-gluon harmonics. For this we define     
\begin{equation}
	\frac{dN^{\rm odd} (\un{k}) }{d^3 k} \big[\rho_p,\rho_T\big]  = 
	\frac{1}{2} \left( 
	\frac{dN (\un{k}) }{d^3 k} \big[\rho_p,\rho_T\big] - 
	\frac{dN (-\un{k}) }{d^3 k} \big[\rho_p,\rho_T\big] \right) 
	\,; 
	\label{Eq:NumNOdd}
\end{equation}
with the explicit expression presented in Appendix~\ref{Sec:Comparing}.   
We thus can extract  
\begin{equation}
	V_{n}^{\rm odd} (k) = 
	\int\limits_0^{2 \pi} d\phi  \, e^{i n \phi} \,
	\frac{dN^{\rm odd} (\un{k}) }{d^3 k} \big[\rho_p,\rho_T\big]
\end{equation}
and the angular-averaged 
\begin{equation}
	V_{0} (k) = 
	\int\limits_0^{2 \pi}  d\phi
	\frac{dN^{\rm odd} (\un{k}) }{d^3 k} \big[\rho_p,\rho_T\big]\,. 
\end{equation}
$V_{0} (k)$ contributes to the normalization of the cumulants of the
azimuthal anisotropy and as such has to be computed to the leading
order only. Note that in this case $V_{0} (k)$ is manifestly real.

The two-gluon cumulants for odd harmonics are then 
\begin{equation}
	v^2_n\{2\} (|{\un k}_1|, |{\un k}_2|)   = \left\langle \frac{V_n^{\rm odd} (|\un {k}_1|)}{V_0 (|\un {k}_1|)} 
	\, 
	 \left[ \frac{V_n^{\rm odd} (|\un {k}_2|)}{V_0 (|\un {k}_2|)} \right]^* 
	\right\rangle_{\rho_p,\rho_T}\,.
\end{equation}

The averages with respect to the ensembles $\langle \dots
\rangle_{\rho_p,\rho_T} $ are performed in the Gaussian MV model. The
target MV configurations are generated as described in
\cite{Lappi:2007ku,Dumitru:2014vka} and are complemented by the MV
configurations for the projectile, which are computed for a single
slice in $x^-$.  We also assume an infinite target with the color
charge density defined by a single number describing the color charge
fluctuations, $\mu^2={\rm const}$. For the projectile we assume a
finite size $R_p = 1/Q_{s p}$ with a Gaussian profile, that is
\begin{equation}
	\mu^2_p(\un{x}) = c_p \, \mu^2 \exp\left(-\frac{|\un{x}|^2}{R_p^2} \right). 
	\label{Eq:mup_num}
\end{equation}
Strictly-speaking, for our analytic approach to be valid one has to
have $c_p \ll1$, see e.g. \myref\cite{Kovchegov:2012nd}. In our
numerical simulations we fixed $c_p =1/2$. Our choice of the free
parameters is driven by the predisposition to simplify the problem as,
in the current paper, we have no ambitions to describe the
experimental data quantitatively.

\begin{figure}[ht]
\begin{center}
\includegraphics[width= 0.49 \textwidth]{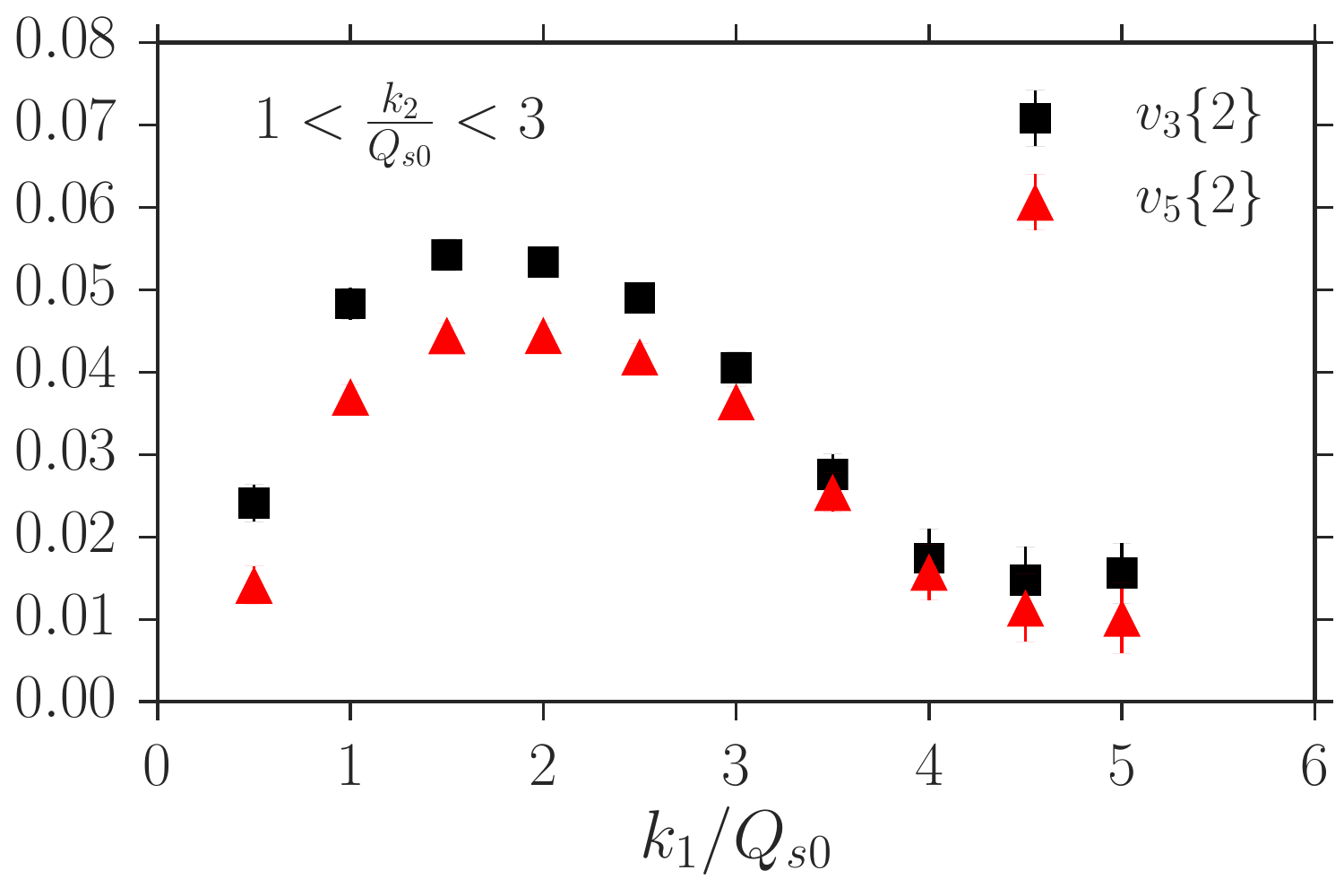} 
\caption{ The odd harmonic coefficients for gluons plotted as
  functions of the transverse momentum $k_1/Q_{s0}$ with $k_2$
  integrated over the $k_2 \in [Q_{s0}, 3 \, Q_{s0}]$ interval.  }
\label{vodd}
\end{center}
\end{figure}

The odd harmonic coefficients $v_3$ and $v_5$ resulting from our
numerical simulations are shown in Fig \ref{vodd}. We warn the reader
that the calculations were performed without a high multiplicity bias;
the gluon fragmentation was not accounted for; the Glauber
fluctuations were neglected.  Additionally, the parameters we used to
model the projectile wave function are not very
realistic. Nevertheless Fig.~\ref{vodd} can be viewed as a proof of
the concept demonstrating the presence of the odd azimuthal harmonics
in the saturation/CGC formalism; the magnitude of $v_3\{2\}$ is of the
same order as observed experimentally in pA collisions at LHC.

Conducting phenomenologically relevant calculations would require a
significant numerical effort and will be reported elsewhere
\cite{phemen2018}.

\begin{figure}[ht]
\begin{center}
\includegraphics[width= 0.49 \textwidth]{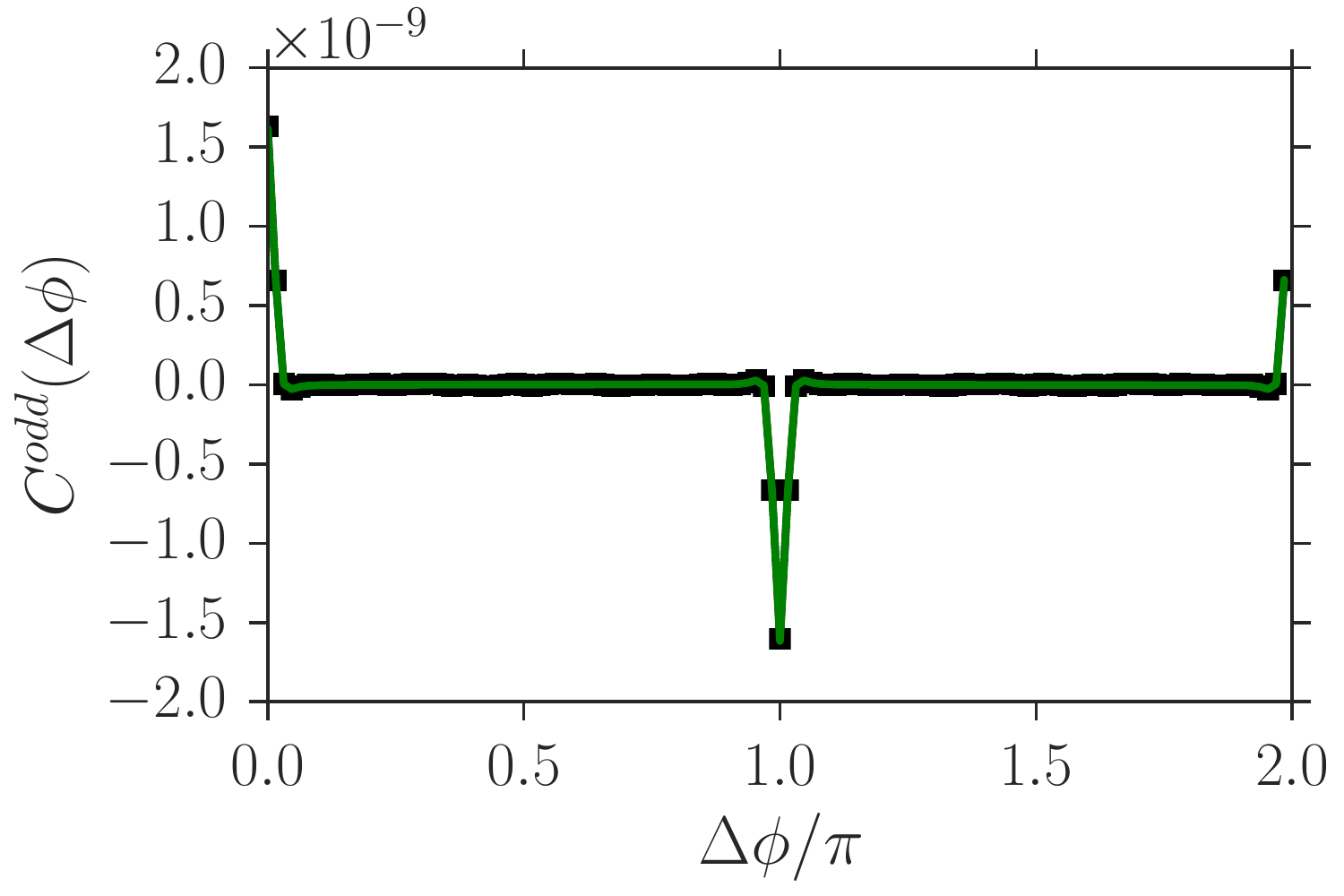} 
\includegraphics[width= 0.49 \textwidth]{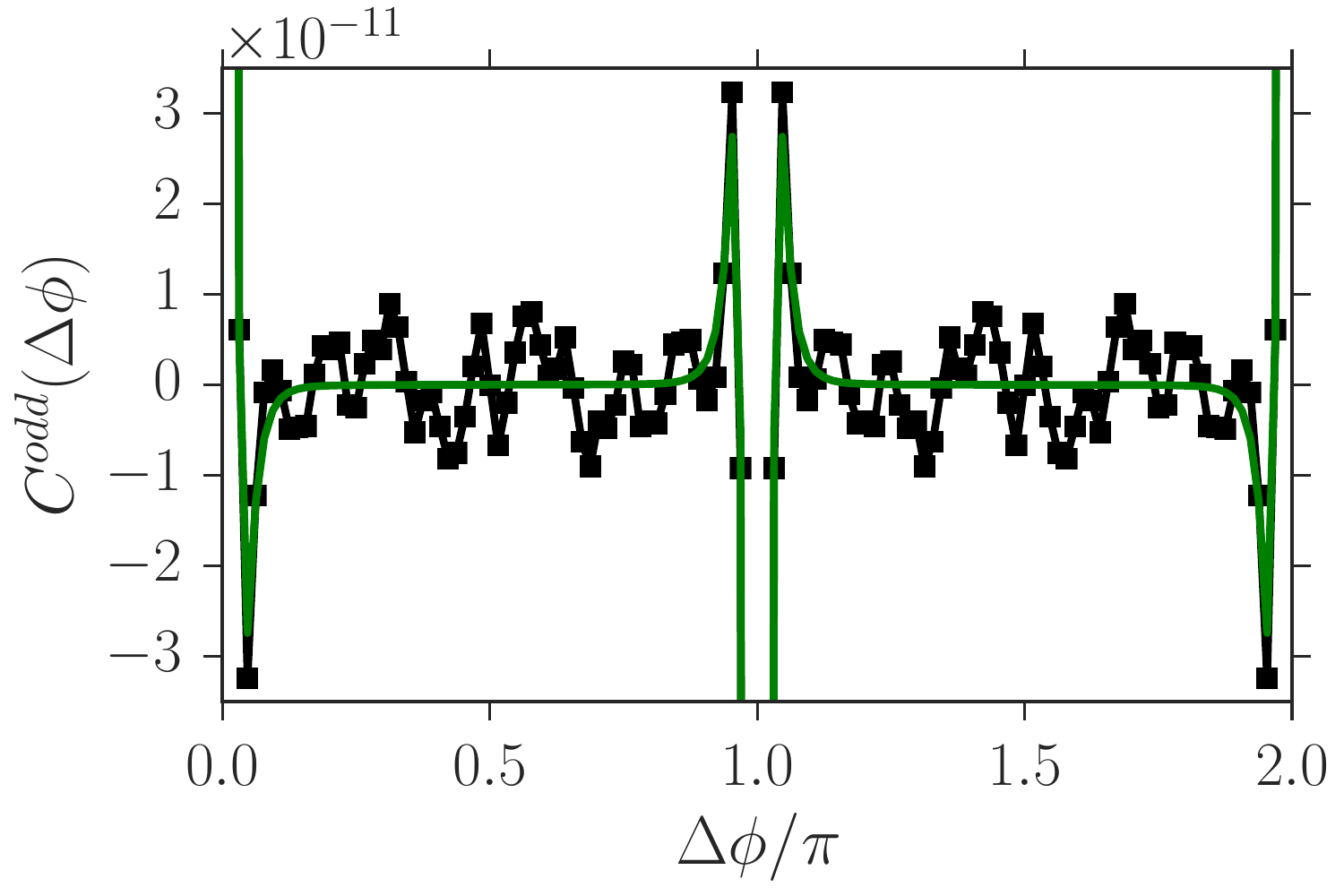} 
\caption{ Left panel: the odd correlation function $C^{\rm odd}$
  defined by Eq.~\eqref{Eq:NumCorrFunct} as a function of the
  azimuthal angle $\Delta \phi = \phi_1 - \phi_2$ for $|\un{k}_1| =
  |\un{k}_2| \approx 5 \, Q_{s0} $. Right panel: the same as in the
  left panel, but zoomed in. The green solid line shows the fit
  \eqref{Eq:AnzatzFit} inspired by the analytical result
  \eqref{full_result}.  The prominent peaks at $\Delta \phi /\pi =0$,
  $1$ and $2$ illustrate the HBT-type correlation from the diagram
  C. The peaks disappear for $|\un{k}_1| \ne |\un{k}_2|$ (not shown).
}
\label{Codd}
\end{center}
\end{figure}

By numerically computing the two-gluon correlation function, we are
able to reproduce the main features of Eq.~\eqref{full_result},
including the (anti-)HBT peaks and the contribution from the diagram
A. Consider the odd part of the angular correlation function $C^{\rm
  odd} (|\un{k}_1|, |\un{k}_2|, \Delta \phi) $ defined by
\begin{align}
	&C^{\rm odd} (|\un{k}_1|, |\un{k}_2|, \Delta \phi)
	\notag \\ &=  
	\frac{1}{4} E_1 E_2 
	\int\limits_0^{2 \pi} \frac{d \phi_1}{2\pi}  
	\int\limits_0^{2 \pi} \frac{d \phi_2}{2\pi}  
	\, \delta(\Delta\phi - \phi_1 + \phi_2) 
	\left( 
	\frac{d^2N(\un{k}_1,\un{k}_2)}{d^3 k_1 d^3 k_2} - 
	\frac{d^2N(\un{k}_1,-\un{k}_2)}{d^3 k_1 d^3 k_2} - 
	\frac{d^2N(-\un{k}_1,\un{k}_2)}{d^3 k_1 d^3 k_2} +
	\frac{d^2N(-\un{k}_1,-\un{k}_2)}{d^3 k_1 d^3 k_2}  
	\right) 
	\notag \\
	& = 
	E_1 E_2 
	\int\limits_0^{2 \pi} \frac{d \phi_1}{2\pi}  
	\int\limits_0^{2 \pi} \frac{d \phi_2}{2\pi}  
	\, \delta(\Delta\phi - \phi_1 + \phi_2) 
	\left\langle  
	\frac{dN^{\rm odd} (\un{k}_1) }{d^3 k_1} \big[\rho_p,\rho_T\big] \,\,  
	\frac{dN^{\rm odd} (\un{k}_2) }{d^3 k_2} \big[\rho_p,\rho_T\big] 
	\right\rangle _{\rho_p,\rho_T}   .
	\label{Eq:NumCorrFunct}
\end{align}
Performing numerical configuration-by-configuration analysis it is
possible to extract $C^{\rm odd} (|\un{k}_1|, |\un{k}_2|, \Delta \phi)
$.  The numerical results are depicted in Fig.~\ref{Codd} for
$|\un{k}_1| = |\un{k}_2| \approx 5 \, Q_{s0} $ with $Q_{s0}$ the
target saturation scale; the black points connected by the straight
lines are the results of the numerical calculations, the green curve
represents the fit inspired by the analytical result in
Eq.~\eqref{full_result}, namely
\begin{equation}
	C^{\rm odd}_{\rm fit} = 
	a_C \exp \left( - \frac{ (1+\cos(\Delta \phi) )^2 } { b_C^2 } \right)
	- a_A \frac{ (2+\cos(\Delta \phi))^2 }  { ( 1+\cos(\Delta \phi) +b_A^2)^3 } + (\Delta \phi \to  \Delta \phi -\pi),  
	\label{Eq:AnzatzFit}
\end{equation}
with positive $a_C$ and $a_A$.  Here the first term corresponds to the
HBT peak, which, for our Gaussian projectile wave function in
Eq.~\eqref{Eq:mup_num} is a Gaussian itself. This replaces the Dirac
$\delta$-function peaks, originating in \eq{full_result} from the
Fourier transform with respect to the impact parameter over an
infinite projectile (cf.~\cite{Kovchegov:2012nd}). The second term in
\eq{Eq:AnzatzFit} corresponds to the contribution from the diagram A
in which the denominator was regularized by the scale of order
$1/R_p$. As shown in Fig.~\ref{Codd}, the fit reproduced the numerical
calculations quite well, confirming our analytical findings from
\eqref{full_result}. We believe that the disagreement between the
green solid line and some of the numerical data points in the right
panel of \fig{Codd} is either due to the higher-order corrections
originating from the MV ensemble (which are outside the precision of
the classical approximation employed here) or is caused by the
unavoidable discretization errors. Note that the contribution of the
diagram B is not seen in our numerical analysis, which appears to be
consistent with the possibility that the contribution of the diagram B
is zero, as outlined above near \eq{BB17}.

\section{Conclusions and Outlook}
\label{sec:conc}

In this paper we have demonstrated analytically that the classical
gluons fields of the saturation/CGC approach to heavy ion collisions
do generate odd azimuthal harmonics in the two-gluon correlation
function. Since the classical fields are the leading-order
contribution to the two-gluon production cross section, we conclude
that the odd azimuthal harmonics are an inherent property of particle
production in the saturation framework. This conclusion is consistent
with the results of numerical simulations for the classical gluon
fields of two colliding heavy ions carried out in
\cite{Lappi:2009xa,Schenke:2015aqa}. The difficulty in identifying the
odd-harmonics contribution analytically is related to the fact the
analytic expressions for the classical single- and double-gluon
production cross sections in heavy ion collision do not exist:
instead, as explained in the Introduction, to obtain analytic results
one has to assume that one of the nuclei is dilute and expand in the
interactions with this dilute projectile order-by-order.  As we have
shown in this work, odd azimuthal harmonics appear only in the terms
contributing at least three interactions with the projectile to the
two-gluon production cross section. The part of this term giving the
odd harmonics was found above and is given by \eq{full_expr}.

We evaluated this odd-harmonics contribution to the two-gluon
production cross section analytically in the GBW model by expanding
the interaction with the target to the lowest non-trivial order (six
gluon exchanges). The result is given in \eq{full_result} and is
non-zero: hence the classical gluon fields do generate odd
harmonics. In Sec.~\ref{sec:num} we evaluate the same odd-harmonics
correlation function numerically in the full MV model, taking into
account the full interaction with the target. The resulting odd
harmonics coefficients are plotted in \fig{vodd} while the correlation
function is given by \fig{Codd}.

Both the analytic expression \eqref{full_result} and the correlation
function in \fig{Codd} appear to be dominated by the
$\delta$-function-like peaks at $\Delta \phi =0$ and $\Delta \phi =
\pi$. These peaks result from the so-called gluon HBT diagrams in the
notation of \cite{Kovchegov:2012nd}. We believe the dominance of these
peaks is responsible for the $v_3$ and $v_5$ in \fig{vodd} being so
close to each other in their values. To see this imagine a toy
two-particle distribution given by
\begin{align}\label{eq:toy}
  \frac{d N_{toy}}{d \Delta \phi} \sim A \, \delta (\Delta \phi) + B
  \, \delta (\Delta \phi - \pi)
\end{align}
with some coefficients $A$ and $B$. Clearly the expectation values of
$\langle \cos (n \, \Delta \phi) \rangle$ averaged with the
distribution \eqref{eq:toy} are independent of $n$ for even and odd
$n$ separately, implying that all $v_{2n+1}$ are equal to each other,
and all the $v_{2n}$ are equal to each other (but different from
$v_{2n+1}$). Something similar happens in \fig{vodd} due to the
dominance of the $\delta$-function contribution to the
correlator. Certainly, in the actual collisions, fragmentation will
turn gluons into hadrons, in the process broadening these
$\delta$-function peaks: while a detailed investigation of the
fragmentation effects on $v_n$'s is left for further work, one could
hope that part of the toy model mechanism suggested here remains,
contributing to the similarity of all $v_{2n+1}$ and of all $v_{2n}$
coefficients. (The importance of the $\Delta \phi =0$ and $\Delta \phi
= \pi$ peaks for $v_n$ values was also stressed in
\cite{Ray:2013fza}.)

Let us stress further that the gluon correlation function in
\fig{Codd} should not be compared directly to the data on the
di-hadron correlation function. At the very least we expect the
fragmentation functions to modify the shape of the correlator, most
probably broadening the $\delta$-function peaks. Note also that even
in the numerical part of this work, we expand the interaction with the
projectile nucleus to the lowest order needed for odd harmonics:
higher-order interactions with the projectile may need to be included
for the comparison with the experimental data. Unfortunately, at this
point, this appears possible to do only numerically. In addition,
further theoretical work should include the small-$x$ evolution
effects
\cite{Balitsky:1995ub,Balitsky:1998ya,Kovchegov:1999yj,Kovchegov:1999ua,Jalilian-Marian:1997dw,Jalilian-Marian:1997gr,Iancu:2001ad,Iancu:2000hn}
into the two-gluon production cross section. Only after all of the
above effects are included can one try comparing the resulting
correlation function to the experimental data.  Finally, we would like
to point out that while our MV-model power counting in the
calculations presented in this paper assumed a collision of two nuclei
(with one of them being more dilute than the other, $A_1 \ll A_2$),
the results of our calculations can also be applied to describe the
data on di-hadron correlations reported in high-multiplicity pp and pA
collisions (with all the caveats listed above) if the saturation
scales of both the target and the projectile are perturbatively large,
with the target saturation scale being larger than the projectile one.

%
\section*{Acknowledgments}
%

The authors are grateful to Larry McLerran for the interesting
discussions of the mechanism for generating odd harmonics originally
suggested in \myref\cite{McLerran:2016snu} and further developed in the
present paper.  YK would like to thank Giovanni Chirilli and Douglas
Wertepny for their collaboration on \myref\cite{Chirilli:2015tea}: that
work was essential for the derivation of the main result of this
paper.  The authors thank Kevin Dusling, Alex Kovner, Michael Lublinsky, Mark Mace,  Matthew Sievert and Raju
Venugopalan for illuminating discussions concerning various aspects of
this problem.  VS gratefully acknowledges the ExtreMe Matter Institute
EMMI (GSI Helmholtzzentrum f\"ur Schwerionenforschung, Darmstadt,
Germany) for partial support and hospitality.  This material is based
upon work supported by the U.S. Department of Energy, Office of
Science, Office of Nuclear Physics under Award Number DE-SC0004286
(YK).


\appendix 


\section{Odd harmonics in the classical field language}
\label{Sec:CFL}

The same conclusion as in Sec.~\ref{sec:disc} about the phase
difference between the leading and higher-order amplitudes being the
mechanism for generating the odd harmonics can be obtained from the
classical gluon field approach. One starts with the contribution of a
single gluon field to the LSZ formula: it can be cast as
\begin{align}\label{ampl}
\int d^4 x \, e^{i k \cdot x} \, \Box A_\mu = - \int d^4 x \, \pd_0 \left[ A_\mu \, \overset{\leftrightarrow}{\pd_0} \, e^{i k \cdot x}  \right] = \int d^3 x \, \left[ (\pd_0 - i E_k) \, A_\mu \right] \, e^{i k \cdot x} \Big|_{t \to + \infty} ,
\end{align}
where one assumes that $A_\mu =0$ at $t = - \infty$, as is the case
for the classical field, which diagrammatically can be constructed out
of retarded Green functions instead of Feynman propagators. The
produced two-gluon multiplicity in $A^+ =0$ light-cone gauge is
\begin{align}
\frac{d N}{d^2 k_1 \, d y_1 \, d^2 k_2 \, dy_2} & \sim \int d^3 z_1 \, d^3 w_1 \, d^3 z_2 \, d^3 w_2 \, e^{-i {\vec k}_1 \cdot ({\vec z}_1 - {\vec w}_1) - i {\vec k}_2 \cdot ({\vec z}_2 - {\vec w}_2) } \\ & \times \left\langle  (\pd_0 - i E_1) \, A_\mu^\perp (z_1) \,  (\pd_0 + i E_1) \, A^{\perp \, \mu} (w_1) \, (\pd_0 - i E_2) \, A_\nu^\perp (z_2) \,  (\pd_0 + i E_2) \, A^{\perp \, \nu} (w_2) \right\rangle\Bigg|_{z_1^0, w_1^0, z_2^0, w_2^0 \to + \infty}  , \notag 
\end{align} 
where we have employed the fact that the contributions of infinite
times cancel in the exponent. Next consider flipping ${\vec k}_1 \to -
{\vec k}_1$: this corresponds to ${\un k}_1 \to - {\un k}_1, \, y_1
\to - y_1$. However, since the classical particle production is
rapidity-independent, we conclude that the ${\vec k}_1 \to - {\vec
  k}_1$ substitution only affects the transverse momentum, ${\un k}_1
\to - {\un k}_1$, and, therefore, is the right substitution if one
searches for odd harmonics. Since ${\vec k}_1 \to - {\vec k}_1$ is
equivalent to ${\vec z}_1 \leftrightarrow {\vec w}_1$ (and ditto for
${\vec k}_2$) we conclude that the part of the cross section that may
give odd harmonics is
\begin{align}\label{2part_dist}
& \frac{d N_{\rm odd}}{d^2 k_1 \, d y_1 \, d^2 k_2 \, dy_2}  \sim - E_1 \, E_2 \, \int d^3 z_1 \, d^3 w_1 \, d^3 z_2 \, d^3 w_2 \, e^{-i {\vec k}_1 \cdot ({\vec z}_1 - {\vec w}_1) - i {\vec k}_2 \cdot ({\vec z}_2 - {\vec w}_2) } \\ & \times \left\langle  \left[ (\pd_0  A_\mu^\perp (z_1)) \,  A^{\perp \, \mu} (w_1) - A_\mu^\perp (z_1) \, (\pd_0  A^{\perp \, \mu} (w_1)) \right] \, \left[ (\pd_0  A_\nu^\perp (z_2)) \,  A^{\perp \, \nu} (w_2) - A_\nu^\perp (z_2) \, (\pd_0  A^{\perp \, \nu} (w_2)) \right] \,   \right\rangle\Bigg|_{z_1^0, w_1^0, z_2^0, w_2^0 \to + \infty}    \notag 
\end{align} 
(see a  similar discussion in \myref\cite{Lappi:2009xa}, Eqs. (3.11)-(3.14)). 
This equation can be  rewritten  by introducing 
real $B_\mu$ and $\theta$ according to 
\begin{align}\label{simple_phase}
 A_\mu (t, {\vec k}) = \int d^3 x \, e^{- i {\vec k} \cdot {\vec x}} \, A_\mu (x) \equiv B_\mu (t, |{\vec k}|) \, e^{i \, \theta (t, {\vec k})} .
\end{align} 
Reality of $A_\mu (x)$ implies that $\theta (t, - {\vec k}) = - \theta (t, {\vec k})$. Then \eq{2part_dist} becomes
\begin{align}\label{2part_dist5}
& \frac{d N_{odd}}{d^2 k_1 \, d y_1 \, d^2 k_2 \, dy_2}  \sim 4 E_1 \, E_2 \, [\pd_0 \theta (t, {\vec k}_1)] \, [\pd_0 \theta (t, {\vec k}_2)] \, B_\mu^\perp (t, |{\vec k}_1|) \,  B^{\perp \, \mu} (t, |{\vec k}_1|) \, \, B_\nu^\perp (t, |{\vec k}_2|) \,  B^{\perp \, \nu} (t, |{\vec k}_2|) \bigg|_{t \to +\infty} .
\end{align} 
Just as in Sec.~\ref{sec:disc}, we see that the odd harmonics are generated by the phase of the gluon field $A_\mu (t, {\vec k})$ taken in the mixed representation (Fourier-transformed into 3-momentum space, but also time-dependent). The exact relation between this phase and the phase difference between $M_1$ and $M_3$ in Sec.~\ref{sec:disc} is not clear at this point (see more on the relation between the phase difference and the classical gluon fields below).

For $\theta (t, - {\vec k}) = - \theta (t, {\vec k})$  to be true, one has to have $\theta (t, {\vec k}) \sim ({\un k} \cdot {\un b})^{2 m +1}$ where $\un b$ represents various vectors determining the positions of the valence quarks in the nuclei and $m \ge 0$ is an integer. In addition, the phase $\theta (t, {\vec k})$ has to be time-dependent.

Specifically for the problem at hand, it is convenient to use the
light cone coordinates; we thus return to \eq{ampl} and apply an
ultra-boost using
\begin{align}
 x^+ = e^\Delta \, x'^+, \ \ \ x^- = e^{-\Delta} \, x'^-  
\end{align}
with $\Delta \gg 1$. In $A^+ =0$ gauge the gauge condition is
preserved by the boost. We are interested in the transverse components
of the field. For those we get (after dropping the primes on the new
coordinates $x'^+, x'^-$)
\begin{align}\label{ampl_LC}
  \int d^4 x \, e^{i k \cdot x} \, \Box A^\perp_\mu = \int d^2 x_\perp
  \, d x^- \, \left[ (\pd_- - i k^+) \, A^\perp_\mu \right] \, e^{i k
    \cdot x} \Big|_{x^+ \to + \infty} .
\end{align}
Now  define
\begin{align}
  A_\mu^\perp (x^+, k^+, {\un k}) = \int d^2 x_\perp \, dx^- \, e^{i
    k^+ x^- - i {\un k} \cdot {\un x}} \, A_\mu^\perp (x).
\end{align}
Using this in \eq{ampl_LC} one gets
\begin{align}\label{ampl_LC2}
  \int d^4 x \, e^{i k \cdot x} \, \Box A^\perp_\mu = - 2 i k^+ \,
  e^{i k^- x^+} \, A^\perp_\mu (x^+, k^+, {\un k}) \Big|_{x^+ \to +
    \infty} .
\end{align}

With the help of \eq{ampl_LC2} and using the fact that the gluon field
is real,
\begin{align}\label{real_sym}
A_\mu^\perp (x^+, k^+, {\un k})^* = A_\mu^\perp (x^+, -k^+, -{\un k}),
\end{align}
we arrive at
\begin{align}\label{2part_dist6}
  & \frac{d N}{d^2 k_1 \, d y_1 \, d^2 k_2 \, dy_2} \sim (2 k_1^+)^2
  \, (2 k_2^+)^2 \, \left\langle A_\mu^\perp (x^+, k_1^+, {\un k_1})
    \, A^\mu_\perp (x^+, -k_1^+, -{\un k_1}) \, A_\nu^\perp (x^+,
    k_2^+, {\un k_2}) \, A^\nu_\perp (x^+, -k_2^+, -{\un k_2})
  \right\rangle \Big|_{x^+ \to + \infty} .
\end{align} 
Naively one can argue that since the two-particle distribution in the
classical approximation is rapidity-independent, nothing should depend
on $k_1^+$ and $k_2^+$.  We then seem to conclude that the
distribution is indeed ${\un k}_i \to - {\un k}_i$ symmetric. Then the
question arises: how could one get odd harmonics in the classical
approximation?

The resolution to this is that classical field may have a $\mbox{Sign}
(k^+)$ dependence, schematically
\begin{align}
	\label{Eq:sign}
        A_\mu^\perp (x^+, k^+, {\un k}) = A_\mu^{(1) \perp} (x^+, k^+,
        {\un k}) + \mbox{Sign} (k^+) \, A_\mu^{(2) \perp} (x^+, k^+,
        {\un k}),
\end{align}
such that
\begin{align}\label{Aconj}
  A_\mu^\perp (x^+, k^+, {\un k})^* = A_\mu^{(1) \perp} (x^+, - k^+, -
  {\un k}) - \mbox{Sign} (k^+) \, A_\mu^{(2) \perp} (x^+, - k^+, -
  {\un k}).
\end{align}
Here $A_\mu^{(1) \perp} (x^+, k^+, {\un k})$ and $A_\mu^{(2) \perp}
(x^+, k^+, {\un k})$ are assumed to be functions of $k^+$ without jump
discontinuities. Using these in \eq{2part_dist6}, we arrive at
\begin{align}\label{2part_dist8}
  \frac{d N_{\rm odd}}{d^2 k_1 \, d y_1 \, d^2 k_2 \, dy_2} \sim & (2
  k_1^+)^2 \, (2 k_2^+)^2 \ \mbox{Sign} (k_1^+) \, \mbox{Sign} (k_2^+)
  \\ & \times \, \left\langle \left[ A_\mu^{(1)\perp} (x^+, k_1^+,
      {\un k_1}) \, A^{(2) \mu}_\perp (x^+, -k_1^+, -{\un k_1}) -
      A_\mu^{(2)\perp} (x^+, k_1^+, {\un k_1}) \, A^{(1) \mu}_\perp
      (x^+, -k_1^+, -{\un k_1}) \right] \right. \notag \\ & \times \,
  \left. \left[ A_\nu^{(1) \perp} (x^+, k_2^+, {\un k_2}) \, A^{(2)
        \nu}_\perp (x^+, -k_2^+, -{\un k_2}) - A_\nu^{(2) \perp} (x^+,
      k_2^+, {\un k_2}) \, A^{(1) \nu}_\perp (x^+, -k_2^+, -{\un k_2})
    \right] \right\rangle \Big|_{x^+ \to + \infty} , \notag
\end{align} 
which is odd under ${\un k}_i \to - {\un k}_i$ if we assume that all
the $k_1^+, k_2^+$ dependence cancels everywhere in the expression
with the exception of the sign functions.

Additionally the representation in \eq{Eq:sign} is useful to
illustrate the phase difference discussed in
Sec.~\ref{sec:disc}. Performing the Fourier transformation into
transverse coordinate space, we rewrite
\begin{align}
	\label{Eq:signX}
        A_\mu^\perp (x^+, k^+, {\un x}) = A_\mu^{(1) \perp} (x^+, k^+,
        {\un x}) + \mbox{Sign} (k^+) \, A_\mu^{(2) \perp} (x^+, k^+,
        {\un x})\,.
\end{align}
The reality of the gluon field requires  
\begin{align}
  A_\mu^{*\perp} (x^+, k^+, {\un x}) = A_\mu^\perp (x^+, -k^+, {\un
    x})
\end{align}
and thus we get 
\begin{align}
  (A_\mu^{(1) \perp} (x^+, k^+, {\un x}))^* &= A_\mu^{(1) \perp} (x^+, -k^+, {\un x})\,, \\
  (A_\mu^{(2) \perp} (x^+, k^+, {\un x}))^* &= - A_\mu^{(2) \perp}
  (x^+, -k^+, {\un x}).
\end{align}
For classical fields we can neglect the $k^+$ dependence in
$A_\mu^{(1,2) \perp}$ and conclude that $A_\mu^{(2) \perp}$ is
imaginary while $A_\mu^{(1) \perp}$ is real; this results in the same
phase difference as between $M_1$ and $M_3$ in Sec.~\ref{sec:disc}.


\section{Color sums} 
\label{Ap:Color}
In order to compute the diagram A, the following color sums were used 
\begin{align}
	\delta^{\alpha \gamma} \delta^{\alpha'  \gamma'} \delta^{\beta \beta'}\,  C^{\alpha \beta \gamma \ \alpha' \beta' \gamma'} 
	&= \frac18 N_c^4 (N_c^2-1) , \\ 
	\delta^{\alpha  \gamma' } \delta ^{\alpha' \beta' } \delta ^{\beta \gamma }\,  C^{\alpha \beta \gamma \ \alpha' \beta' \gamma'} 
	& = 0 , \\ 
	\delta ^{\alpha  \beta' } \delta ^{\alpha' \gamma'} \delta ^{\beta \gamma }\,  C^{\alpha \beta \gamma \ \alpha' \beta' \gamma'}
	& = \frac18 N_c^4 (N_c^2-1), \\
	\delta ^{\alpha \gamma } \delta ^{{\alpha'}{\beta'}} \delta ^{\beta {\gamma'}}\, C^{\alpha \beta \gamma \ \alpha' \beta' \gamma'} 
	& = \frac14 N_c^4 (N_c^2-1), \\
	\delta ^{\alpha \gamma } \delta ^{\beta {\alpha'}} \delta ^{{\beta'}{\gamma'}}\, C^{\alpha \beta \gamma \ \alpha' \beta' \gamma'} 
	&= \frac14 N_c^4 (N_c^2-1) ,\\
	\delta ^{\alpha {\alpha'}} \delta ^{\beta \gamma } \delta ^{{\beta'}{\gamma'}}\, C^{\alpha \beta \gamma \ \alpha' \beta' \gamma'} 
	&= \frac18 N_c^4 (N_c^2-1), \\
	\delta ^{\alpha {\gamma'}} \delta ^{\gamma {\alpha'}} \delta ^{\beta {\beta'}}\, C^{\alpha \beta \gamma \ \alpha' \beta' \gamma'} 
	&= 0 ,\\ 
	\delta ^{\alpha {\beta'}} \delta ^{\gamma {\alpha'}} \delta ^{\beta {\gamma'}}\, C^{\alpha \beta \gamma \ \alpha' \beta' \gamma'} 
	&= -\frac14 N_c^4 (N_c^2-1), \\ 
	\delta ^{\alpha \beta } \delta ^{\gamma {\alpha'}} \delta ^{{\beta'}{\gamma'}}\, C^{\alpha \beta \gamma \ \alpha' \beta' \gamma'} 
	& = \frac14 N_c^4 (N_c^2-1) ,\\
	\delta ^{\alpha \beta } \delta ^{{\alpha'}{\gamma'}} \delta ^{\gamma {\beta'}}\, C^{\alpha \beta \gamma \ \alpha' \beta' \gamma'} 
	& = 0 ,\\ 
	\delta ^{\alpha {\gamma'}} \delta ^{\beta {\alpha'}} \delta ^{\gamma {\beta'}}\, C^{\alpha \beta \gamma \ \alpha' \beta' \gamma'} 
	& = 0, \\ 
	\delta ^{\alpha {\alpha'}} \delta ^{\beta {\gamma'}} \delta ^{\gamma {\beta'}}\, C^{\alpha \beta \gamma \ \alpha' \beta' \gamma'} 
	&= 0 ,\\ 
	\delta ^{\alpha \beta } \delta ^{{\alpha'}{\beta'}} \delta ^{\gamma {\gamma'}}\, C^{\alpha \beta \gamma \ \alpha' \beta' \gamma'} 
	&= -\frac14 N_c^4 (N_c^2-1), \\ 
	\delta ^{\alpha {\beta'}} \delta ^{\beta {\alpha'}} \delta ^{\gamma {\gamma'}}\, C^{\alpha \beta \gamma \ \alpha' \beta' \gamma'} 
	& = -\frac14 N_c^4 (N_c^2-1), \\ 
	\delta ^{\alpha {\alpha'}} \delta ^{\beta {\beta'}} \delta ^{\gamma {\gamma'}}\, C^{\alpha \beta \gamma \ \alpha' \beta' \gamma'} 
	& = -\frac18 N_c^4 (N_c^2-1),  
\end{align}
where 
\begin{equation}
	C^{\alpha \beta \gamma \ \alpha' \beta' \gamma'} 
	=
	f^{abc} f^{\gamma'  a d} f^{a' b' c'} f^{\gamma  a' d'} f^{\alpha b d} f^{\alpha' b'e} f^{\beta c e} f^{\beta' c' d'}.
\end{equation}
These color sums and those appearing in the diagrams B-C were
evaluated by using the definition
\begin{equation}
	f^{abc} = \frac{i}{2} t^a [t^b, t^c] 
\end{equation}
and the Fierz identity
\begin{equation}
  t^a_{ij} t^a_{kl}  = \frac12 \left( \delta^{il} \delta^{jk} - \frac{1}{N_c} \delta^{ij} \delta^{kl}  \right). 
\end{equation}


\section{Angular integrals}
\label{Ap:Int}

Here we list the angular integrals which were used to compute the
diagram B:
\begin{align}
	\int\frac{l^{i}}{({\un k} - {\un l})^{2}}\mathrm{Sign}({\un k}\times {\un l})d\varphi & 
	=\frac{2}{k^2}\ln\frac{q_{-}}{q_{+}}\epsilon_{im}k_{m}, \\
\int\frac{l^{i}}{({\un k}-{\un l})^{4}}\mathrm{Sign}({\un k} {\times} {\un l})d\varphi & 
=-\frac{4   l  }{(k^2-l^2)^{2}  k   }\epsilon_{im}k_{m}, \\
\int\frac{l^{i}l^{j}}{({\un k}-{\un l})^{2}}\mathrm{Sign}({\un k} \times  {\un l})d\varphi &
=-\frac{1}{k^4}\left(2k l+(k^2+l^2)\ln\frac{q_{-}}{q_{+}}\right)(k_{i}\epsilon_{jm}k_{m}+k_{j}\epsilon_{im}k_{m}), \\
\int\frac{l^{i}l^{j}}{({\un k}-{\un l})^{4}}\mathrm{Sign}({\un k}\times {\un l})d\varphi 
& =-\frac{1}{k^4}\left(\frac{2 k l  (k^2+l^2)}{(k^2-l^2)^{2}}+\ln\frac{q_{-}}{q_{+}}\right)(k_{i}\epsilon_{jm}k_{m}
+k_{j}\epsilon_{im}k_{m}), \\
\int\frac{l^{i}l^{j}l^{n}}{({\un k}-{\un l})^{4}}\mathrm{Sign}({\un k}\times {\un l})d\varphi & 
=-\frac{1}{k^{6}}\left(\frac{2 k l  (k^4+l^4)}{(k^2-l^2)^{2}}
+(k^2+l^2)\ln\frac{q_{-}}{q_{+}}\right)(k_{i}k_{j}\epsilon_{nm}k_{m}
+k_{j}k_{n}\epsilon_{im}k_{m}+k_{i}k_{n}\epsilon_{jm}k_{m})\notag\\
 & +\frac{1}{k^{6}}\left(2 k l +(k^2+l^2)\ln\frac{q_{-}}{q_{+}}\right)\epsilon_{im}k_{m}\epsilon_{jp}k_{p}\epsilon_{nr}k_{r}, 
 \end{align}
where $q_\pm = q_> \pm q_<$ and  $q_<={\rm min} (  k , l  )$ and $q_>={\rm max} (  k, l )$.


\section{Continuous charge density}
\label{Sec:Comparing}

The goal of this Appendix is to express the amplitudes in terms of the
continuous charge density and to prove the equivalence between this
calculation and the result of \myref\cite{McLerran:2016snu}.  Finally
we will also discuss which representation of the amplitudes is the
most convenient for numerical simulations.

\begin{figure}[ht]
\begin{center}
\includegraphics[width= 0.45 \textwidth]{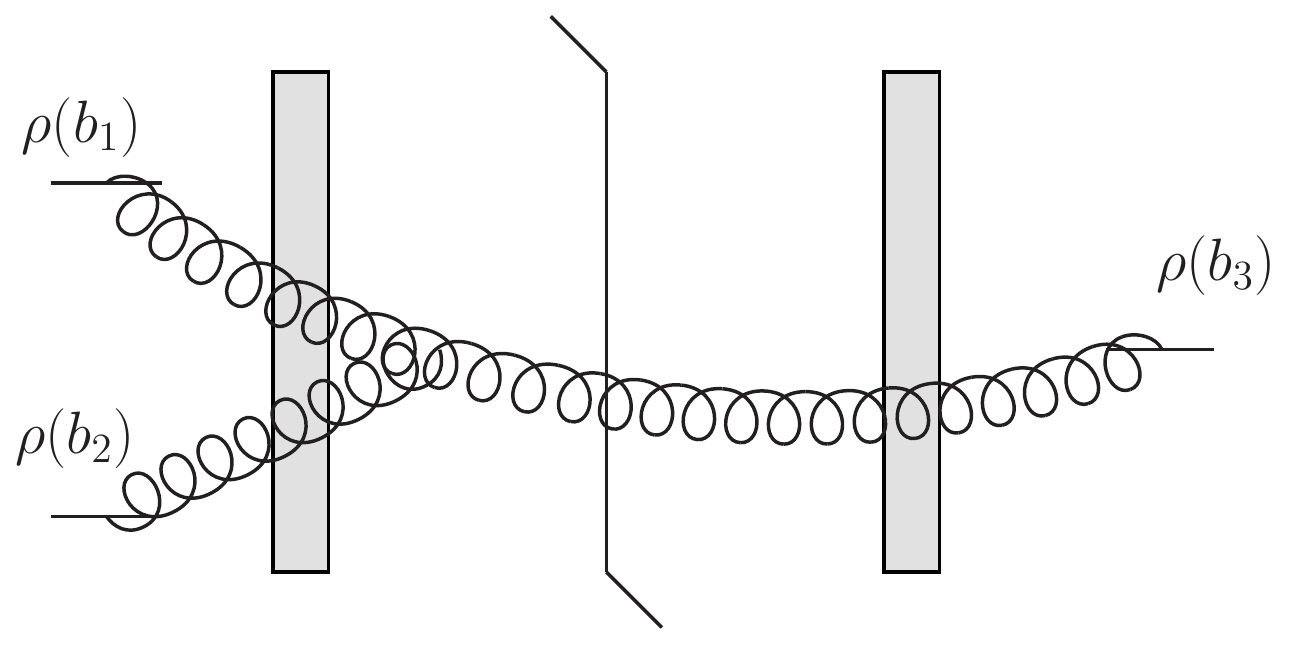} 
\caption{The key diagram for the numerical calculations of the odd
  correlation function. Together with its complex conjugate, it
  defines the configuration-by-configuration contribution to the
  operator in Eq.~\eqref{Eq:Di1}.  }
\label{all_graphs_num}
\end{center}
\end{figure}

Translating the amplitudes ${\un M}_1$ and ${\un M}_3$ to the language
of ``sources'' (continuous charge density) boils down to replacing
$V_{\un b} t^a \to \rho^a({\un b})$. Thus in momentum space we have
\begin{align}
  \label{eq:1gluon}
  {\un \epsilon}_\lambda^* \cdot  {\un M}_1 ({\un k}) &= 2 g \, 
  \int d^2 z
  e^{-i {\un k}\cdot  {\un z}} 	
  \int d^2 b
   \int \frac{d^2 q}{(2 \pi)^2} 
	\, e^{i {\un q} \cdot ({\un z} - {\un b})} \, \frac{{\un \epsilon}_\lambda^* \cdot
    {\un q}}{{q}^2} \, \left[ U^{ag}_{{\un z}} -
    U^{ag}_{{\un b}} \right] \, 
	\rho^g({\un b}) = \\
	& = 
 2 g \, \int \frac{d^2 q}{(2 \pi)^2} 
 {\un \epsilon}_\lambda^* \cdot \left(
 \frac{{\un q}}{{q}^2}  
 - \frac{{\un k}}{{k}^2}  
 \right)
 U^{a g} (\un{k}-\un{q}) \rho^g(\un{q})\notag
\end{align}
and 
\begin{align}
	{\un \epsilon}_\lambda^* \cdot {\un M}_3 ({\un k})  
	 \\ = \notag  & - 2 \, i \, g^3 
	\, \int d^2 x_1 \, d^2 x_2 \,
 d^2 b_1 \, d^2 b_2 \,
	\int
   \frac{d^2 l}{(2 \pi)^2} \, \frac{d^2
    q_1}{(2 \pi)^2} \, \frac{d^2 q_2}{(2 \pi)^2} \, 
	\\ & e^{i {\un
      q}_{1} \cdot ({\un x}_{1} - {\un b}_{1 }) + i
    {\un q}_{2} \cdot ({\un x}_{2} - {\un b}_{2}) +
    i {\un l}\cdot ({\un x}_{2} - {\un x}_{1}) +
    i {\un k}\cdot ({\un z}- {\un x}_{2 })}
  \notag \\ & \times \frac{1}{q_{1}^2 \, q_{2}^2} \, \left(
    - {\un q}_{1} \cdot {\un q}_{2} \, \frac{{\un
        \epsilon}_\lambda^* \times {\un
        k}}{k^2} + {\un \epsilon}_\lambda^*
    \cdot {\un q}_{1} \, \frac{{\un q}_{2} \times ({\un
	k}- {\un l})}{({\un k}- {\un l})^2} + {\un \epsilon}_\lambda^* \cdot
    {\un q}_{2} \, \frac{{\un q}_{1} \times {\un
        l}}{{l}^2} \right) \, \mbox{sign} ({\un
    k}\times {\un l}) \notag \\ & \times \, f^{abc} \,
  \left[ U^{bd}_{{\un x}_{1}} - U^{bd}_{{\un b}_{1}}
  \right] \, \left[ U^{ce}_{{\un x}_{2}} - U^{ce}_{{\un
        b}_{2}} \right] \,
\rho^d({\un b}_{1})
\rho^e({\un b}_{2}) = \notag \\  \notag
& =  
 -2 ig^2 \int \frac{d^2 l}{(2 \pi)^2} \, \frac{d^2
    q_1}{(2 \pi)^2} \, \frac{d^2 q_2}{(2 \pi)^2} \, 
	{\rm Sign}({\un k}\times \un{l}) f^{abc} 
	U^{bd}(\un{l}-\un{q}_1) 
	\rho^d(\un{q}_1) 
	U^{ce}(\un{k}-\un{l}-\un{q}_2)
	\rho^e(\un{q}_2)
	\times 
	\\& \notag
	\Bigg( 
	-  \frac{{\un
        \epsilon}_\lambda^* \times {\un
        k}}{k^2} 
		\left( 
		\frac{{\un q}_{1}}{q_{1}^2} 
		- \frac{{\un l}}{l^2} 
		\right)
		\cdot
		\left( 
		\frac{{\un q}_{2}}{q_{2}^2} 
		- \frac{{\un {k}}  - {\un {l}}}{|{\un {k}}  - {\un {l}}|^2} 
		\right) + {\un \epsilon}_\lambda^* \cdot  
		\left(
		\frac{{\un q}_{1}}{q_{1}^2}  
 		- \frac{{\un l}}{l^2}  
 		\right)
		\frac{ {\un q}_{2}  \times ({\un k}  - {\un l} )  }   {q_{2}^2 |{\un k}- {\un l}|^2}
\\ \notag
		&
		+ {\un \epsilon}_\lambda^* \cdot  
		\left(
		\frac{{\un q}_{2}}{q_{2}^2}  
		- \frac{{\un k}- {\un l}}{   |{\un k}- {\un l}|^2 }  
 		\right)
		\frac{ {\un q}_{1}  \times {\un l}   }   {q_{1}^2 l^2} 
	\Bigg)  \,.
 \end{align}

 The combination of interest, see Fig.~\ref{all_graphs_num}, 
\begin{align}
	\label{Eq:Di1}
	\frac12  
	{\un M}_3 ({\un k}) \cdot 
	{\un M}^*_1 ({\un k}) + {\rm c.c.} = &
	- 2 ig^4
	\int \frac{d^2 l}{(2 \pi)^2}
	 \frac{d^2 q}{(2 \pi)^2}
	\, 
	\frac{d^2
    q_1}{(2 \pi)^2} \, \frac{d^2 q_2}{(2 \pi)^2} \, 
	\\ & \notag  
	{\rm Sign}({\un k}\times {\un l}) f^{abc} 
	U^{bd}(\un{l}-\un{q}_1) 
	\rho^d(\un{q}_1) 
	U^{ce}(\un{k}-\un{l}-\un{q}_2)
	\rho^e(\un{q}_2)
	\left[  U^{a g} (\un{k}-\un{q}) \rho^g(\un{q})
	\right]^* 
	\times
	\\& \notag  
	\Bigg( 
	-  \frac{{\un q} \times {\un k}}{q^2 {\un k}^2} 
		\left( 
		\frac{{\un q}_{1}}{q_{1}^2} 
		- \frac{{\un l}}{l^2} 
		\right)
		\cdot
		\left( 
		\frac{{\un q}_{2}}{q_{2}^2} 
		- \frac{{\un {k}}  - {\un {l}}}{|{\un {k}}  - {\un {l}}|^2} 
		\right)\\
		& \notag
		+ 	\frac{ {\un q}_{2}  \times ({\un k}   - {\un l}  )  }   {q_{2}^2 |\un{k} - \un{l} |^2}
		\left( 
		\frac{{\un q}}{q^2} 
		- \frac{{\un k}}{{\un k}^2} 
		\right)
		\cdot  
		\left(
		\frac{{\un q}_{1}}{q_{1}^2}  
 		- \frac{\un{l}}{l^2}  
 		\right)
\\
		&\notag
		+ 
		\frac{ {\un q}_{1}  \times {\un l}    }   {q_{1}^2 l^2 }
			\left( 
		\frac{{\un q}}{q^2} 
		- \frac{{\un k}}{{\un k}^2} 
		\right)
		\cdot  
		\left(
		\frac{{\un q}_{2}}{q_{2}^2}  
		- \frac{\un{k} - \un{l}}{   |\un{k} - \un{l} |^2 }  
 		\right)
	\Bigg) + 
{\rm c.c.}  ,
\end{align}
defines the odd contribution to the functional  
\begin{align}
	\label{Eq:Di2}
	&\frac{E_k}{ 2} \left( \frac{d N (\un{k})}{d^3 k} [\rho_p, \rho_T]  
	-	\frac{d N (-\un{k})}{d^3 k}  [\rho_p, \rho_T]  
 \right) = \frac{1}{16 \pi^3}  
	{\un M}_3 ({\un k}) \cdot 
	{\un M}^*_1 ({\un k}) + {\rm c.c.} 
\end{align}


The calculations performed in \myref\cite{McLerran:2016snu} resulted in
\begin{align}
	\label{Eq:FinalAssymitry}
	& E_k  \frac{d N^{\rm odd} (\un{k})}{d^3 k} [\rho_p, \rho_T] 
    = 
	\frac{E_k}{ 2} \left( \frac{d N (\un{k})}{d^3 k}  [\rho_p, \rho_T]  
	-	\frac{d N (-\un{k})}{d^3 k} 
	 [\rho_p, \rho_T] 
	\right)
	= \notag \\   
    &=  
	{ \frac{1}{8\pi} }
	{\rm Im}
	\left\{
		\frac{2g}{\pi^2 {\un k}^2} 
		\int \frac{d^2 l}{(2\pi)^2} 
				\frac{  {\rm Sign}({\un{k}\times \un{l}}) }{l^2 |\un{k}-\un{l}|^2 } 
		f^{abc}
			\Omega^a_{ij} (\un{l}) 
			\Omega^b_{mn} (\un{k}-\un{l})
			\Omega^{c\star}_{rp} (\un{k})
		\notag
		\times \right.  \\ & \quad 
		\left.
		\left[
			\left( 
			{\un k}^2 \epsilon^{ij} \epsilon^{mn}
		-\un{l} \cdot (\un{k} - \un{l} ) 
		(\epsilon^{ij} \epsilon^{mn}+\delta^{ij} \delta^{mn}) 
		\right) \epsilon^{rp}+ 
		2 \un{k} \cdot (\un{k}-\un{l}) { \epsilon^{ij} \delta^{mn}} \delta^{rp}
		\right]
	\right\} ,
\end{align}
where 
\begin{equation}
	\Omega_{ij}^a(\un{k}) = 
	g \int \frac{d^2 p_\perp}{(2\pi)^2}
	\frac{p_{i} (k-p)_{j} }{p^2}
	\rho_b(\un{p}) U_{ab} (\un{k}-\un{p})\,. 
	\label{Eq:Omega}
\end{equation}
Using the definition \eq{Eq:Omega} we can rewrite  
\begin{align}
	\label{Eq:My}
	&
	\frac{E_k}{ 2} \left( \frac{d N (\un{k})}{d^3 k}  [\rho_p, \rho_T] 
	-	\frac{d N (-\un{k})}{d^3 k} 
	 [\rho_p, \rho_T] 
	\right)  = \\ 
	\notag &
	{\frac{1}{8\pi} } {\rm Im}
 \int \frac{d^2 l}{(2 \pi)^2} \,
  \frac{d^2 q}{(2 \pi)^2} \,
 \frac{d^2
    q_1}{(2 \pi)^2} \, \frac{d^2 q_2}{(2 \pi)^2} \,
	\frac{2 g^4 	{\rm Sign} ( \un{k} \times \un{l}) 
} { \pi^2 k^2 l^2 |\un{k}-\un{l}| ^2} 
	f^{abc} 
	U^{ad} (\un{l} - \un{q}_1) \rho^d(\un{q}_1)
	U^{be} (\un{k} - \un{l} - \un{q}_2) \rho^e(\un{q}_2)
	\left[ U^{cg} (\un{k} - \un{q}) \rho^g(\un{q}) \right]^* 
	\times \\\notag &
	\Bigg[
		(k^2 - \un{k} \cdot \un{l} + l^2)  
		\frac{\un{q}\times \un{k} }{q^2} 
		\frac{ \un{q}_1\times \un{l} } {q_1^2}
		\frac{ \un{q}_2\times (\un{k}-\un{l}) } {q_2^2}
- 
\frac{\un{q}\times \un{k} }{q^2} 
\frac{ \un{q}_1\cdot (\un{l}-\un{q}_1) } {q_1^2}
\frac{ \un{q}_2\cdot (\un{k}-\un{l}-\un{q}_2) } {q_2^2}
\un{l}\cdot (\un{k}- \un{l})
\\\notag &+ 
\frac{\un{q}_1\times \un{l} }{q_1^2} 
\frac{ \un{q}_2\cdot (\un{k}-\un{l}-\un{q}_2) } {q_2^2}
\frac{ \un{q}\cdot (\un{k}-\un{q}) } {q^2}
\un{k}\cdot (\un{k}- \un{l})
+ 
\frac{\un{q}_2\times (\un{k}-\un{l}) }{q_2^2} 
\frac{ \un{q}_1\cdot (\un{l}-\un{q}_1) } {q_1^2}
\frac{ \un{q}\cdot (\un{k}-\un{q}) } {q^2} 
\un{k}\cdot \un{l}
	\Bigg] \,,
\end{align}
where the last two terms originate from the last term in
Eq.~\eqref{Eq:FinalAssymitry} and the symmetry $q_1\leftrightarrow
q_2$ and $l\to k-l$:
\begin{align}
	&
 \int \frac{d^2 l}{(2 \pi)^2} \,
  \frac{d^2 q}{(2 \pi)^2} \,
 \frac{d^2
    q_1}{(2 \pi)^2} \, \frac{d^2 q_2}{(2 \pi)^2} 
	\frac{{\rm Sign} (\un{k} \times \un{l}) 
} { \pi^2 k^2 l^2 |\un{k}-\un{l}| ^2} 
	f^{abc} 
	U^{ad} (\un{l} - \un{q}_1) \rho^d(\un{q}_1)
	U^{be} (\un{k} - \un{l} - \un{q}_2) \rho^e(\un{q}_2)
	\left[ U^{cg} (\un{k} - \un{q}) \rho^g(\un{q}) \right]^* 
	\\
	&\times \Bigg[
		\frac{\un{q}_1\times \un{l} }{q_1^2} 
		\frac{ \un{q}_2\cdot (\un{k}-\un{l}-\un{q}_2) } {q_2^2}
		\frac{ \un{q}\cdot (\un{k}-\un{q}) } {q^2}
		\un{k}\cdot (\un{k}- \un{l})
	-
	\frac{\un{q}_2\times (\un{k}-\un{l}) }{q_2^2} 
	\frac{ \un{q}_1\cdot (\un{l}-\un{q}_1) } {q_1^2}
	\frac{ \un{q}\cdot (\un{k}-\un{q}) } {q^2} 
	\un{k}\cdot \un{l} 
\Bigg]  =0  \notag \, . 
\end{align}

\subsection{Equivalence} 

In order to prove the equivalence between Eqs. \eqref{Eq:Di1},
\eqref{Eq:Di2} and \eq{Eq:My} we have to prove the following identity
\begin{align}
	\label{Eq:Id}
	&
	-  \frac{{\un q} \times {\un k}}{q^2 { k}^2} 
		\left( 
		\frac{{\un q}_{1}}{q_{1}^2} 
		- \frac{{\un l}}{l^2} 
		\right)
		\cdot
		\left( 
		\frac{{\un q}_{2}}{q_{2}^2} 
		- \frac{{\un {k}}  - {\un {l}}}{|{\un {k}}  - {\un {l}}|^2} 
		\right)
		+ 	\frac{ {\un q}_{2}  \times ({\un k}   - {\un l}  )  }   {q_{2}^2 |\un{k} - \un{l} |^2}
		\left( 
		\frac{{\un q}}{q^2} 
		- \frac{{\un k}}{{\un k}^2} 
		\right)
		\cdot  
		\left(
		\frac{{\un q}_{1}}{q_{1}^2}  
 		- \frac{\un{l}}{l^2}  
 		\right)
\\ \notag 
		&
		+ 
		\frac{ {\un q}_{1}  \times {\un l}    }   {q_{1,}^2 l^2 }
			\left( 
		\frac{{\un q}}{q^2} 
		- \frac{{\un k}}{{\un k}^2} 
		\right)
		\cdot  
		\left(
		\frac{{\un q}_{2}}{q_{2}^2}  
		- \frac{\un{k} - \un{l}}{   |\un{k} - \un{l} |^2 }  
 		\right)
= \\ \notag  &	
\frac{1}{{k}^2 |\un{k}-\un{l}|^2 \un{l}^2 } 
\Bigg(	
(k^2 - \un{k} \cdot \un{l} + l^2)  
\frac{\un{q}\times \un{k} }{q^2} 
\frac{ \un{q}_1\times \un{l} } {q_1^2}
\frac{ \un{q}_2\times (\un{k}-\un{l}) } {q_2^2}
	- 
	\frac{\un{q}\times \un{k} }{q^2} 
	\frac{ \un{q}_1\cdot (\un{l}-\un{q}_1) } {q_1^2}
	\frac{ \un{q}_2\cdot (\un{k}-\un{l}-\un{q}_2) } {q_2^2}
	\un{l}\cdot (\un{k}- \un{l})
\\ \notag &+ 
\frac{\un{q}_1\times \un{l} }{q_1^2} 
\frac{ \un{q}_2\cdot (\un{k}-\un{l}-\un{q}_2) } {q_2^2}
\frac{ \un{q}\cdot (\un{k}-\un{q}) } {q^2}
\un{k}\cdot (\un{k}- \un{l})
	+ 
	\frac{\un{q}_2\times (\un{k}-\un{l}) }{q_2^2} 
	\frac{ \un{q}_1\cdot (\un{l}-\un{q}_1) } {q_1^2}
	\frac{ \un{q}\cdot (\un{k}-\un{q}) } {q^2} 
	\un{k}\cdot \un{l}
\Bigg)\,.	
\end{align}
First define 
\begin{equation}
	A(\un{q}_1, \un{l} , \un{q}_2, \un{p}) 
	 = 
	\left( 
		\frac{{\un q}_{1}}{q_{1}^2} 
		- \frac{{\un l}}{l_{\perp}^2} 
		\right)
		\cdot  
		\left(
		\frac{{\un q}_{2}}{q_{2}^2}  
 		- \frac{\un{p}}{p^2}  
 		\right)
\end{equation}
so that the left hand side of Eq.~\eqref{Eq:Id} is 
\begin{equation}
	\label{Eq:LH}
		-  \frac{{\un q} \times {\un k}}{q^2 {k}^2} 
		A(\un{q}_1,\un{l},\un{q}_2,\un{k}-\un{l})
		+ 	\frac{ {\un q}_{2}  \times ({\un k}   - {\un l}  )  }   {q_{2}^2 |\un{k} - \un{l} |^2}
		A(\un{q},\un{k},\un{q}_1,\un{l})
		+ 
		\frac{ {\un q}_{1}  \times {\un l}    }   {q_{1}^2 l^2 }
		A(\un{q},\un{k},\un{q}_2,\un{k}-\un{l})\, .
\end{equation}
Using the following identities 
\begin{align}
	{\un q}_{1} \cdot \un{p} &= 
	\frac{1}{l^2} \left( 
	- {\un q}_{1} \times \un{l} \ \un{l} \times \un{p} 
	+ {\un q}_{1} \cdot \un{l} \ \un{l} \cdot \un{p} 
\right)\,,\\
{\un q}_{2} \cdot \un{l} &= 
	\frac{1}{p^2} \left( 
	- {\un q}_{2} \times \un{p} \ \un{p} \times \un{l} 
	+ {\un q}_{2} \cdot \un{p} \ \un{l} \cdot \un{p} 
\right)\,,
\end{align}
which can be derived starting  from 
\begin{equation}
	(\un{a}\times \un{b}) (\un{c}\times \un{d}) = (\un{a}\cdot \un{c}) (\un{b}\cdot \un{d})  -  (\un{a}\cdot \un{d}) (\un{b}\cdot \un{c}), 
\end{equation}
and 
\begin{align}
	{\un q}_{1} \cdot \un{l}\  
	{\un q}_{2} \cdot \un{p} \ 
	\un{l} \cdot \un{p}  = 
	{\un q}_{1} \cdot {\un q}_{2} \, p^2 l^2
	&+ \frac12 	{\un q}_{1} \times   \un{l}
	\left(  
	{\un q}_{2} \cdot  \un{p} 
	\  \un{l} \times  \un{p} 
 - 
	{\un q}_{2} \times  \un{l} \ p^2
	\right) 
	\\
	&- \frac12 	{\un q}_{2} \times   \un{p}
	\left(  
	{\un q}_{1} \cdot  \un{l} 
	 \ \un{l} \times  \un{p} 
 + 
	{\un q}_{1} \times  \un{l}\  p^2
	\right)\notag \,, 
\end{align}
we get 
\begin{align}
	A(\un{q}_1, \un{l} , \un{q}_2, \un{p}) 
	& =
	\frac{ \un{l} \cdot   \un{p}\
	{\un q}_{1}\cdot ( 	\un{l} - 	{\un q}_{1}  )
	\ {\un q}_{2}\cdot ( 	\un{p} - 	{\un q}_{2}  )
	} { q_{1}^2  q_{2}^2 l^2 p^2}
-
	\frac{ {\un q}_{2} \times   \un{p}
	\ \un{l} \times  \un{p} } {q_{2}^2 l^2 p^2}
	+
	\frac{ {\un q}_{1} \times   \un{l} \
	\un{l} \times  \un{p} } {q_{1}^2 l^2 p^2} 
	\\ \notag & - 
	\frac{1}{2} 
	\left(
	\frac{ {\un q}_{1} \times   \un{l} \
	{\un q}_{2} \cdot   \un{p} \
	\un{l} \times   \un{p}
    }  {   q_{1}^2  q_{2}^2 l^2 p^2} 
-
	\frac{ {\un q}_{2} \times   \un{p} \
	{\un q}_{1} \cdot   \un{l}\ 
	 \un{l} \times   \un{p}
    }  {   q_{1}^2  q_{2}^2 l^2 p^2} 
	\right)
	\\ \notag & + 
	\frac{1}{2} 
	\left(
	\frac{ 
		{\un q}_{1} \times   \un{l} \
		{\un q}_{2} \times   \un{l}
    }  {   q_{1}^2  q_{2}^2 l^2 } 
+
	\frac{ 
		{\un q}_{1} \times   \un{p}\ 
		{\un q}_{2} \times   \un{p}
    }  {   q_{1}^2  q_{2}^2 p^2 } 
	\right)\,. 
\end{align}

Substituting $A$ into \eq{Eq:LH} one recovers Eq.~\eqref{Eq:Id} as
most of the terms cancel.

This concludes the proof of the identity between the result of this
paper and \myref\cite{McLerran:2016snu}.

\subsection{Representation for numerics} 

We finish this Appendix with a short discussion on what representation
of the functionals is the most convenient for numerical
implementations. We also define the functionals we used in the actual
numerical calculation.

To the leading order, the functional describing gluon production is
given by
\begin{equation}
  E_k \frac{d N}{d^3k} [\rho_p, \rho_T] = \frac{1}{2(2\pi)^3} \, {\un M}_1 ({\un k}) \cdot  {\un M}^*_1 ({\un k})
\end{equation}
or using the explicit form for the amplitude from \eq{eq:1gluon}
\begin{equation}
		\label{Eq:functSI}
	E_k \frac{d N}{d^3k} [\rho_p, \rho_T] = \frac{2}{(2\pi)^3} 
	g^2 
	\int \frac{d^2 q}{(2 \pi)^2} 
	\frac{d^2 q'}{(2 \pi)^2} 
\left(
 \frac{{\un q}}{{\un q}^2}  
 - \frac{{\un k}}{{\un k}^2}  
 \right)
\cdot 
 \left(
 \frac{{\un q}'}{ { {\un q}' } ^2}  
 - \frac{{\un k}}{{\un k}^2}  
 \right)
 \rho^{* a}(\un{q}')
 [ U^\dagger (\un{k}-\un{q}')  U(\un{k}-\un{q})  ]^{a b}  \rho^{b}({\un q})\, . 
\end{equation}
This equation contains two momentum integrals which are not obviously
factorizable and therefore numerically challenging.  As was shown in
\myref\cite{McLerran:2016snu}, using the identity
\begin{equation}
	\left(
 \frac{{\un q}}{{\un q}^2}  
 - \frac{{\un k}}{{\un k}^2}  
 \right)
\cdot 
 \left(
 \frac{{\un q}'}{ { {\un q}' } ^2}  
 - \frac{{\un k}}{{\un k}^2}  
 \right)
 = \frac{ \delta_{ij} \delta_{lm}  +  \epsilon_{ij} \epsilon_{lm} }{k^2} 
 \frac{q_i (k-q)_j }{q^2} 
 \frac{q'_i (k-q')_j }{ {q'}^2}\,. 
\end{equation}
Equation \eqref{Eq:functSI} can be rewritten as follows
\begin{equation}
	\label{Eq:SIO}
		E_k \frac{d N}{d^3k} [\rho_p, \rho_T] =  \frac{2}{(2\pi)^3} 
		\frac{ \delta_{ij} \delta_{lm}  +  \epsilon_{ij} \epsilon_{lm} }{k^2} 
		\Omega^a_{ij} ({\un k})
		\left( \Omega^a_{lm} ({\un k}) \right)^*\,,
\end{equation}
where $\Omega^a_{ij} ({\un k})$ was defined in \eq{Eq:Omega}. In the
transverse coordinate space,
\begin{equation}
  \Omega^a_{ij} ({\un x}) = g^2 \frac{\partial_i}{\partial^2_\perp} \rho^b(\un{x}) 
	\partial_j U^{ab} (\un{x})\,.
\end{equation}
The advantage of the representation \eqref{Eq:SIO} is that the
momentum integrals are factorized explicitly and can be numerically
computed by performing a Fast Fourier transform of
$\Omega^a_{ij}(\un{x})$. This requires $(N_c^2-1)\times2\times N \log
N$ operations~\footnote{Note that we need to compute the Fourier
  transform only for the combinations $\delta_{ij}
  \Omega^a_{ij}(\un{x})$ and $\epsilon_{ij} \Omega^a_{ij}(\un{x})$,
  and not for all four spatial components of $\Omega^a_{ij}(\un{x})$.
}, where $N$ is the number of the lattice sites.  While direct
numerical implementation of \eq{Eq:functSI} would require $N^4$
operations and thus is prohibitively computationally expensive.

The same logic, but with greater effect, applies to
Eqs. \eqref{Eq:Di1}, \eqref{Eq:Di2} and \eq{Eq:FinalAssymitry} because
Eq.~\eqref{Eq:Di1} involves four two-dimensional integrals! We thus
opt to numerically compute its alternative but fully equivalent form
given by \eq{Eq:FinalAssymitry}.


\bibliography{v3,references}


\end{document}